\documentclass[useAMS,usenatbib,preprint]{elsarticle}
\usepackage{pifont}
\usepackage{amssymb}
\usepackage{amsmath}
\usepackage{natbib} 
\usepackage{geometry} 
\usepackage{graphicx}
\usepackage{color}
\usepackage{epsfig}
\usepackage{bm}
\usepackage{dcolumn}
\usepackage{journals}
\newcommand{\atlas}{ATLAS$^{\rm 3D}$}
\newcommand{\etal}{et~al.~}

\newcommand{\be}{\begin{equation}} 
\newcommand{\ee}{\end{equation}} 
\newcommand{\kms}{\ifmmode\,{\rm km}\,{\rm s}^{-1}\else km$\,$s$^{-1}$\fi} 
\newcommand{\magarc}{\ifmmode {{{{\rm mag}~{\rm arcsec}}^{-2}}} 
             \else {{{mag}$~${arcsec}$^{-2}$}} 
             \fi} 
\newcommand\ltsima{$\; \buildrel < \over \sim \;$}
\newcommand\ltsim{\lower.5ex\hbox{\ltsima}}
\newcommand\gtsima{$\; \buildrel > \over \sim \;$}
\newcommand\gtsim{\lower.5ex\hbox{\gtsima}}

\newcommand{\ml}{$M^*$/$L$}
\newcommand{\mlb}{$M$/$L_B$}

\newcommand{\br}{{\rm ($B$--$R$)}}
\newcommand{\ik}{{\rm ($I$--$K$)}}
\newcommand{\ha}{{\rm H$\alpha$ }}

\newcommand{\HdA}{{\rm H$\delta_{A}$}}

\newcommand{\msun}{\textrm{M}$_\odot$} 
 
\newcommand{\Msol}{\msun}

\def\mseld{M$_{*,{\rm Ein}}^{\rm LD}$} 
\def\msesed{M$_{*,{\rm Ein}}^{\rm SPS}$} 
\newcommand{\mre}{M$_{r_{\rm e}/2}$} 

\def\Mstar{M_*}

\def\msun{{M_\odot}}
\def\re{r_e}
\def\mre{\Mstar/\re}
\def\omit#1{}

\def\ltsima{$\; \buildrel < \over \sim \;$}                     
\def\lsim{\lower.5ex\hbox{\ltsima}}   
\def\gtsima{$\; \buildrel > \over \sim \;$}                     
\def\gsim{\lower.5ex\hbox{\gtsima}}  

\def \littleprime{\ifmmode{\scriptscriptstyle \prime } 
     \else{\hbox{$\scriptscriptstyle \prime$ }}\fi} 
\def \arcmin{\raise .9ex \hbox{\littleprime}} 
\def \arcsec{\raise .9ex \hbox{\littleprime\hskip-3pt\littleprime}}

\def \ion#1#2{#1{\footnotesize{#2}}\relax}   
 
\def \ha       {H$\alpha$} 
\def \hbeta    {H$\beta$}

\def \hi       {\ion{H}{I}} 
\def \hii      {\ion{H}{II}}

\def \nii      {[\ion{N}{II}]} 
\def \oii      {[\ion{O}{II}]} 
\def \oiii     {[\ion{O}{III}]} 
\def \sii      {[\ion{S}{II}]} 
\def \Vobs     {V_{\rm obs}}
\def \Vlos     {V_{\rm los}}

\def\Eq#1{Eq.~(\ref{eq:#1})} 
\def\eq#1{eq.~(\ref{eq:#1})} 
\def\Fig#1{Figure~\ref{fig:#1}}
\def\Table#1{Table~\ref{tab:#1}}
\def\se#1{\S\ref{sec:#1}}

\begin{document}
\date{\today}

\title{Galaxy Masses}

\author[Queens]{St\'ephane Courteau}
\address[Queens]{Queen's University, Department of Physics, Engineering
  Physics and Astronomy, Kingston, Ontario, Canada}

\author[Oxford]{Michele Cappellari}
\address[Oxford]{
Sub-department of Astrophysics, Department of Physics, University of
Oxford, Denys Wilkinson Building, Keble Road, Oxford OX1 3RH, UK}

\author[AIP]{Roelof S. de~Jong}
\address[AIP]{Leibniz-Institut f\"{u}r Astrophysik Potsdam (AIP),
  An der Sternwarte 16, 14482 Potsdam, Germany}

\author[MPIA]{Aaron A. Dutton}
\address[MPIA]{Max-Planck-Institut f\"{u}r Astronomie, K\"{o}nigstuhl 17,
  69117 Heidelberg, Germany} 

\author[ESO]{Eric Emsellem}
\address[ESO]{European Southern Observatory, 
  Karl-Schwarzschild-Strasse 2, 85748, Germany and
  Universit\'e Lyon 1, Observatoire de Lyon, Centre de Recherche
  Astrophysique de Lyon and Ecole Normale Sup\'erieure de Lyon, 9 avenue
  Charles Andr\'e, F-69230 Saint-Genis Laval, France}

\author[Leiden]{Henk Hoekstra}
\address[Leiden]{Leiden Observatory, Leiden University, P.O. Box 9513,
 NL-2300 RA Leiden, The Netherlands}

\author[Groningen]{L.V.E. Koopmans}
\address[Groningen]{University of Groningen, Kapteyn Astronomical Institute, 
  P.O.Box 800, 9700 AV, Groningen, The Netherlands}

\author[IAP]{Gary A. Mamon}
\address[IAP]{Institut d'Astrophysique de Paris
 (UMR 7095: CNRS \& UPMC), 98 bis Bd Arago,
  F-75014 Paris, France}

\author[ICG]{Claudia Maraston}
\address[ICG]{University of Portsmouth, Institute of Cosmology
 and Gravitation, Dennis Sciama Building, Burnaby Road, Portsmouth, UK}

\author[UCSB]{Tommaso Treu}
\address[UCSB]{University of California, Santa Barbara, 
Department of Physics, Santa Barbara, CA}

\author[Queens]{Lawrence M. Widrow}

\begin{abstract}
Galaxy masses play a fundamental role in our understanding of structure
formation models.  This review addresses the variety and reliability
of mass estimators that pertain to stars, gas, and dark matter.  The
different sections on masses from stellar populations, dynamical
masses of gas-rich and gas-poor galaxies, with some attention paid
to our Milky Way, and masses from weak and strong lensing methods,
all provide review material on galaxy masses in a self-consistent manner. 
\end{abstract}


\maketitle
\tableofcontents
 
\newpage


\section{Introduction}\label{sec:intro}

The distribution of matter in cosmological structures is a fundamental
property of nature as the mass of a system is likely the  major driver
of its evolution.
This is especially true for stars whose evolution depend almost fully
on their initial mass (and chemical composition) on the main sequence,
as embodied by the (idealistic) Vogt-Russell theorem.  Mass also plays
a fundamental role in galaxy evolution.  Galaxies have largely been
shaped through mergers and galaxy interactions in hierarchical fashion
whereby small systems merged into bigger ones.  At early times, star
formation was most effective in massive galaxies but as the Universe
aged, star formation was likely quenched in those massive systems but
continued in smaller galaxies, a phenomenon now called ``downsizing''.
Oldest stars are thus found in the most massive systems.  The complex
interplay between star formation efficiency and quenching is likely
modulated by a galaxy's total mass. 

Measurements of the distribution of matter in the Universe enable 
a variety of tests of structure formation models on different scales. 
For instance, the distribution of galaxy masses on all scales enables
the closest possible, though not direct, comparison of predicted mass
functions for baryonic and non-baryonic matter in the Universe.  The
relative fraction of baryonic to non-baryonic matter is also indicative
of fundamental, yet poorly understood, processes in galaxy formation which
typically give rise to tight scaling relations based on the stellar
and dynamical masses of galaxies.

Because galaxy masses play such a critical role in our understanding
of the formation and evolution of cosmic structures, we wish to review
the variety and reliability of mass estimators for gas-poor and gas-rich
galaxies and discuss our ability to derive from those estimators meaningful
constraints of theoretical galaxy formation models.  While certain techniques
enable only the measurement of galaxy masses on large scales, others allow
the decomposition of individual mass components such as gas, stars and dark
matter at different galactocentric radii.  The latter methods probe the
gravitational potential through the dynamics of
visible tracers where baryons are (sub-)dominant.  Although many galaxies
may be safely assumed to be virialized, uncertainties in their mass
estimates remain, for instance due to anisotropies in the velocity
distributions.  Furthermore, baryon-dominated regions remain poorly
understood, which complicates a direct comparison of galaxy formation
models to observational data.  

Many techniques exist for the determination of galaxy masses.  The most
popular involves the measurement of Doppler shifts of nebular and/or
stellar atomic lines due to internal dynamics.  Stellar motions can also
be resolved in the closest galaxies, such as our The Milky Way, Andromeda,
and other Local Group stellar systems; galaxy masses of more distant
systems otherwise rely on integrated spectra. 
Another mass estimator consists of converting the galaxy light profile
into a mass profile using a suitable stellar mass-to-light ratio
(usually derived from stellar population models).  A more global
approach has also involved the mapping of gravitational lensing
effects, both strong and weak.  This list is not meant to be complete,
as we review below.  However, in all cases, galaxy mass estimates
account for matter encompassed within a specified radius and are thus
always a lower limit to the total galaxy mass.

This review has evolved from discussions which took place during the
celebrations of Vera Rubin's career at Queen's University in June
2009\footnote{See {\tt http://www.astro.queensu.ca/GalaxyMasses09}
for workshop presentations and photographs.}.  All the authors of
this review were indeed present at that conference.  While each
section of this review was initially written by separate teammates,
the final product reflects the full team's imprimatur.  This review
was inspired by, and is meant as a modern revision of, early
treatises on the masses and mass-to-light ratios of galaxies
by \cite{Burbidge75} and \cite{Faber79}, respectively.

The review is organized as follows: we first present in \se{stellpop}
the central topic of stellar $M/L$ determinations from stellar population
models.  This is followed by a discussion of the mass estimates
for gas-rich galaxies in \se{gasrich}, including the special (resolved)
case of the Milky Way in \se{milkyway}.  Gas-poor galaxies are addressed
in \se{gaspoor} and weak and strong lensing techniques are presented
in \se{weak} and \se{strong}, respectively.  Conclusions, with a view
towards future developments, are presented at the end of each section.

This review is naturally incomplete; conspicuously missing topics
include the measurement of stellar and dynamical masses of high
redshift galaxies \citep[e.g.][]{Forster06,Bezanson11,Alaghband12},
the direct comparison of stellar and dynamical mass estimates
\citep[e.g.][]{deJong07,Taylor10}, mass function determinations
\citep[e.g. stellar mass functions:][]{bundy06,pozzetti10,maraston13}
\citep[e.g. dynamical mass functions:][]{TrujilloGomez11,Papastergis11,
Papastergis12}, constraints on halo masses by statistical
techniques such as those involving satellite kinematics
\citep{More11,WM13}, group catalogs \citep{Yang09}, and
abundance matching \citep{behroozi12}, to name a few. 

Furthermore, this review is restricted to mass analyses based
on Newtonian dynamics.  Alternatives exist, the most popular
being MOND \citep[e.g.][]{milgrom83}, but a proper treatment
of them is beyond the scope of this review.  Readers interested
in alternative models, MOND or others, are referred to the review
by \cite{Famaey12}.


\newpage

\section{From light to mass: modelling the stellar \ml\ ratio}
\label{sec:stellpop}
\subsection{Modelling galaxies and their Stellar Populations, 
a Historical Introduction.}\label{sec:stellpopintro}
The stellar mass $M^{*}$ of a galaxy is a key physical parameter
of galaxy formation and evolution studies as it traces the galaxy
formation process.  The stellar mass of a galaxy grows through processes
such as the internal conversion of gas and dust into stars via star
formation, or external events like major interactions with other
galaxies and subsequent merging which may induce further
star formation, as well as minor events such as accretion of
satellites.  Moreover, knowledge of the galaxy
stellar mass is crucial in order to decompose the contributions
from stars and dark matter to the dynamics of galaxies. 
Modern galaxy formation models embedded in a $\Lambda$-Cold Dark Matter
universe can also predict the evolution of the
galaxy mass assembly over cosmic time \citep[e.g.][]{delucia07}. 

Galaxies shine because their stars radiate the energy they produced
via nuclear reactions in their cores.  The theory of stellar evolution
describes the amount of energy released by a star given its initial
mass. Hence, by modelling the light emitted by all the stars 
in a galaxy over all wavelengths - the so-called ``integrated spectral energy
distribution (SED)'' - one can in principle derive the stellar mass that is
responsible for such radiation.  However, a certain fraction of evolved stars 
no longer shine yet still contribute to the galaxy mass budget in the
form of stellar remnants such as white dwarfs, neutron stars and black
holes.  The sum of {\it living} stars plus remnants makes up the 
``stellar mass'', $M^{*}$, of a galaxy. 

Despite our detailed knowledge of stellar evolution, the modelling
of a galaxy spectrum - which is the superposition of all spectra from individual
stars - is a challenging exercise since the exact stellar composition
of a galaxy and its overall stellar generations are unknown {\it a priori}.
These depend on the history of star formation, chemical enrichment, accretion
and interaction.  Unlike stellar clusters whose vast majority of stars are
coeval and share the same chemical composition, galaxies are a complicated
ensemble of stellar generations. Recent extensive reviews of SED modelling
of galaxies have appeared in \citet{walcher11} and \citet{ConroyARAA13}.

As recognised early on by \cite{Oort26} and \cite{Baade44}, our own Milky
Way is composed of various populations of stars, each featuring different
dynamics, chemical properties, and formation epochs. Thus, from a stellar
content viewpoint, galaxies can be broken into {\it stellar populations}
with shared definable properties. The ``simple stellar population'' (SSP)
is defined as a group of coeval stars with homogeneous chemistry (at birth)
and similar orbits/kinematics. A recent, comprehensive text book on
stellar populations in galaxies is due to \citet{greggio11}. 

\begin{figure}[t]
\centering
\includegraphics[width=0.6\textwidth]{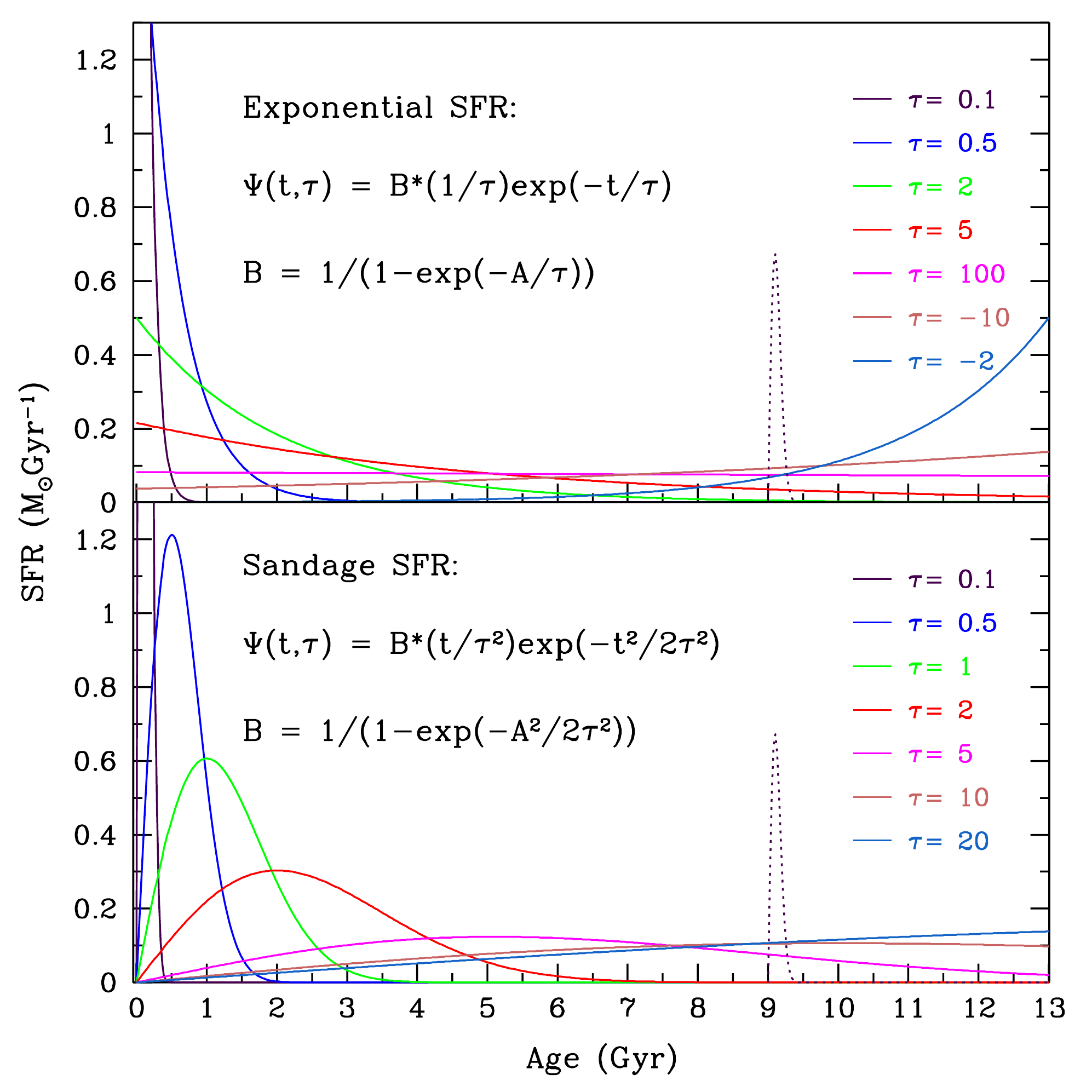}
\caption{
Time evolution of the exponential (upper panel) and \cite{Sandage86}
(lower panel) star formation histories (solid curves). The dotted
curve is a Sandage-style burst of star formation in which 10\% of
the total mass of stars are formed. 
From \citet{macarthur04}.}
\label{fig:SFHs}
\end{figure}

Star clusters, either open or globular, are the closest realisation
of SSPs in nature.  The main unknown of an SSP is the stellar
Initial Mass Function (IMF), which gives the mass spectrum of the
stellar generation at birth.  
The latter is not known from first principles.  Empirical determinations
of the IMF based on solar neighborhood data were first modeled by
\citet{Salpeter55} as a power-law with exponent of $\sim -2.35$. 
An IMF must be assumed when calculating the properties of population models. 
While galaxies are not SSPs, they can be viewed as a sum of all present SSPs: 
that is, Galaxies = $\sum_j SSP_j $.  The distribution of stellar generations
in time and chemical enrichment is called the ``Star Formation History'' (SFH). 
Several analytical laws describe plausible SFHs which depend on the timescale
of the Star Formation Rate (SFR), such as exponentially-declining models,
or $\tau$-models, models with constant star formation, models with
time-increasing star formation, etc.  Examples of such SFHs are shown
in \Fig{SFHs}. 

Ultimately, the stellar content of a galaxy over time, $t$, may
be thought of as: 
\be
{\rm Galaxy} = \sum_{\rm time} {\rm SFR(t)} \times {\rm SSP(t,Y,Z,IMF)},
\ee
with Y the Helium abundance and Z the abundance of heavier elements
(metallicity).  Note that Y, Z, and the IMF may vary between different
stellar generations, i.e. amongst different SSPs, but they do not vary
within an SSP by definition.

It is useful to note that little is known about the physical processes that
drive the rate of star formation and the emerging mass spectrum (the IMF).
We know that stars form from dense, cold gas in gas that is shock compressed
(e.g. in disks, during galaxy interactions or dynamical instabilities), but
a theory which predicts the SFR and the IMF in different galaxies and
as a function of time has yet to be written. For these reasons, these two
physical quantities are parametrized in population models and observations
to guide ongoing developments. Indeed, our limited knowledge about the SFR
and IMF is a major problem in the precise determination of a galaxy's stellar
mass. 

Historically, the problem of modelling a galaxy spectrum has been
approached in two ways. In the so-called ``optimised population synthesis''
\citep{spinrad71,faber72,oconnell76,pickels85,bicall86}, empirical stellar
spectra are
combined in proportions such that the resulting composite spectrum can best
reproduce the galaxy spectrum. These proportions can be ad hoc, hence neither
necessarily obeying stellar evolution timescales nor a realistic stellar IMF.
The obtained best-fit model can provide an excellent representation of the
galaxy spectrum, but it cannot be evolved with time. Hence the optimised
spectral fitting does not allow to study galaxy evolution in a cosmological
context.  Still, it can provide important insights on the types of stars
which are effectively present in a stellar system \citep[e.g.][]{macarthur09}.
Optimised synthesis can also be used to obtain an instantaneous description
of a galaxy spectrum in order to achieve accurate estimates of broadening
of absorption lines for velocity dispersion measurements (see \se{gaspoor}). 

The alternative approach makes use of stellar evolutionary models, which
describe the detailed time evolution of the luminosity and temperature of
stars of different mass. Integrated spectra for galaxies are calculated
by adding up the contributions of the individual model stars after assuming
an IMF and a SFH. These so-called ``evolutionary population synthesis
models'', based on stellar evolution theory, can be evolved in time
back and forth and galaxy evolution can be studied at arbitrary cosmic
distances with the same underlying theory.   The comparison between
these models and observational data provides estimates for the average
formation epoch, metallicity and SFH of a galaxy, thus enabling an
evaluation of the stellar mass through the model mass-to-light, or \ml,
ratio. These models, pioneered by \citet{tinsley72}, \citet{tingun76},
\citet{renvol81}, and \citet{bruzual83}, from which galaxy stellar
masses are derived, will be described extensively in the next
section (\se{stellpopbasics}), focusing on the quantities that affect
directly the stellar mass derivation, using various models available
in the literature.  We also address the impact of different model
fitting techniques on extracted results and conclude with an assessment
of the accuracy of galaxy stellar mass estimates.

\subsection{Basics of Stellar Population Models}\label{sec:stellpopbasics}

Evolutionary Population Synthesis (EPS) models provide the expected
spectral energy distribution (SED) of a stellar population as a function
of key parameters, such as:
\begin{enumerate}
\item the formation epoch, or the time elapsed since the beginning
 of star formation, normally referred to as the {\it age}, $t$, of the
 population, measured in years;
 \item the star formation history (SFH), often parametrized with
  analytic functions,  e.g. $SFR\propto e^{-t/\tau}$ (see \Fig{SFHs}); 
\item the chemical composition, often referred to as metallicity and expressed
 as the fractional abundance $Z$~of elements heavier than He and H ([Z/H]), or
 as the fractional abundance of iron ([Fe/H]); 
\item the chemical abundance ratios, or the ratios of all key elements with respect
 to those values measured in the Sun: e.g. the ratio of Magnesium to Iron [Mg/Fe],
 the ratio of Oxygen to Iron [O/Fe]; etc.; 
\item the Initial Mass Function (IMF).
\end{enumerate}

These are the main parameters controlling the time evolution of the population.
Further assumptions need to be made for stars evolving over specific evolutionary
phases, which will be mentioned below.

The model ingredients are: the stellar evolutionary tracks and/or isochrones,
the stellar spectral libraries, the parametrization for the mass-loss which
affects several late stages of evolution such as the Thermally-Pulsing Asymptotic
Giant Branch (TP-AGB), the Red Giant Branch (RGB), the Horizontal Branch (HB),
and also the Main Sequence (MS) in young populations. 

A further model feature is the computational procedure, which may be
an integration by mass of the luminosity contributions (so-called
isochrone synthesis technique, \citealp{brucha93})
or a technique based on Renzini's {\it fuel consumption theorem}
\citep{renzini81,buzzoni89,maraston98,maraston05}. 
This theorem states that the number of stars at each burning stage
is proportional to the time it takes to exhaust the nuclear fuel burnt
at that stage. This can be interpreted as the conservation of energy
for stellar populations.
In population synthesis models, it is a useful tool to quantify the contribution
of rapid, and very luminous, stellar phases as found at the tip of the RGB,
the AGB and the RGB bump \citep[cf. discussion in][]{maraston05}.
Numerical experiments have shown that calculations based on the fuel
consumption theorem and isochrone synthesis agree well \citep{chabru91},
provided the mass bin of the mass integration in the isochrone synthesis
case is small (\citealp{maraston98} finds 10$^{-6}\Msol$ for the tip of
the RGB).  Moreover, the fuel consumption theorem is useful for including
in synthetic integrated models those stellar phases for which a complete
isochrone may not be available, such as the AGB, the Horizontal Branch
with different morphologies, and the hot stars responsible for producing
the UV-upturn in ellipticals.

A detailed description of the individual population models can be found
in the corresponding papers, e.g. \citet[][hereafter, Vazdekis models]{vazdekis96},
\citet[][PEGASE models]{pegase}, \citet[][BC03]{brucha03},
\citet[][Maraston]{maraston05} and \citet[][FSPS]{conroygunn09}, 
in addition to the reviews cited in the previous section. 
  
The basic model EPS unit is the SSP. In the following SSPs are first used
to illustrate the fundamental dependencies of \ml\ on age and metallicity,
with composite models treated later on.

\begin{figure}
\centering
\includegraphics[width=0.6\textwidth]{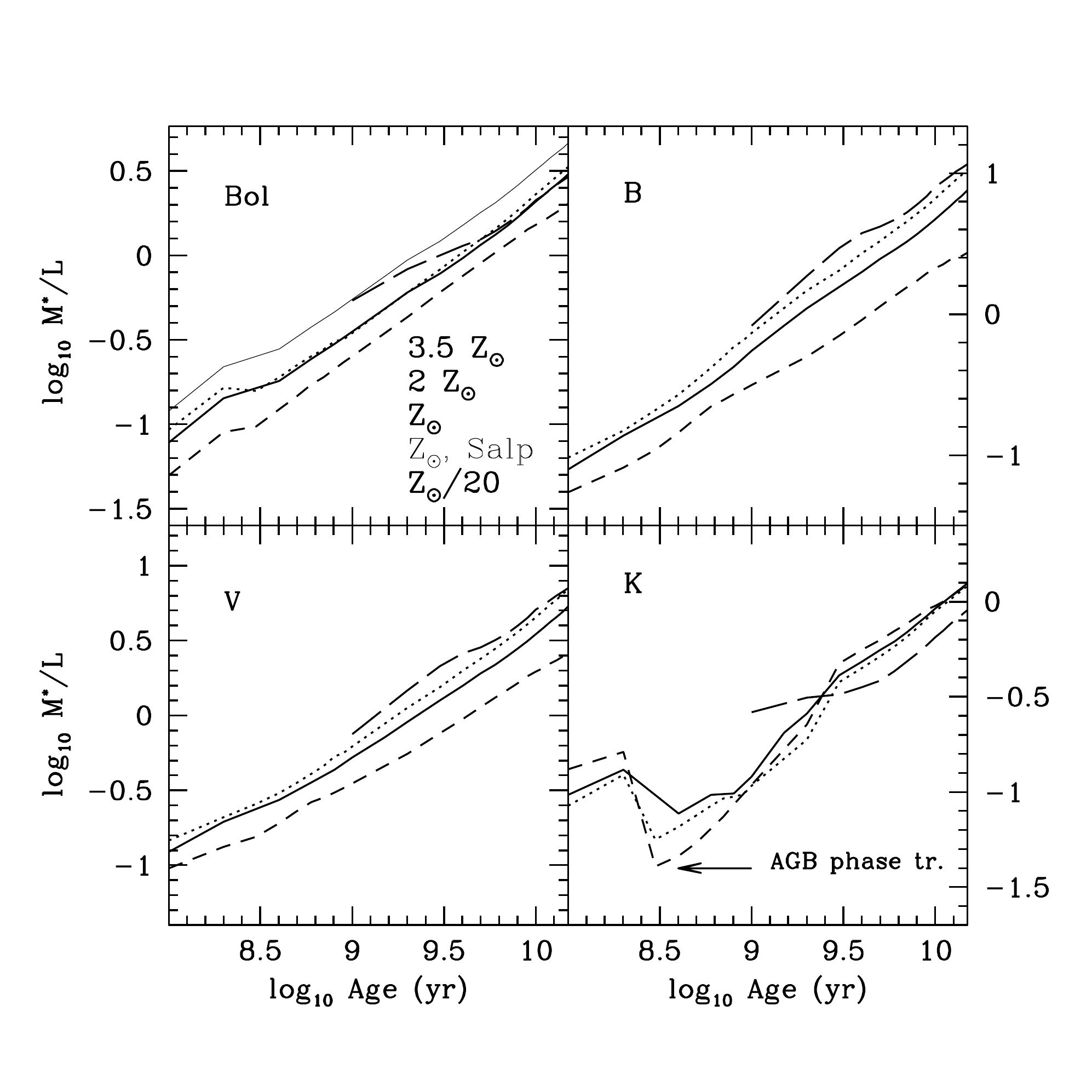}
\caption{
The stellar bolometric \ml\ and \ml\ in various bands
($B$, $V$, $K$) as function of age for SSPs (single burst models) and
the different indicated metallicities. All models for Kroupa IMF, except
for the thin solid line in the top left diagram for a Salpeter IMF. The
stellar mass $M^{*}$~in this figure accounts for stellar mass losses, as
in Maraston (1998; 2005). See also text for details.
From \citet{maraston05}.}
\label{fig:colML_ageZM05}
\end{figure}

The most important driver of an SSP's luminosity evolution is its {\it age},
since the most massive stars live quickly but are orders of magnitude more
luminous than smaller mass stars. For most IMFs the mass of a stellar population
is dominated by the faintest stars and changes relatively
little with time after the first Gyr of age (see \Fig{colML_ageZM05}),
but the luminosity of a population is dominated by its brightest stars
showing large changes over time.  Besides the Main Sequence, which
provides a substantial contribution to the light at virtually every
age, the brightest stars are found in different post-Main Sequence (PMS)
evolutionary phases. Which phase dominates depends upon the age of the
stellar population and the wavelength of observation.
In young populations ($t\ltsim$200 Myr),
Helium burning stars dominate the light, while at intermediate-age
($200~{\rm Myr}<t\ltsim$2~Gyr), the Thermally-Pulsing Asymptotic
Giant Branch (TP-AGB) stars take over (in some models, see below); at old
ages, Red Giant Branch (RGB) stars outshine all other stars. 
AGB and RGB phases are mostly bright in the near-IR (NIR), while MS stars
contribute mostly to optical bands \citep[see Fig.~11 in][]{maraston98}.

The luminosity of an SSP is therefore a strong function of time.
This rate varies with wavelength given the contributions from different
evolutionary phases. The overall luminosity evolution is more significant
in the blue/optical spectral range where MS stars dominate, scaling roughly
logarithmically with time \citep{tinsley72}, as compared to the NIR
light of old populations ($t\gtsim~2-3$~Gyr), where the slowly evolving
RGB stars dominate.  The rate of luminosity change at NIR wavelengths
is large near 0.3-1 Gyr in models including the TP-AGB phase
\citep[e.g.][]{maraston98,maraston05,marigo08}, since the onset of
this phase implies a rapid increase of the NIR luminosity due to the
cool and luminous TP-AGB stars. This effect is model-dependent.  Global
age and metallicity effects on the \ml\ in various bands for
SSP models is shown in \Fig{colML_ageZM05}.

The stellar population's metallicity also affects stellar evolution time scales
and mostly the stellar SEDs. Metal-rich stars are cooler (because of a higher
opacity in their stellar envelopes) and fainter (because the turnoff mass
is smaller and most outgoing photons are trapped into their envelope), hence
the higher \ml\ of a metal-rich population due to a lower luminosity
(while $M^*$ changes little). This trend is visible in
\Fig{colML_ageZM05}.  Metallicity effects are significantly milder at NIR
bandpasses; at very high metallicity (see the 3.5 solar metallicity
track in \Fig{colML_ageZM05}, long dashed line), the trend reverses in NIR
bands, since the luminosity of such a metal-enriched population is concentrated
at longer wavelengths \citep[see][for a full discussion]{maraston05}. 
As is well known, ageing
of a stellar population has the same effect as increasing metals since,
as the most massive stars die out, the temperature distribution skews
towards cooler values.  The combined age and metallicity effects on dwarf
and giant stars result in the so-called ``age-metallicity (A/Z) degeneracy''
in the optical region of their spectrum
\citep[e.g.][]{faber72,renzinibuzzoni86,worthey94,maraston00}.
At optical wavelengths, the effect is such that a population of stars that
is three times more metal rich mimics a population twice its age; this
is the ``3/2 rule'' of \citet{worthey94}.  This A/Z degeneracy at optical
wavelengths obviously holds for \ml\ ratios as well. The optical A/Z
degeneracy can be lifted by including data at longer wavelengths where
giants dominate the spectrum and increasing metallicity results in redder
colors, with a small dependence on age (with the exception of the AGB time). 
\cite{worthey94} proposed that the spectral region around 1 micron
(i.e. between the $I,J$~bandpasses, depending slightly on age and IMF)
is the most insensitive to metallicity. This notion has been exploited
with color-color diagrams involving optical and at least one NIR
passbands to analyse the stellar populations of integrated stellar
systems  
\citep[e.g.][]{deJong96,Bell00,maraston01,macarthur04,roediger11}.
This concept of using an extended wavelength range has also been
exploited for the study of high-redshift galaxies where the
time spanned since the Big Bang is short and the age dependencies
can be disregarded. In particular, the TP-AGB phase appears in galaxy
spectra and the inclusion of the NIR allows galaxy ages to be better
constrained \citep{maraston06}. 

\begin{figure}
\centering
\includegraphics[width=0.6\textwidth]{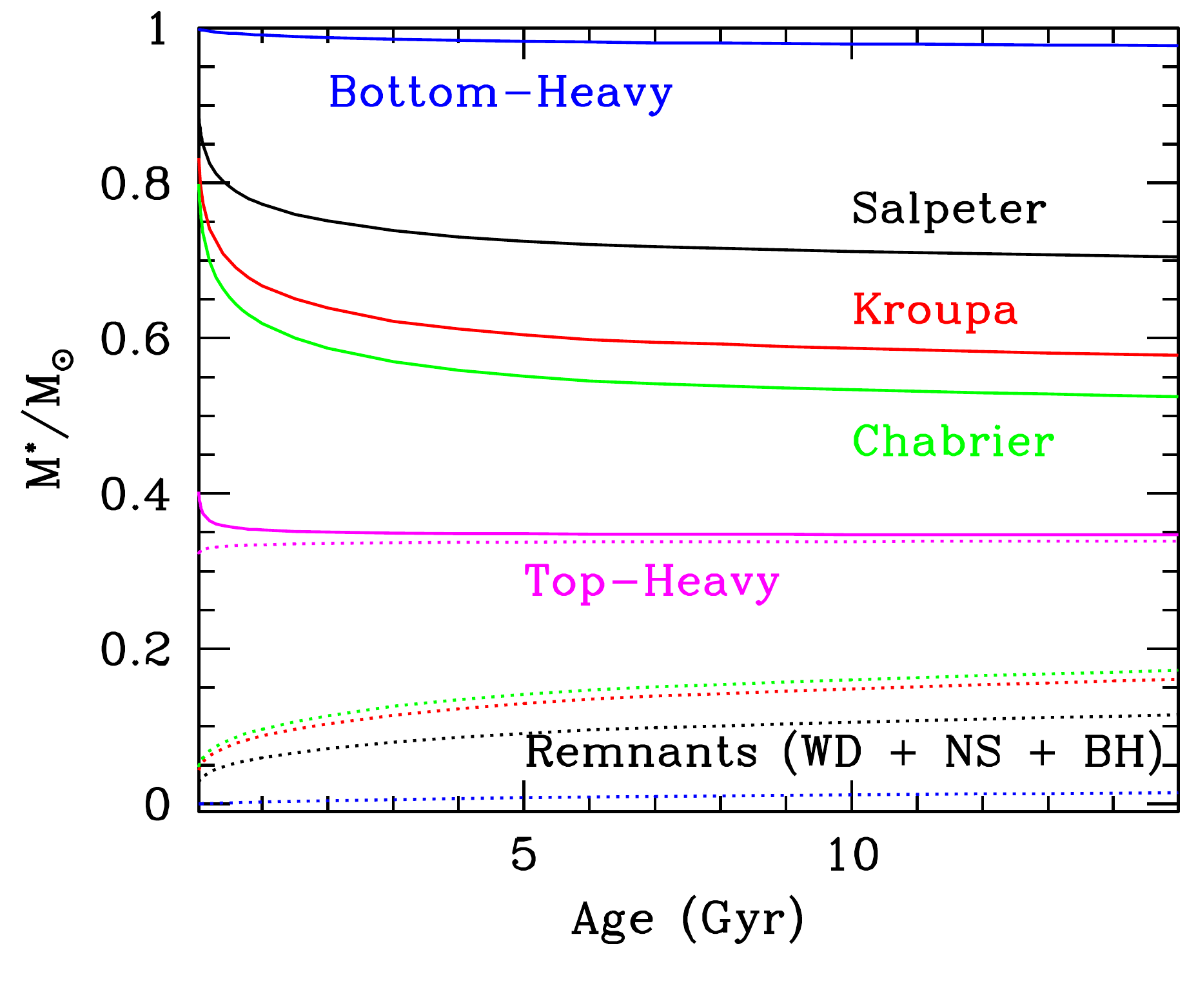}
\caption{
Evolution of the stellar mass fractions for stellar populations
with the same total initial mass (normalized to $1 \rm M_{\odot}$)
and different initial mass functions.  ``Bottom-Heavy'' and ``Top-Heavy''
are extreme cases of single-sloped IMFs with exponents 3.5 and 1 in
the notation in which the Salpeter slope is 2.35.  $M^{*}$ evolves
because stars die progressively and leave remnants with mass lower
than the initial mass. Also shown are the fractions of $M^{*}$ in
remnants, namely White Dwarfs (WD), Neutron Stars (NS) and Black Holes (BH).
Based on \citet{maraston05} models.}
\label{fig:mstar_ageM05}
\end{figure}

The total stellar mass $M^*$ of an SSP also evolves with time. It
typically decreases with time since the most massive stars progressively
die leaving stellar remnants with mass smaller than the initial ones.
$M^*$ is a strong function of the IMF. \Fig{mstar_ageM05} shows the
evolution of $M^*$ for several widely-used empirically-based IMFs, namely
\citet[][black]{Salpeter55}, \citet[][red]{kroupa01}, 
\citet[][green]{chabrier03}. These IMFs follow the same Salpeter
power law slope for stellar masses larger than 0.6 $M_{\odot}$,
but have, to varying degrees, less stars than predicted by this
Salpeter law slope below this mass limit (hence bottom-light).
It should be reminded that
\citet[][black]{Salpeter55}, \citet[][red]{kroupa01},
\citet[][green]{chabrier03} were all based on solar neighborhood data.

Also shown are two additional, not empirically-based IMFs meant to
illustrate extreme cases of a dwarf-dominated (labeled ``bottom-heavy'')
and a giant-dominated (labeled ``top-heavy'') IMF. These are single-sloped
IMFs with exponents 3.5 and 1 in the notation in which the Salpeter's
one is 2.35, and are meant to illustrate galaxies dominated by low-mass
and high-mass stars, respectively. Evidence for these extreme IMFs has
been advocated in the literature. For example, \cite{vDCon12} suggest
a dwarf-dominated IMF in massive early-type galaxies (hereafter ETGs)
to explain the strength of near-IR lines.  Similarly, dynamical modeling
studies \citep[][see also \se{gaspoor}]{Cappellari+12} and gravitational
lensing studies \citep[][see also \se{strong}]{Treu10,Auger10,Brewer12}
of ETGs find evidence for the same type of IMF, with larger mass-to-light
ratios than those predicted by Chabrier-like IMFs.

At the other end of the mass spectrum, \citet{Baugh07} find that
a top-heavy IMF in high-redshift bursts helps in explaining the
colours of massive dusty and bursty distant galaxies (so-called
``sub-millimiter'' galaxies).
  
For the same total initial mass, the stellar masses of top-heavy
IMFs evolve faster with time because of their larger proportion
of massive stars. Most of the evolution occurs within the first Gyr,
following the much faster evolution timescales of stars more massive
than roughly $2~M_{\odot}$. Over a Hubble time, the amount of mass
loss averages 30 to 40\%\ of the initial mass.  For composite population
models with ongoing star formation, the decrement is reduced to
$\sim 20\%$ \citep[see][]{maraston06}. In the Maraston models,
the total remnant mass is budgeted amongst white dwarfs,
neutron stars and black holes, following the analytical prescription
of \citet{rencio93}. \citet{maraston98} has explored IMFs with various
exponents to show that $M^*$ is maximally large for both dwarf-dominated
as well as top-heavy IMFs, while the minimum $M^*$ is achieved with a
bottom-light\footnote{Note that bottom-light refers to an IMF that is
not as rich in dwarf stars as the Salpeter one, namely has a different
slope above 0.6 $M_{\odot}$. This is different from a top-heavy IMF,
which is strongly dominated by giant stars due to a smaller exponent
value all over the mass range}. IMFs such as the \citet{scalo86}, 
\citet{kroupa01} or \citet{chabrier03}-type IMFs \citep[see
Figs 16, 17 in][]{maraston98}. A population born with a
dwarf-dominated (or ``bottom-heavy") IMF has a large $M^*$
since most stars have a small mass hence their extended lifetime and
they contribute their total mass to $M^*$. A giant-dominated (or
``top-heavy'') IMF has a large $M^*$ given by the large number of
massive remnants left by the evolved massive stars. 
These considerations are important as the value
of $M^*$ and the assumptions regarding the IMF impact directly the 
evaluation of the total stellar mass and the dark matter content in galaxies.

It should be noted that the predicted $M^*$ may differ amongst different
population synthesis models (see model comparisons between
\citet{brucha03} and \citet{maraston05} in Fig.~3 of \citet{macarthur10}).
This discrepancy may reflect a different accounting of stellar remnants.
Hence, in comparing stellar masses obtained with different EPS models,
one should also consider if and how the remnant masses are accounted for.
For example, the \citet{worthey94} models consider the mass contribution
of the sole living stars, while \citet{brucha93} consider a constant $M^*$,
etc.

Other factors may complicate the
broad-brush stellar evolution picture painted above. For example, stars
in exotic evolutionary phases - such as Hot Horizontal Branch (HHB) at
high metallicity and Blue Straggler (BS) stars - may alter the luminosity
in the blue spectral range.  These events probably affect mostly globular
cluster studies, as the relative contribution of these phases to the total
light of a galaxy should not be very significant. Also, one should not
rely on a single band - especially a blue band - to determine stellar
masses.  The reddest side of the spectrum is equally challenging with
the evolution along the TP-AGB being woefully uncertain due to the unknown
mass loss.  A lively debate pertaining to the reliability of optical/NIR
stellar mass estimates is currently ongoing (e.g., \citet{bruzual07},
\citet{marigo08}, \citet{conroygunn09}, \citet{zibcha09},
\citet{macarthur10}, \citet{conroygunn10}, \citet{lyubenova12}, 
and \citet{ConroyARAA13}).

A resolution of the significance of the TP-AGB phase to stellar mass
estimates cannot be achieved in this review.  However, a sound
estimate of the stellar mass appears to be provided by the $g-r$
color \citep{zibcha09,taylor11}.
The effect of the TP-AGB phase on stellar mass derivations stemming
from the use of models with and without a substantial TP-AGB contribution
is highlighted when appropriate.

Extinction from dust also affects broadband luminosities hence \ml\
ratios and stellar population colors. However, in optical passbands
the expected dust effects on \ml\ and population color run parallel
to the expected stellar population color-\ml\ relations and the mass
estimates are only weakly affected \citep[e.g.][]{Bell01}. 
Likewise, line indices are only slightly affected by
dust \citep{macarthur05} and can help disentangle extinction from
population effects. For full stellar mass estimation dust extinction
should be included in analyses, but dust is here not further addressed.
Finally, most stellar systems will not be approximated by single-burst
or SSP models, rather by composite populations with multiple ages and
metallicities. The unknown galaxy star formation histories complicate
the interpretation of integrated light observations and will be discussed
in \se{basicSFH}).

In the following we focus on mass determinations in relation to
fitting of the broad-band spectrum, as this situation is common to
both low as well as high-redshift studies, where high resolution
spectral fitting is presently unfeasible. It should also be noticed
that comparison of galaxy mass determinations obtained via broad-band
or spectral fitting agree well when the signal-to-noise of the
spectrum is high \citep{chen12}. 

\subsection{Stellar Mass from \ml\ vs Colour Diagnostics}

\begin{figure}
\centering
\includegraphics[width=1.0\textwidth]{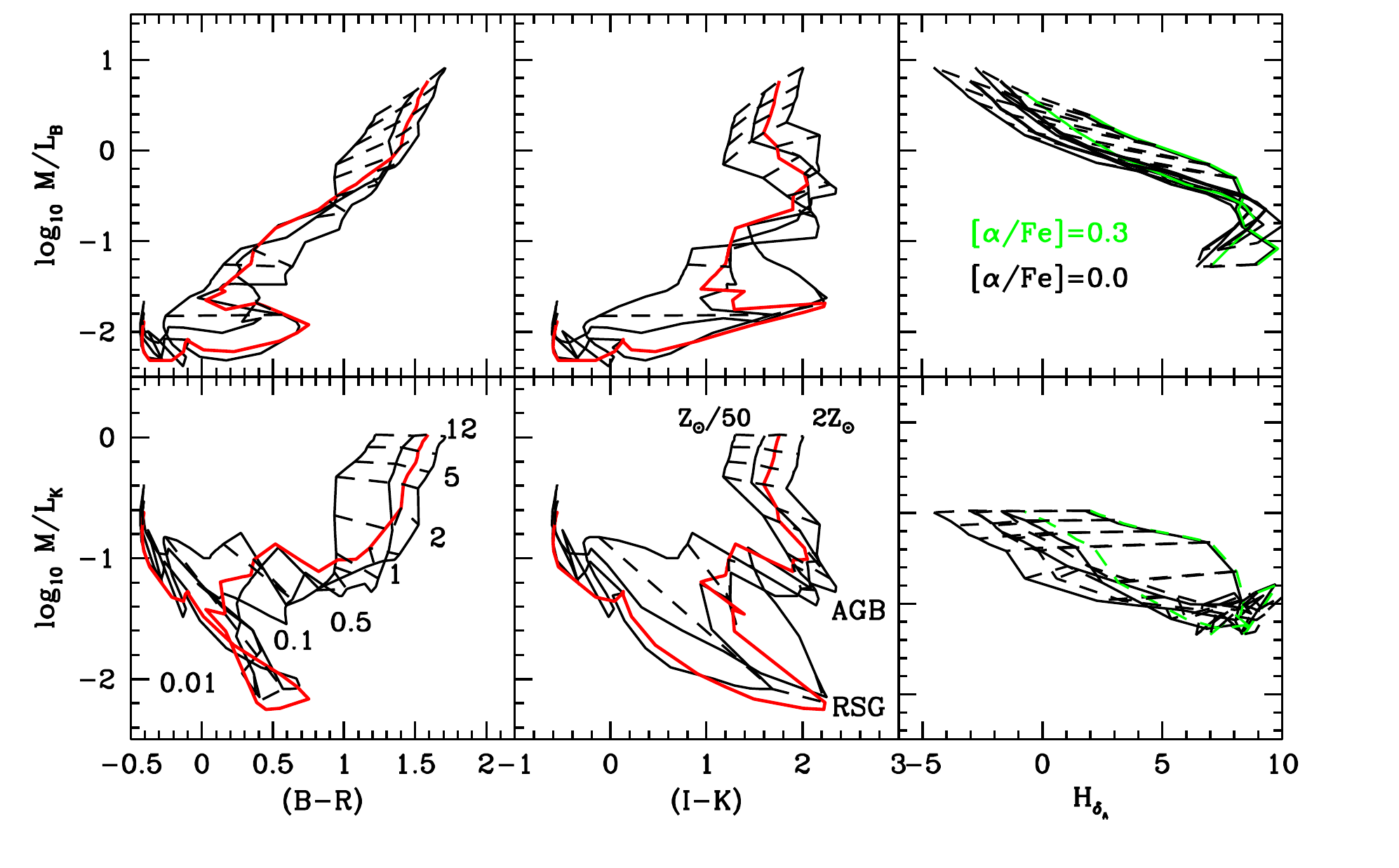}
\caption{
Trends of \citet{maraston05} SSP models with various ages
and metallicities, and a Salpeter IMF. \ml\ values in $B$ and
$K$-band (top-to-bottom) are shown versus \br, \ik, and the
\HdA\ line index (left-to-right). The latter models are from
\citet{tmj11}; green lines connect models for $[\alpha/{\rm Fe}]$~ratio
of $+0.3$, black lines connect solar-scaled models. 
Models with a same age (in Gyr) are connected with dashed lines, 
models with a same metallicity with solid lines (Z=0.001, 0.01, 0.02,
0.04). Red lines highlight solar metallicity (Z=0.02).}
\label{fig:colML_SB_cla}
\end{figure}

The list at the beginning of \se{stellpopbasics} makes clear that the
"Big 4" variables that stand between the observed photometric distribution
of star-light and our interpretation in terms of stellar mass are: 
(i) the correctness of stellar evolution models along stellar phases; 
(ii) the SFH; 
(iii) the chemical enrichment history; and 
(iv) the IMF.  These are all inter-connected astrophysically
and phenomenologically.  Part of this interconnection could be labeled
the 5th variable of "environment". In most cases when analysing galaxies,
little information is available on the big 4, with perhaps the exclusion
of (i) which can be calibrated and tested in local stars. Hence, usually
a variety of models are assumed including different formalizations for
(ii), (iii) and (iv). The significance of these assumptions is shown below. 

Following \citet{Bell01}, we illustrate how \ml\ ratios in the $B$
and the $K$-band correlate with ($B$--$R$) and \ik\ colors (as tracers
of age and/or metallicity),
and the mostly age sensitive \HdA\ index. \Fig{colML_ageZM05} shows that
\ml\ values in other optical passbands behave similarly as the \mlb\ plots,
just with a slightly different slope. 

\Fig{colML_SB_cla} shows grids of model colors, \ml\ 
and a spectroscopic index for SSPs of various ages and metallicities.
Once a population is older than about 0.1\,Gyr,
\ml$_B$ versus \br\ displays a good correlation, which is fairly independent of
metallicity. The \ml$_K$ versus \br\ is more ambiguous with age as \ml$_K$ evolves
slowly after the TP-AGB phase-transition (meaning, after some Gyr of age). The
relations of \ml\ with \ik\ color cannot be used alone to derive $M^*$ because
the model loci are nearly vertical after 0.1 Gyr, meaning that at a given
metallicity \ml\ is nearly independent of color.  As mentioned earlier, the
luminosity in the NIR after a few 100 Myr is dominated by evolved (AGB and
later RGB) stars. The lower NIR \ml\ at early ages ($\sim$10\,Myr) is due
to short-lived, bright red supergiants. Note that this caveat is only relevant
for very recent star formation, basically from \hii\ regions.

Perhaps the best approach to measuring \ml\ is to fit SEDs simultaneously
in at least three passbands, with one in the NIR, 
hence breaking the age-metallicity degeneracy.  Modulo model
uncertainties as detailed in the following sections, convergence
to the right \ml\ value may be achieved (as shown by \cite{zibcha09}
or by \cite{maraston01} for massive star clusters and
\cite{maraston06} for high-redshift galaxies).


Finally, the rightmost column of \Fig{colML_SB_cla} shows
an example of \ml\ trends with an absorption line index. The \HdA\
Balmer line index \citep{wortheyotta97} was chosen as it has been
used specifically for the estimation of $M^*$ for the SDSS galaxies
\citep[e.g.][]{kauffmann03}. This index is sensitive to the age of
the stellar population, being the strongest around 0.3-0.5 Gyr
(depending on metallicity) when A-type stars dominate the spectrum,
and smaller at both older and younger ages
\citep[see Fig.~8 in][]{maraston01}. This degeneracy for \HdA\ can
obviously be broken (for a single burst population) by using an
extra spectral indicator with a monotonic behavior with age. 
Two further complications should be noted. Because of its
temperature dependence, the index is also affected by old, though
hot, stars such as metal-poor blue HB stars \citep{maraston03}.
Moreover, the index has been shown to be sensitive to the
[$\alpha$/Fe] ratio, because of strong Fe lines present in the
pseudo-continua \citep{tmk04}. For instance, the index is stronger
in [$\alpha$/Fe]-enhanced populations (such as those of massive ETGs)
than in solar-scaled ones. The trend with abundance ratios is shown
in \Fig{colML_SB_cla}, where the grid in green displays models for
the same total metallicity $Z$, but enhanced [$\alpha$/Fe]=0.3.
The consideration of these effects when dealing with massive
galaxies is important, as a strong index can otherwise only be explained by
a lower age, which in turn would induce a mismatch in the derivation of the
\ml\ ratio. Ideally in the future it should be possible to complete the last
panel of \Fig{colML_SB_cla} by considering the abundance-ratio effects on
colors and mass-to-light. Broadbrush, the model grid distribution
shown in \HdA\ versus \ml\ is similar to the \br\ versus \ml\
distributions, except for the very youngest ages. The same
degeneracies are therefore also present in composite age models and we
will no longer show the \HdA\ models separately.

Single burst models are ideal cases that apply well to star clusters, but
not to galaxies, where a prolonged star formation is in general more
appropriate.  The determination of $M^*$ in these cases is much more
difficult, as the latest generations dominate the light and drive down
the \ml\ hence $M^*$, which leads to underestimate of the stellar mass
\citep{Bell01,macarthur09,maraston10}.  These issues are discussed
in the next section.

\subsubsection{Effect of Star Formation History}\label{sec:basicSFH}

\begin{figure}
\centering
\includegraphics[angle=-90,width=\textwidth,clip=True]{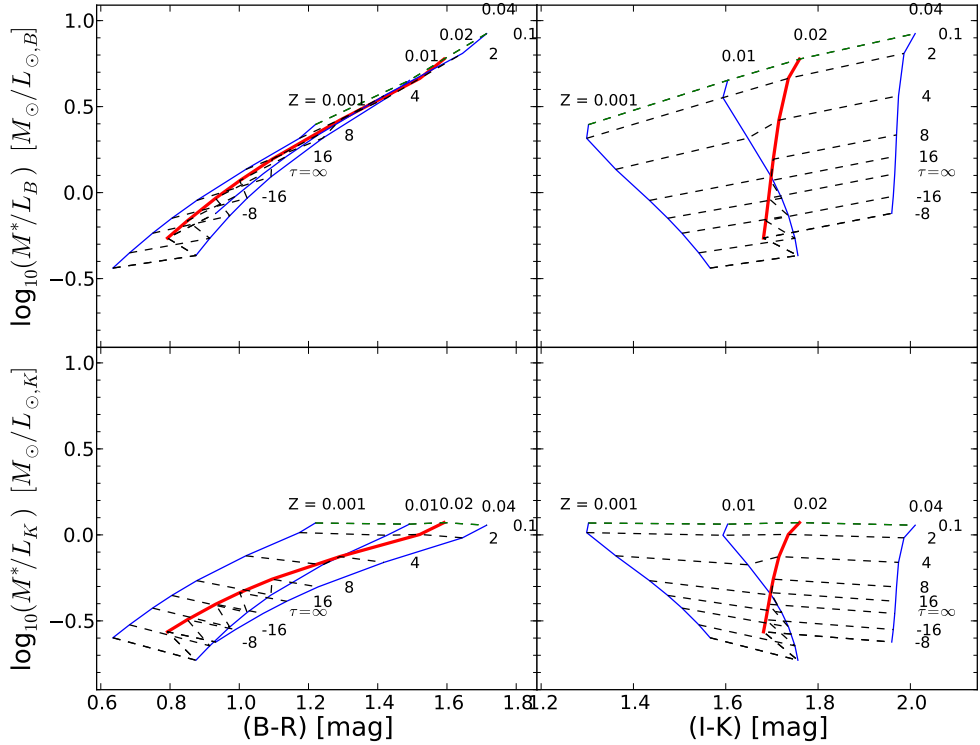}
\caption{Trends in \ml\ ratios using exponentially declining star
  formation rate models. The models are the same as in \Fig{colML_SB_cla},
  except that instead of a single burst an exponentially declining
  star formation rate is used observed at age 12\,Gyr after star
  formation began. 
  Models with the same $e$-folding timescale $\tau$ are connected with
  dashed lines, where positive $\tau$ stands for decreasing star formation
  rate with time, negative $\tau$ for increasing star formation rate, and
  $\tau=\infty$ is a constant star formation rate model. Solid lines connect
  models with equal indicated metallicity, where the red line highlights
  solar metallicity.  Based on \citet{maraston05} models.
\label{fig:colML_tau}
}
\end{figure}

Stellar systems like galaxies are expected to have a wide range of SFHs.
The effect of prolonged star formation history on the model grids
is visualised in \Fig{colML_tau} by using exponentially declining SFR models or
$\tau$-models \citep[as introduced by][]{bruzual83},
with SFR$(t) \propto \exp(-t/\tau)$, where $\tau$ indicates
the $e$-folding timescale of star formation
and can be both positive and negative.  
This model is a reasonable first approximation of the star formation history of
a spiral galaxy or, for very low $\tau$'s, of a passive
system. Negative $\tau$ values represent galaxies which have
increasing SFRs, especially galaxies with recent star bursts
\citep[e.g.][]{Bell00}.

The model \ml\ ratios in the optical are nearly degenerate versus \br\ in
\Fig{colML_tau}. This degeneracy is somewhat broken in \br\
versus the near-IR \ml$_K$, especially for the lowest metallicities ($Z<0.004$).
However, realizing that chemical evolution caused by modest amounts of
star formation raises the system metallicity rapidly to at least 0.1
solar (or Z=0.002; in a closed box, conversion of $\sim$20\% of gas mass
into stars raises the average metallicity to over 0.1 solar), the
range of relevant metallicities becomes narrower in most applications,
making the color-\ml$_K$ relation tighter. 

Tracing mainly evolved stellar populations, \ik\ is largely metallicity
sensitive, weakly sensitive on age, and \ik\ on its own
(without any other passbands) is not useful for mass estimation
as \Fig{colML_tau} clearly shows.  Moreover, the
details of evolved stellar evolution stages are still relatively
poorly modeled and the exact shapes of these \ik\ diagrams are highly
dependent on models used, as shown in \se{ingredients}.

\begin{figure}
\centering
\includegraphics[angle=-90,width=\textwidth,clip=True]{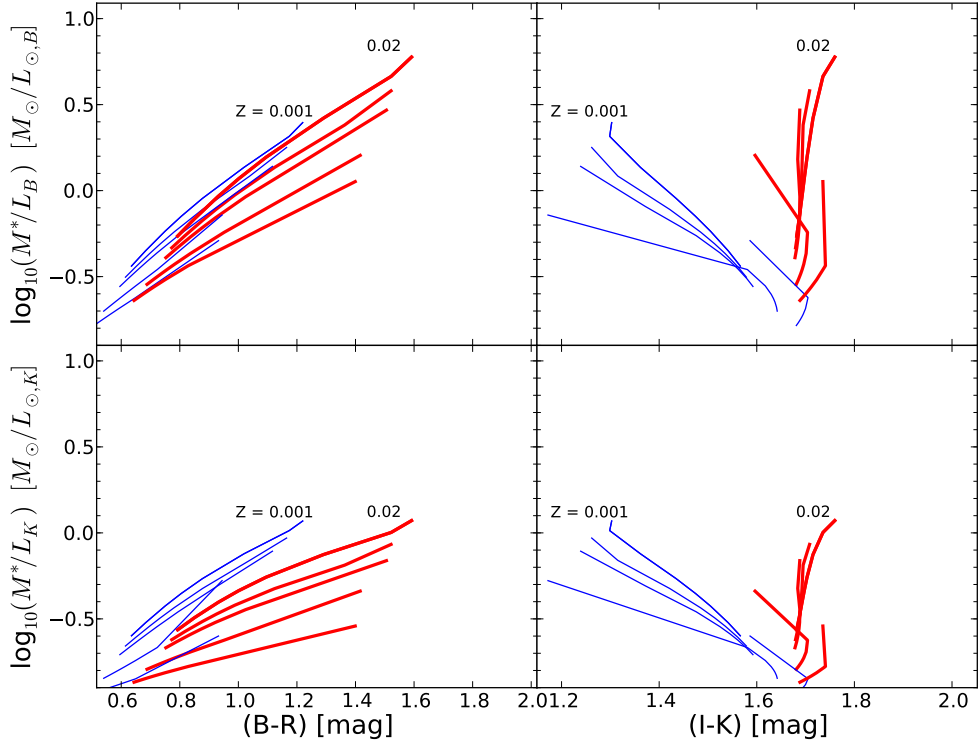}
\caption{\ml\ ratios for exponentially declining star formation rate
  models with different final age. The models are the same as in \Fig{colML_tau}, except
  that the populations are now observed 12, 8.5, 6, 3, 2\,Gyr
  (top-to-bottom) after star formation commenced. The top ends of the
  lines are the ends with the smallest positive $\tau$ values,
  i.e. the oldest average age.
  For clarity, only the solar metallicity (Z=0.02; thick red lines)
  and the 1/20th solar metallicity (Z=0.001; thin blue lines) are plotted
  and the dashed lines connecting the
  same $\tau$ values are not shown as in \Fig{colML_tau}.  In principle
  these models form similar grids with different offsets.
  Based on \citet{maraston05} models.
\label{fig:colML_tau_age} 
}
\end{figure}

Not all stellar systems are $\sim$12\,Gyr old, either because they are
observed at higher redshift when the Universe was much younger or because
they may have had their major epoch of star formation significantly delayed.
The effect of a younger final age on exponential SFR models is shown in
\Fig{colML_tau_age}. This figure shows the same range of $\tau$ values
as in \Fig{colML_tau}, though now after 12, 8.5, 6, 3, and 2\,Gyr (and
for clarity only for solar and 1/20 solar models). For concordance
cosmology, this corresponds to roughly redshifts z=0, 0.3, 0.7, 1.5,
and 2 when starting star formation 12\,Gyr ago. Model colors are shown
in rest-frame. 

The main conclusion from the color-\ml\ diagrams in \Fig{colML_tau_age}
is that a final age change mainly results in a simple offset
since the slope of the relation stays nearly constant especially for solar
metallicities. The age-metallicity degeneracy stays intact in the
optical relations. Only at the youngest ages and lowest metallicities, and
predominantly in \ml$_K$, does the slope of the relation change.

\begin{figure}
\centering
\includegraphics[angle=-90,width=\textwidth,clip=True]{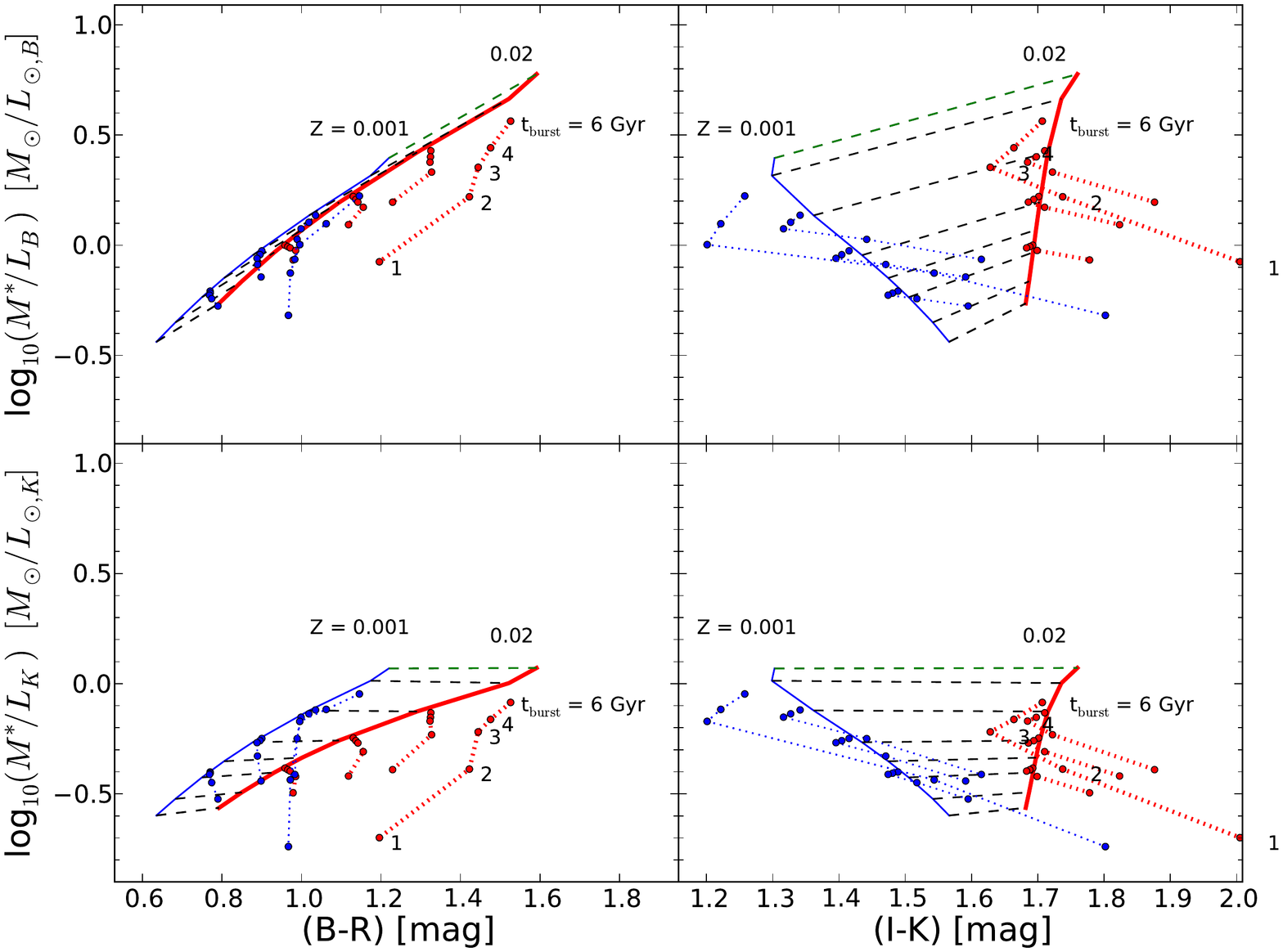}
\caption{\ml\ ratios for exponentially declining star formation rate
  models with an additional star burst. The starting models without
  starburst are the same 12 Gyr old exponentially decaying SFR models
  as in \Fig{colML_tau}, except that for
  clarity only the solar metallicity (Z=0.02; thick red) and the
  1/20 solar (Z=0.001; thin blue) are plotted. For $\tau$ values $\infty$, 8, 4,
  1 a 10\% final mass fraction starburst of 0.2\,Gyr duration is added
  that occurred 1, 2, 3, 4, 6\,Gyr ago (solid circles connected with
  dotted lines). 
  Based on \citet{maraston05} models.
\label{fig:colML_tau_sb}
}
\end{figure}

So far only smooth SFHs have been considered in this section, but star
formation will be more bursty in nature for especially smaller systems. A recent burst of star
formation will dramatically lower the \ml\ of a total
stellar system as well as change its SED to make it look much younger,
because a young population is so much more luminous. The size of the
effect will depend on the size of the star burst relative to the
underlying population and the age difference between the
populations. This effect may be most relevant for small galaxies,
where any burst of stars is significant, and for stellar masses
dominated by an old population, where any trickle of young stars will
dramatically alter their properties. For instance, the rejuvenation
caused by a ``frosting'' of young stars has been used to
explain the apparent line indices based young ages of morphological
ETGs that are thought to be mostly old \citep{deJongDavies97,trager00},
though hot horizontal branch stars produce the same effect on the
observed line strengths \citep{maraston00}.

\Fig{colML_tau_sb} shows the effect of adding 10\% of
the final mass of the system in a star burst of 0.2 Gyr duration. The
models start with the canonical set of 12\,Gyr old exponential SFR
models of \Fig{colML_tau} to which 1, 2, 3, 4, or 6 Gyr old star bursts
are added. For clarity, only the solar and 1/20 solar results are shown.

A number of features are apparent from \Fig{colML_tau_sb}.
Firstly, the effects of starbursts are largest in both color and \ml\
when starting with an a old, small $\tau$ population
(independent of metallicity). Secondly, almost any burst
of star formation will bias models towards lower \ml\ values
at a given color compared to smooth star formation models.
At a given \br\ color, the maximum offset from a smooth SFH
due to a late starburst  is less than 0.5\,dex, but in most
combinations it is less than 0.3\,dex. This effect is stronger for
higher metallicities. Finally, when starting off from a fairly young
underlying population ($\tau>5$\,Gyr) the effects of a the modeled
starbursts are only larger than 0.1\,dex in \ml\ for bursts younger
than 1--2\,Gyr. In this case the population may actually become
redder than the underlying population in \br\ after a few Gyr
after the burst, because the starburst will increase
the luminosity weighted average age of the total population.
When starting from mostly old populations ($\tau<5$\,Gyr),
the effect of a starburst on the \ml\ may be long-lived (4--6\,Gyr).

If most galaxies had irregular star formation in their last 2--5\,Gyr
with variations in star formation rate of factors greater than two,
most galaxies are expected to lie below the smooth exponential SFR
model SED-\ml\ relations. One can calculate this offset and increase
in scatter by using models that include a varying amount of star
formation and incorporate this in the derived SED-\ml\ relations.
Alternatively, one can reduce the scatter induced by these recent
star burst by including an SED tracer
of a stellar population younger than 2--3\,Gyr when fitting the SED.
One has to chose a tracer that is not degenerate in age and
metallicity with the other SED tracers used.  Some options include
a ($U$--$B$) color or a higher order Balmer line.  The combination
of such recent star formation indicators and SED templates
with an irregular SFR should yield a scatter reduction
in \ml\ by $\sim$0.1\,dex if many objects with irregular
SFR are contained in the sample.

More bandpasses (at least four) can provide additional
degeneracy lift. For example, in the Maraston models a very red \ik\
color only corresponds to the TP-AGB dominated ages of $\sim$1 Gyr,
which is more pronounced in a small tau-model. Most models behave
similarly in the optical bands, while the NIR is driven by the
treatment of evolved phases, for which there is a strong variance
within existing models (see also \se{ingredients}).


\subsubsection{Stellar Initial Mass Function}\label{sec:IMF}

\begin{figure}
\centering
\includegraphics[width=0.7\textwidth]{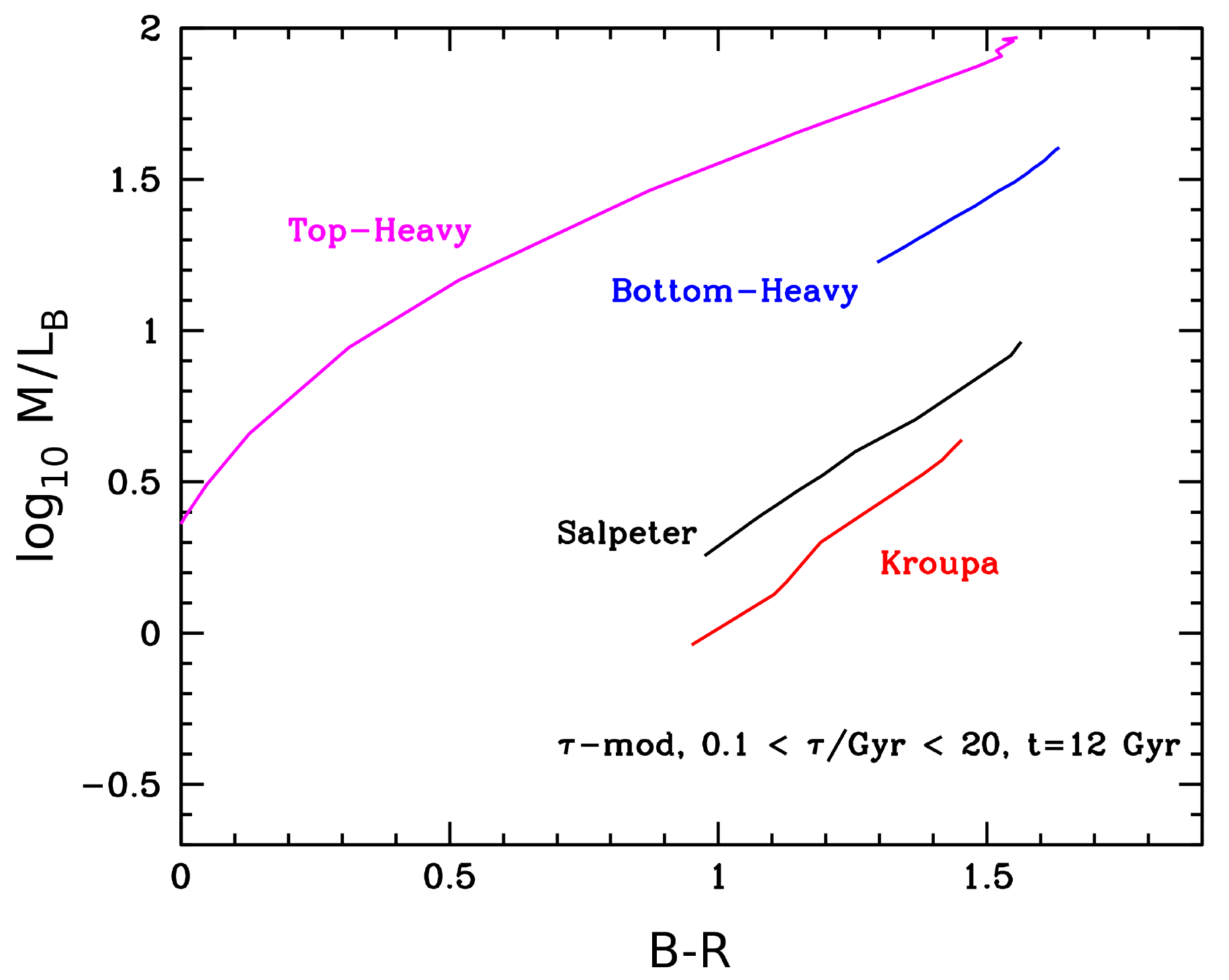}
\caption{\ml\ ratios vs \br\ for exponentially declining star formation
rate models of age 12\,Gyr and solar metallicity, with a Salpeter,
Kroupa and the same top-heavy and bottom-heavy IMFs as in \Fig{mstar_ageM05}.
}
\label{fig:colML_IMF_cla} 
\end{figure}
As noted in \se{stellpopintro}, the IMF is a major uncertainty
in SP modelling and \ml\ ratios are strongly dependent on it. 
The IMF of external galaxies is in principle unknown
\citep[though see][]{tortora12}, but most IMFs determined for
the Milky Way in the Solar neighborhood show a very similar
behavior, namely a turnover in their logarithmic slope
at about 0.6 Solar mass. The exact details of this turnover in IMF
slope around this mass scale are poorly constrained 
\citep[see discussions in e.g.][]{scalo86,kroupa01,chabrier03}, 
but critically determine the total mass of the system.
\Fig{colML_IMF_cla} visualises the solar metallicity \ml\ versus
color relation using different IMFs, specifically the same as in
\Fig{mstar_ageM05}. 
The predicted relations are clearly very similar in shape
for exponential SFR models and the plausible Salpeter or Kroupa
IMF (the Chabrier and Kroupa IMF's behave very similarly).
These different IMFs result primarily in offsets in zero-point of the
\ml\ versus color relations.  These offsets are independent of metallicity
and nearly the same whether one measures \ml$_B$ or \ml$_K$. 

The \ml\ ratio versus color slope remains unchanged even assuming a rather
extreme bottom-light IMF, but the offset is much larger (a factor of $\sim
10$) for the latter (\Fig{colML_IMF_cla}). However, this IMF is ruled
out by strong gravitational lensing and stellar dynamics which only
permit \ml\ ratios up to a factor of $\sim 2$ higher than that
predicted for a Salpeter IMF \citep[e.g.][]{Brewer12,Cappellari+12}.

\subsubsection{Model ingredients}\label{sec:ingredients}

\begin{figure}
\centering
\includegraphics[angle=-90,width=\textwidth,clip=True]{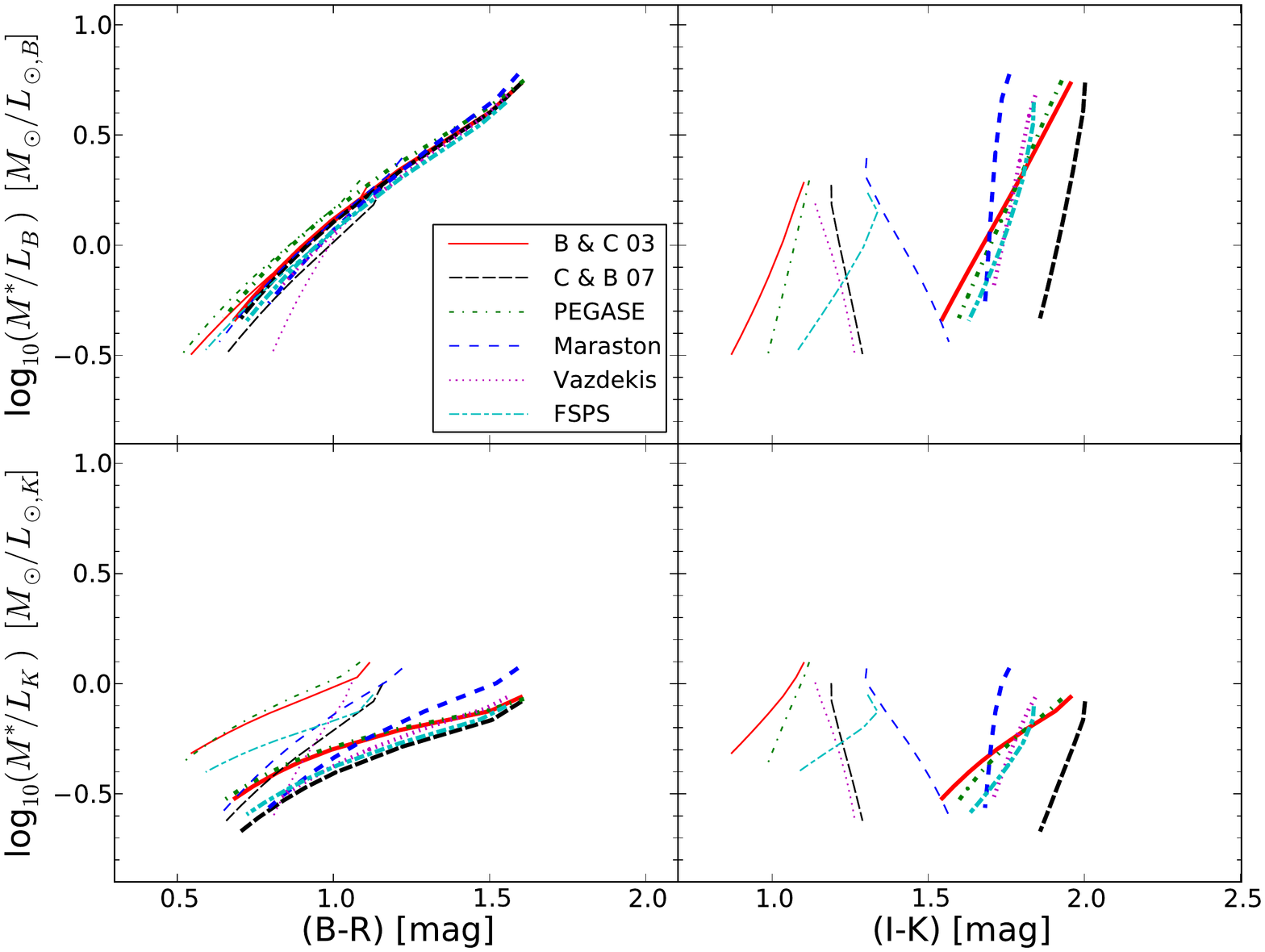}

\caption{\ml\ for exponential SFR models of age 12\,Gyr using SSPs from different
  authors. The models compared are GALAXEV 2003 \citep{brucha03} with Padova 1994
  tracks, GALEXEV (2007, an updated version of \citep{brucha03} with different
  treatment of the TP-AGB phase), PEGASE 2.0 \citep{pegase}, and the FSPS
  models \citep{conroygunn09} all based on Padova stellar models and a
  \citet{chabrier03} IMF, and the \citet{maraston05} and the \citet{vazdekis96}
  models using a \citet{kroupa01}. Solar metallicity (Z=0.02; thick lines)
  and 1/5 solar metallicity (Z=0.004; thin lines) models are used, except
  for the Maraston models, where this metallicity is not available and a lower
  metallicity model of Z=0.001 is plotted.  The Chabrier and Kroupa IMFs give
  essentially the same results for broadband colors, hence by using these IMFs
  all the model sets can be compared even though some models are only available
  with either Chabrier or Kroupa IMFs.
}
\label{fig:colML_models} 
\end{figure}

As noticed in \se{stellpopintro}, EPS models include several ingredients and
assumptions (stellar evolutionary tracks, stellar spectral libraries,
IMF, etc.), which may be treated differently in different models.
It is beyond the scope of this review to investigate the effects
of each ingredient and its uncertainty on the models 
\citep[for reviews see, e.g.][]{cwb96,maraston05,conroygunn09,leitherer11}. 
Rather, \ml\ versus color relations for the different
EPS are compared here using exponential SFR models.
The GALAXEV, PEGASE and FSPS models are used with a Chabrier IMF,
while the Maraston models use a Kroupa IMF. 

Not surprisingly, the GALAXEV (BC03) and PEGASE results are very similar
in all color-\ml\ diagrams, since they use the same Padova
stellar evolutionary tracks and similar spectral libraries.
At optical wavelengths, nearly all models agree to within
0.1\,dex in \br\ versus an \ml\ measured in an optical band.

As discussed in \citet{cwb96} and \citet{maraston05}, the treatment
of TP-AGB and RGB stars lead to the largest discrepancies in the NIR
which is rehashed here with the \ik\ versus \ml\ diagram. The \ik\ colors
of old (small $\tau$) populations with the same metallicity can differ
by 0.2\,mag, and up to 0.5\,mag for young populations. The reason 
for this discrepancy is two-fold. Firstly, they are due to a different
treatment of the (uncertain)
late stages of stellar evolution (red supergiant, AGB, RGB).  In these
late stages, stars suffer mass loss that cannot be connected to the basic
stellar parameters from first principles and must be parametrized and
calibrated with data. This uncertainty involves energetics, stellar
temperatures and stellar spectra for the AGB, and mostly stellar
temperatures and spectra for the RGB.  The importance of the TP-AGB
phase of stellar evolution became clear, when it was shown that the
BC03/Pegase/Starburst99 and \citet{maraston05} models yielded systematic
differences of several tenths of magnitudes in the NIR at intermediate
ages of 0.2 to 2 Gyr \citep{maraston05,maraston06,bruzual07}, mostly
attributable to differences in the treatment of TP-AGB stars.
Secondly, numerical instabilities of luminosity integration
along short-lived phases in the isochrone synthesis approach
\citep[see][]{maraston98,maraston05} may explain why models based
on the same stellar evolution tracks exhibit large fluctuations.

The models in \Fig{colML_models} with a small 
TP-AGB contribution (i.e. BC03, Pegase) display bluer values of \ik\ colors
at young ages (low \ml). The Vazdekis models would behave similarly,
but they are redder only because they only include ages $> 0.1$ Gyr.
Models with a substantial TP-AGB contribution
\citep[][GALEXEV 2007]{maraston05} display redder colors
at young ages/large or negative $\tau$'s.  At solar metallicity,
old ages/small $\tau$'s (upper right corner of the diagram),
models based on Padova tracks (B\&C, FSPS, Pegase) are redder
than the Maraston models based on the \cite{cassisi97} tracks
because the former have a redder RGB
\citep[Fig.~9 in ][and discussion therein]{maraston05}. 

\subsection{Data Fitting Techniques}
The variety of techniques to derive \ml\ values by fitting spectra has increased
dramatically in the last decade \citep[e.g.][]{walcher11}. Methods depend on
the data available, ranging from two bandpasses \citep[e.g.][]{Bell01},
multiple broadband colors \citep[e.g.][]{bell03,maraston13}, a few line indices
\citep[e.g.][]{kauffmann03,Thomas+11} to full spectral
fitting \citep[e.g.][]{blanton07,tojeiro09,chen12}.
Broadband imaging is often preferred over spectroscopy when large
numbers of galaxies are required, when 2D stellar maps
are created \citep{zibcha09}, or in low S/N situations as in high-redshift studies
\citep[e.g.][]{daddi05,shapley05,maraston06,cimatti08}. 
While more data points should in theory increase the accuracy of
the \ml\ estimation, this may not always be true in practice due
to two reasons. 
Firstly, EPS models have larger systematic
uncertainties at certain wavelengths (\Fig{colML_models}). Such
intrinsic model uncertainties should be taken into account while 
fitting the data, however this is rarely done (as it is not easy
to quantify) and typically only the error in the measurement is
used when weighing the different data points in the fit.
Secondly, using more data points can also lead to systematic
biases if the set of model templates is too limited to fit the
complexity of the data.  For example, long-ward $K$ the SED is
dominated by dust re-emission which is very difficult to model. 
Without a good description of dust re-emission, the fitting of
far-IR data is meaningless and can even impede a proper understanding
of shorter wavelength data. Another case is when smooth (e.g.
exponentially declining) SFHs are used with data sets that include
indicators very sensitive to recent star formation. In such cases,
templates should include at least a combination of a smooth SFH
and a late starburst. 

Methods furthermore vary according to the number of templates fitted
to the data, namely the range and type of SFHs and metallicity distributions
(single burst, multiple burst, exponential SFH, etc.). This can lead to
significant offsets whether one assumes that the most
significant SF burst occured 12 Gyr ago \citep[e.g.][]{Bell01}
or more recently \citep[e.g.][]{trager00}. Integrated light SEDs
rarely contain enough information to discriminate between burst
timescales 8--13 Gyr ago, which can lead to offsets as high as
0.2 dex in $M^*$. 

Finally, many mathematically different techniques have been used
to fit models to data, ranging from simple minimum $\chi^2$ fitting
\citep[e.g.][]{bell03}, to maximum-likelihood \citep[e.g.][]{kauffmann03,taylor11},
and Bayesian methods \citep[e.g.][]{Auger09}.
For large data samples, information compression techniques are often used
to reduce computational time, such as principle
component analysis \citep{chen12}, non-negative
matrix factorization \citep{blanton07}, and the linear compression technique used
in MOPED \citep{panter07}. The optimum amount of reduction allowed
while retaining all information available will depend on the data quality
and model used and may be hard to determine \citep{tojeiro07,graff11}.
\subsection{Robustness of Stellar Mass Derivations}
Determining the accuracy and robustness of stellar mass estimation
from SED fitting is non trivial since some of the intrinsic key
properties of galaxies, most notably the SFH and the stellar
IMF, are unknown. Nonetheless, estimates of uncertainties and systematic
biases of a particular fitting method can be obtained by testing
the results on mock galaxies. These mock galaxies
are often based on semi-analytic galaxy formation models, providing
hundreds of thousands of test galaxies with a wide range of (hopefully
realistic) SFHs.  Such comparisons can also provide guidance on the
minimal and optimal data sets to use when fitting real data
\citep{pforr12,wilkins13}.

The sections above make clear that several parameters affect the accuracy
and biases of stellar mass estimates, the most important being: the IMF,
data quality, EPS model, SFHs and chemical evolution, dust, and redshift.
More detailed analyses can be found in, e.g. \citet{conroygunn09},
\citet{gallazzi09}, \citet{pforr12}, \cite{wilkins13}, and
references therein.

Firstly, the IMF is the main systematic uncertainty in \ml\ estimation.
However, as long as the IMF slope does not change for $M > 1~\Msol$,
as is the case for \citet{Salpeter55}, \citet{Kennicutt83}, \citet{kroupa01},
or \citet{chabrier03} IMFs, the incognita from the Initial Mass Function only 
results in a constant offset in stellar mass as the luminosity does not vary
and only the total mass of a stellar population changes. A significant change
of the IMF slope above $M > 1~\Msol$ implies a non constant offset
\citep[e.g.][]{pforr12}. 

Constraints on the IMF normalization can be derived by comparing stellar
population mass estimates to dynamical or lensing mass estimates such as
described in the remainder of this review, with the simple notion that the
mass in stars should not exceed the total measured one. Such a comparison
is done in \citet[e.g.][]{Bell01} using disk galaxy rotation curves. However,
only upper limits can be obtained, since a fraction of the luminous mass can
always be traded off to a (smooth) dark matter component with a similar
mass distribution as the light. Only when there is a dynamical substructure
on scales smaller than those expected for dark matter one can hope to
obtain lower limits to the stellar mass normalizations. Examples
include the dynamics of bar and spiral structure in disk galaxies, the
effect of bulges on galaxy rotation curves, and the vertical velocity
dispersion in disk galaxies (see \se{gasrich}). \citet{deJong07} 
performed such a comparison between dynamical and stellar population
mass estimates and showed that the SED-\ml\ relation normalization
can be constrained to within $\sim$0.2 dex (66\% confidence level).
If the IMF varies amongst galaxy types, as argued for instance by 
\citet{Treu10}, \citet{Auger10}, \citet{Cappellari+12} and
\citet{Conroy12} (see also \se{stellpopimf} and \se{DMfraction}),
the IMF effect becomes much more complicated and unpredictable,
let alone any evolution of the IMF with redshift \citep{vandokkum08}.

Proceeding with random uncertainties, the accuracy of mass estimation
depends on the quality of the data. For example, \citet{gallazzi09}
determine that a spectral signal-to-noise of S/N$>$20 is required
to get an accuracy of 0.1 dex in \ml\, when using a few optical indices.
\citet{tojeiro09} performing full spectral fitting derived a $\sim 0.1$ dex
uncertainty in $M^{*}$~ due to the quality of the data ($S/N\sim10-15$
per pixel). \citet{chen12} exploited low S/N high-$z$ spectra from the
SDSS-III/BOSS survey to argue that spectral fitting relying on a
principal component analysis can lead to reliable results also for
S/N$\sim$5, though the actual parameters of the populations may not
be well determined. Fitting the broad-band signal instead of the
spectrum for the low S/N case may be the best choice \cite{maraston13},
and the two methods converge at high S/N.

The use of different EPS models also affects stellar masses, with
an uncertainty of $\sim$0.2 to 0.3 dex 
\citep[on the logarithimic mass,][]{maraston06,conroygunn09,Ilbert10},
reaching at most 0.6 dex at $z\sim 2$ \citep{conroygunn09} when TP-AGB
stars dominate the spectrum and the models vary most (see \se{ingredients}).

The adopted SFH in the models significantly
contributes to the uncertainty and biases in \ml\ values
\citep[e.g.][]{maraston10} for high-redshift galaxies. The
range in functional forms (if a function is used at all), the
chosen oldest and youngest stars, and the inclusion and number of starbursts
all affect the uncertainty and bias in the \ml\ determination.

For galaxies at low-redshift with late star formation, \citet{gallazzi09}
find that the \ml\ may be overestimated by 0.1 dex when using spectral
indices.  Nearly similar accuracies can be obtained
using one optical color, the choice of which may vary with redshift
\citep{wilkins13}. Using more than one color reduces the offsets
slightly, but even more colors will not reduce the uncertainties
and offsets \citep{gallazzi09,zibcha09}.
This is in apparent contrast with \citet{pforr12}, who argue
that more passbands further reduce uncertainties and stress
the benefits of near-IR data.  This may
be related to the limited set of templates adopted by Pforr et~al.\ 
compared to the star formation histories in their mock catalog
(if the templates do not fully span the range of ``observed''
galaxy parameters, more data help getting closer to the correct result)
and possibly due to their inclusion of dust effects. \citet{pforr12}
also show that - when the template star formation history matches the
real one - as is the case for positive tau models for high-redshift
galaxies \citep{maraston10}, the mass recovery from spectral fit is
excellent and mostly independent of the waveband used in the fit.

Indeed, from the point of view of the SFH, high-redshift
galaxies have less uncertain \ml\ as long as the redshifted
data capture all necessary rest-frame wavelengths. This stems
from a decrease in the number of possible SFHs (less time
has passed since the initial star formation).  Most notably the
age difference between any old, underlying population and a recent
burst is smaller, and the ``frosting'' or outshining effect by
young stars is decreased \citep{maraston10}. For similar quality
multi-passband data, this can reduce biases for dusty galaxies
from 0.5 dex at $z=0.5$ to 0.1 dex at $z=2$ and uncertainties
from 0.5 dex at $z=0.5$ to 0.2 dex at $z=2$ \citep{pforr12}.

Finally, dust in galaxies significantly alters the SED
\citep[e.g.][]{macarthur05}. In case of very dusty systems
spectroscopy clearly becomes the favoured channel for \ml\ estimation.
Access to both photometric and spectroscopic data would enable
a derivation of extinction values by comparing SED expectations
derived from the spectral features to the observed photometry
\citep[e.g.][]{kauffmann03}.  \citet{pforr12} show that including
dust prescription in the SED fit dramatically increases the
uncertainties and biases in stellar \ml\ estimates especially
at low redshifts, with offsets being as large 0.5 dex and similarly
sized rms uncertainties. When focussing on mass estimation, the
conservative choice of neglecting dust as an additional parameter
in the spectral fitting may lead to a more robust determination
of stellar mass \citep{maraston13}. 

\newpage
\subsection{Future Prospects}\label{sec:SPsummary}

The stellar mass of galaxies, $M^{*}$, is a key parameter in studying
the formation and evolution of galaxies over the cosmic time, tracing
galaxy dynamics, and disentangling the contribution from dark matter
to the overall galaxy potential. In this Section, we have reviewed
the basic physics entering the derivation of $M^{*}$. This exploits
the theory of stellar evolution to calculate the mass in stars
from the amount of stellar light that galaxies emit. The modelling
of the integrated galaxy light requires several assumptions
regarding the distribution of ages and chemical compositions
of the component stars (i.e. the galaxy star formation history),
the distribution of stellar masses at birth (the stellar initial
mass function) and the attenuation of light from dust.
The fundamental tool to perform such modelling is called a
``Stellar Population model''. We have described such models,
highlighting their main uncertainties and how these affect
the derivation of $M^{*}$. Using models from various authors
we visualised the basic relations between $M^{*}$ and the integrated
spectral energy distribution, including color and spectral indices.
Furthermore, we discussed the part of $M^{*}$ which is composed
of stellar remnants, such as white dwarfs, neutron stars and black
holes, which do not emit light but contribute to the total mass,
and their dependence on the stellar initial mass function. 

Finally, we briefly addressed the main techniques to calculate
$M^{*}$ found in the literature and discussed the typical
uncertainties and biases of SED based on stellar mass estimates. 
While these depend on galaxy type, they are: (i) for star-forming
galaxies, the unknown star formation history and the fact that
a small fraction by mass of newly born stars can outshine the
underlying older population dominating the mass, thus jeopardising
the mass derivation; and (ii) for both star-forming and passively
evolving galaxies, the unknown IMF. Spectro-photometric data
are most crucial for treating star-forming galaxies whereas
the near-IR bands help in constraining the older, outshined
component of the stellar population \citep[see Figure 3 in][]{maraston10}.

Despite many unknowns, and excluding extreme cases of very dusty galaxies
or galaxies with complicated and bursty star formation histories,
relative stellar masses between galaxies can be regarded robust within
$0.2-0.3$~dex. The still poorly constraint IMF normalisation will
shift SED derived \ml\ values of all galaxies up or down by a few
tenth of a dex, and if IMF variations occur across galaxy types
and/or with redshift errors in SED, \ml\ estimation may be as large
as 0.5 dex.  Recent simulations have shown how fundamental model
parameters such as the choice of SFH, adopted wavelength range,
redshift, and inclusion of dust, contribute to the uncertainties
and can be used as a quantitative guide to assess uncertainties
in $M^{*}$ \citep{maraston10,maraston13,pforr12,wilkins13}.

The most urgently required model improvements include constraining
residual uncertainties in stellar evolution, specifically regarding
the temperature of the RGB and the energetics of the TP-AGB stars,
and the effect of non-solar abundance ratios on spectra. 
An improved understanding of star formation history effects
for low-$z$~galaxies would also be beneficial.

This review section is by no means complete, but it provides
the necessary background to understand several statements made
in forthcoming sections.  Other comprehensive reviews which
address stellar mass estimates in galaxies include
\citet{conroygunn09}, \citet{greggio11}, \citet{walcher11},
and \citet{ConroyARAA13}.


\newpage

\section{Dynamical Masses of Gas-Rich Galaxies}\label{sec:gasrich} 
 
Galaxy masses were first inferred from spiral galaxy rotation curves,
which were themselves first measured in the early 1900's
\citep{Scheiner1899,Slipher14,Pease18}.  
The nebular lines from which velocity curves were derived already
showed some evidence of a ``tilt'', indicative of rotation, even
though these early spectra only sampled the inner parts of the galaxy.
In his most original 1922 paper and using Pease's (1918) velocity
curve of the Andromeda galaxy, \citeauthor{Opik22} inferred a mass of
$4.5\times10^9 \Msol$ within 150\arcsec ($\sim{0.6}$ kpc) for M31.  He
did so by requiring that the M31 disk have a mass-to-light (hereafter
M/L) ratio comparable to that of the solar neighborhood.  This is the
first reported measurement of galaxy mass.  That same year,
\cite{Kapteyn22} remarked in his study of the Milky Way's local mass
density that ``We have the means of estimating the mass of dark matter
in the universe.''  This appears to be the very first reference to
the concept of ``dark matter'' in astrophysics.  His dynamical analysis
and determination of the density in the solar neighborhood is also the
first of its kind.  However, Kapteyn failed to find the elusive dark
matter given the limitations of his data and preliminary method,
as reviewed by \citet{Oort32}.  The latter can be credited for the
first discovery of dark matter in galaxies (see \citealt{Zwicky33}
for a similar discovery of dark matter in clusters of galaxies).
In his 1940 study of NGC 3115, \citeauthor{Oort40} remarks: ``the
distribution of mass in the system appears to bear almost no relation
to that of the light''.  This is yet another pioneering report
of large mass-to-light ratios in galaxies.

About our own Galaxy, \citet{Rubin62} remarked ``For $R>8.5$~kpc, the
stellar [rotation] curve is flat, and does not decrease as is expected
for Keplerian orbits.''  Indeed, by the late 1960's, improved detectors
at optical and radio frequencies yielded routine detections of flat
galaxy rotation curves. Using a then-state-of-the-art image-tube
spectrograph at the KPNO 84 inch telescope, Vera Rubin and Kent Ford
(1970) obtained the first extended rotation curve of a galaxy (M31)
out to 120\arcmin ($\sim 27$ kpc).
\citet{Roberts75} confirmed the flatness of the M31 rotation curve
observed by Rubin and Ford with 21cm velocities
extending to 170\arcmin ($\sim 38$ kpc). \citeauthor{Roberts75}
however contended that dwarf M stars are adequate to explain the
required mass and mass-to-light ratio.  They, as well as most
astronomers then, seem to have missed \cite{Freeman70}'s note that
the \hi\ rotation curves that were available at the time did not
turn over at the radius expected from their surface photometry. 
Freeman's 1970 paper, commonly cited for its study of exponential
disks in galaxies (rather than the note about), was likely the very
first to quantify the mismatch at large galactocentric radii between
the observed rotation curve and the rotation curve expected from
the light distribution and constant M/L.

The flatness of observed rotation curves in all galaxy types is now a
well-established fact \citep{Faber79,Rubin85,Sofue01} but it is not by
itself proof of dark matter in galaxies (see the prophetic contribution
by Kalnajs in the 1982 Besancon conference proceedings \citet{IAU100}).
Despite notable efforts by the likes of Kapteyn, Oort, Babcock, Mayall,
de Vaucouleurs, Schwarzschild, the Burbidge's, Roberts, Rubin, and others,
and the realization that luminous galaxies are not a simple Keplerian
environment, the firm manifestation of dark matter through galaxy
rotation curves would await extended dynamical measurements at radio
wavelengths (21cm line of neutral hydrogen), especially with the
Westerbork Synthesis Radio Telescope (WSRT)
\citep[e.g.][]{Rogstad72,Bosma78,vanderKruit78},
and the ability to place upper limits on the contribution of the
baryonic component to the total observed rotation curve
\citep{Bosma78,Carignan85,vanAlbada85,Kent86,Kent87}.

The first detailed and unambiguous demonstration of unseen mass in
galaxy disks from the mass modeling of galaxy rotation curves came
with Albert Bosma's outstanding PhD thesis in \citeyear{Bosma78}.
Using early disk analysis methods by \cite{toomre63}, \citet{shu71},
and \citet{Nordsieck73}, \citeauthor{Bosma78} was able to decompose
the extended rotation curves of 25 spiral galaxies to show for the
first time that the total M/L ratio of galaxies grows with radius.
To our knowledge, \cite{Faber79} were the first to link measurements
for the local mass density \citep{Oort32,Oort65}, the velocity dispersion
in galaxy clusters \citep{Zwicky33}, and the notion of flat
extended galaxy rotation curves \citep{Bosma78}, into a coherent picture
of missing mass on global galactic and extragalactic scales.

While measurements of gas and stellar motions for mass estimates of
gas-rich galaxies are now fairly straightforward, it is of relevance
to discuss the applicability and accuracy of their related mass
estimators.  Modern mass modeling of galaxy rotation curves, and the
ability to disentangle baryonic and non-baryonic components, are being
reviewed below.

A modern review of the structure of galaxy disks can
also be found in \citet[][hereafter vdKF11]{vdKFreeman2011}.
We shall defer to that review in some cases below for more detailed
discussions and/or derivations than can be provided here.

In their 2001 ARAA article, Sofue and Rubin write {\sl ``Babcock
and Oort share credit for uncovering the dark matter problem in
individual spiral galaxies''} for their work in the 1930's.
The pioneering contributions in the 1960's-80's of Toomre,
Kalnajs, Shu, Freeman, Bosma, Carignan, Kent, van Albada,
van der Kruit, Sancisi, and others to the problem of galaxy
mass models should also be underscored.

\subsection{Mass Estimates from Rotation Curves}\label{sec:masses}

The mass distribution in disk galaxies is typically determined from
resolved rotation curves or integrated line profiles extracted from
emission lines such as \ha, CO, and \hi\ lines.  With the current
generation of detectors, the \ha\ and CO lines yield high spectral
resolution spectra over most of the optical disk; greater spatial
coverage (often at the expense of spectral resolution) is usually
obtained with resolved \hi\ velocity curves.  Integrated line widths
only yield an estimate of a total mass within some (uncertain)
isophotal radius.  A more accurate assessment of the extended galaxy
mass profile is obtained from 2D resolved \hi\ velocity fields but the
prohibitive exposure times constrain sample sizes
\citep[e.g.][]{deBlok08}.  For nearby disks, \ha\ velocity fields
(with, e.g., SparsePak, DensePak, PPAK) and CO velocity fields (with,
e.g. CARMA) are just as slow to obtain.
The extent of the neutral gas in spiral galaxies, as traced by
rotation curves, can often exceed twice that of the stars.

There is good agreement between resolved rotation curves extracted from
\ha, \hi\ and CO lines \citep{Sofue01,Simon03,Simon05,Spekkens07}
and from \oii, \oiii, \hbeta, \nii, and \sii\ lines \citep{CourteauSohn03}
within the optical disk of galaxies.

It is often assumed that \hi\ line widths sample the disk to large
galactocentric radii, by analogy to resolved \hi\ rotation curves;
however, \hi\ line widths are a convolution of gas dynamics and
exponentially declining gas surface densities
\citep[e.g.][]{Cayatte94} such that the effective depth of integrated
line widths is likely representative of the gas distribution within
the optical disk of a galaxy.  This is in line with the many linear
transformations that exist between \ha\ rotation measures and \hi\
line widths \citep[][to name a few]{Mathewson92,
  Courteau97,Catinella07}.

The circular velocity of a spherical system in a potential $\Phi$
is given by
\be V^2_{\rm circ}(r) = r {\frac{d\Phi}{dr}} = G {\frac{M(r)}{r}}
\label{eq:vcirc}
\ee 
where $M(r)$ is the enclosed mass within a sphere of radius $r$.
For a flattened disk, as in most spiral galaxies, the right hand side
of \eq{vcirc} must to be replaced by the more exact expression derived
by \citet{Freeman70}.  In the absence of dark matter or bulge,
it should be stated that the exact expression for the rotation curve
of a self-gravitating exponential disk is described by \citep{Freeman70}:
\be V_{\rm circ}^2(R)= 4\pi G \Sigma_0 R_{\rm d} y^2 \bigl [
I_0(y)~K_0(y) - I_1(y)~K_1(y) \bigr ],
\label{eq:Freemandisk}
\ee 
where, $G$ is the gravitational constant, $\Sigma_0$ the central
surface brightness, $R_{\rm d}$ the disk exponential scale length,
$y\equiv R/2R_{\rm d}$, and $I_i(y)$ and $K_i(y)$ are the modified
Bessel functions of the first and second kind \citep[][see also vdKF11
and Fig 2.17 of Binney \& Tremaine 2008a]{Freeman70}.  The rotation
curve of a pure exponential, infinitesimally thin, disk peaks at
$V_{2.2} \equiv V(R=2.15R_{\rm d})$. For disks of finite thickness
(say, $z_0/R_{\rm d}=0.2$, where $z_0$ is the disk scale height), the
rotation curve will have a very similar shape but a $\sim 5$\% lower
peak \citep[e.g.][]{Casertano83}; this will slightly affect the
``shape'' term in the square brackets in \eq{Freemandisk}, but
leave the subsequent scalings untouched.

The reliability of $M(r)$ depends on how the rotational velocities, 
$V_{\rm rot}$, reflect the assumed circular velocities.  Several factors
which involve corrections for observational and physical effects come
into play.  First, the observed line-of-sight velocity, $\Vlos$, must
be corrected for projection via:
\be
 V_{\rm rot} = \Vlos / \sin i.  
\label{eq:vcorr}
\ee
where $i$ is the inclination of the galaxy disk \citep[or a ring along
that disk, for tilted ring solutions;][]{Teuben2002}.  The above equation
applies for velocities along the major axis; for the general case along
any other projection, see \citet{Teuben2002}.

The spatially-resolved rotation curve or integrated line width of
a rotating system is obtained via the Doppler equation: 
\be 
 {\Vlos}(r)= c [\lambda(r) - \lambda_0]/\lambda_0, 
\ee where
$\lambda_0$ is the observed wavelength of the galaxy center
and $r$ is the position along the slit.  A line width $W$, 
usually equal to $2\Vobs$, must also be corrected for internal
turbulence and other effects \citep[e.g.][]{Haynes84,Catinella07}.

The inclination angle, $i$, for an oblate spheroid is given
by the formula (Holmberg 1946):  
\be i = \cos^{-1}
\sqrt{\frac{(b/a)^2 - q^2_0}{1-q^2_0}},
\label{eq:inc}
\ee 
where the semi-major ($a$) and semi-minor ($b$) axes are determined
from isophotal fitting of the galaxy image, and $q_0$ is the axial
ratio of a galaxy viewed edge-on (for late-type disks, $q_0 \simeq
0.13$ \citep[][and references therein]{Hall12}).

\begin{figure*}[t]
\centering
\includegraphics[scale=0.4]{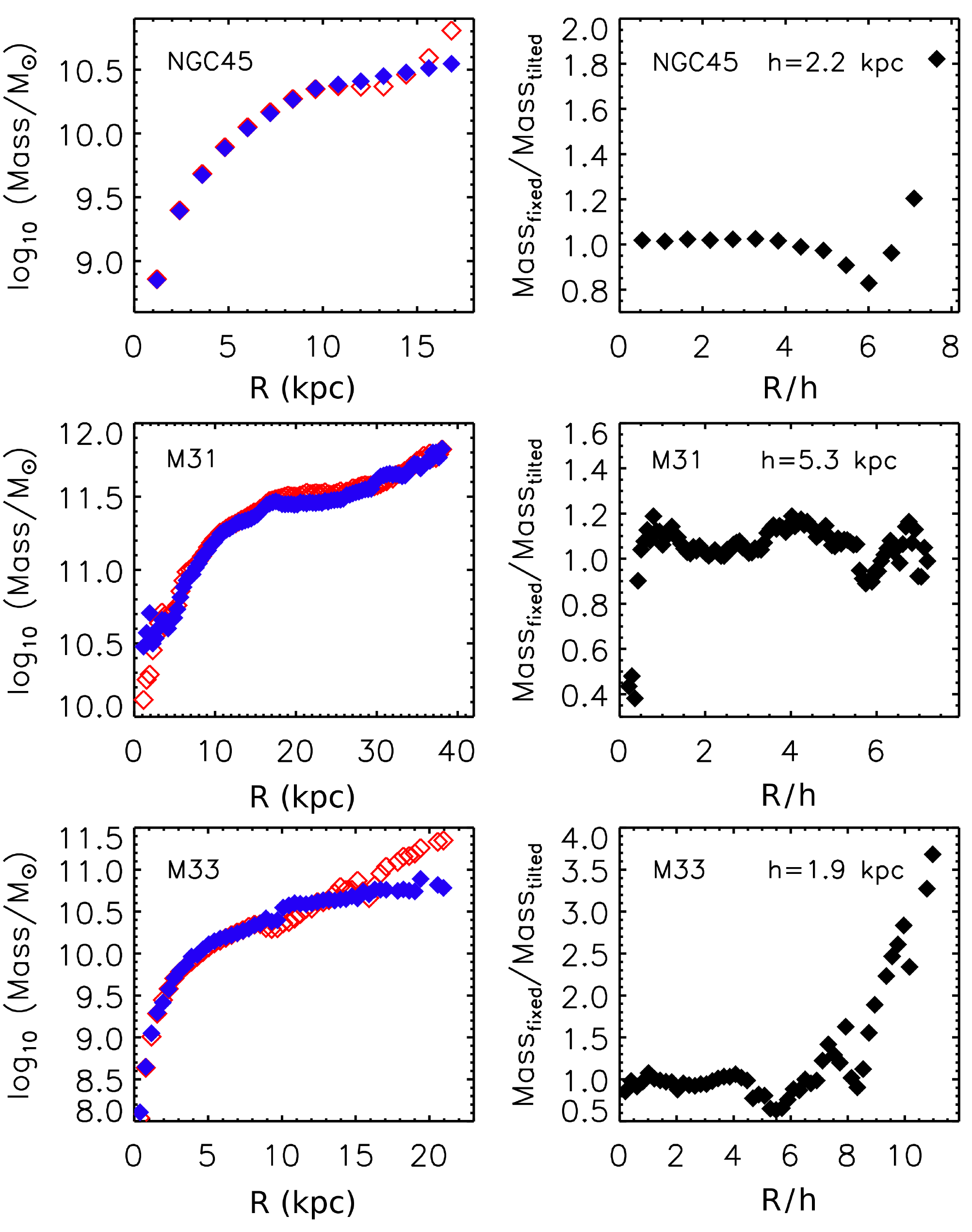}
\caption{Mass profiles from \hi\ rotation curves of NGC 45
\citep{Chemin06a}, M31 \citep{Chemin09a}, and M33 (Chemin et al.,
in prep.).  Mass profiles on the left were obtained from rotation
curves assuming a kinematical model with constant inclination
and constant position angle of the major axis (red open squares)
or a full tilted ring model (blue filled squares).
The adopted distances are NGC 45 ($D=5.9$ Mpc), M31 ($D=785$ kpc),
M33 ($D=800$ kpc).  The ratio of these mass profiles is shown on the
right side where the radial scale has been normalized by the optical
exponential disk scale length, $h$.}
\label{fig:chemintilt}
\end{figure*}

Inclination uncertainties in \Eq{vcirc} can be most significant for
systems with $i<30^\circ$ or for distant galaxies whose disk is poorly
resolved (in such cases, space-based observations or adaptive optics
are needed to overcome the effect of atmospheric
blur). 
Inclination estimates vary as a function of wavelength and are clearly
affected by warps beyond the optical disks \citep{Briggs90}.  Tilt
uncertainty, such as due to warps, can account for a significant
fraction of the mass budget in the outer parts of the disk.  This is
illustrated in \Fig{chemintilt} for the mass profiles of NGC 45, M31,
and M33 obtained assuming a kinematical model with constant
inclination and major axis position angle (red open squares) or a full
tilted ring model (blue filled squares).  The ratio of these curves is
shown on the right side of \Fig{chemintilt} as a function of radius
normalized by the disk exponential scale length, $h$.  While the value
of $h$ depends on the assumed distance to the galaxy and details of
surface brightness profile fitting \citep{Courteau11},
\Fig{chemintilt} makes clear that tilted ring models are required for
rotation curves extending beyond 5-6 optical disk scale lengths.

The effects of slit misalignment (erroneous position angles)
for long-slit spectra on mass estimates are also addressed
in \citet{rhee04}, \citet{spekkens05}, and \citet{Chemin06b}.

Ideally, a total enclosed galaxy mass should be determined using the
most extended rotation curve and probing a regime where the RC is
flat.  Velocity curves can be modeled using a fitting function
\citep{Courteau97,Giovanelli02}.  This is especially useful for low
mass systems whose observed rotation curves rarely reach a plateau
(caution is advised against extrapolations) or for noisy rotation
curves with spiral structure wiggles.  The reliability of mass
estimates depends greatly on the stability of the velocity measure.
For instance, $\Vobs$ is typically measured for one-sided resolved
rotation curves at a fiducial marker such as $V(R_{\rm max})$, where
$R_{\rm max}$ is the radius at which $\Vobs$ reaches its peak value or
$V_{23.5}$, the velocity measured at the 23.5 \magarc I-band isophote
\citep{Courteau97}, or at half of a suitably chosen width of an
integrated line profile \citep{Haynes84,Springob05}.  Different
definitions of rotational velocities or line widths can yield tighter
galaxy scaling relations.  For example, measurements of $\Vobs$ either
at $V_{2.2}$ or $V_{23.5}$ yield the tightest scatter in various
galaxy scaling relations for bright galaxies \citep{Courteau97}.  For
the rising rotation curves of lower surface brightness systems,
\citet{Catinella07} note that the $V_{2.2}$ values may not probe the
RC deeply enough and that these may therefore show a surface
brightness dependence.  $V_{23.5}$ would thus be a safer mass tracer,
provided that the rotation curve is sampled that far.

Optical (typically \ha) rotation curves for bright galaxies may show
extended flattening out to 4-5 disk scale lengths \citep{Courteau97},
such that a maximum $V_{\rm max}=V(R_{\rm max})$ may be estimated.
\hi\ rotation curves routinely extend to that radius, making the
measurement of $V_{\rm max}$ straightforward from these data.

Beyond galaxy disks, little is known about the mass profiles of
individual galaxies. By stacking galaxies of similar masses or
luminosities, it is possible to use weak gravitational lensing (see
\se{weak}) or satellite kinematics \cite[e.g.][]{More11} to measure
total masses within the virial radius of the dark matter halo.  This
mass can be trivially converted into the circular velocity at the
virial radius, $V_{200}$. By comparing with the rotation velocities
within the optical disk, $V_{\rm opt}$, one finds that on average for
late-type galaxies $V_{\rm opt}/V_{200}\simeq 1.2$
\citep{Dutton10,reyes12}. Thus, the dark matter near the virial radius
may have slightly lower circular velocity than that of the inner
baryons, and extended rotation curves are best decomposed into their
major components (bar, bulge, disk, halo) rather than extrapolated.
We discuss such decompositions in \se{modeling}.

The velocity function (or the number of galaxies per unit circular
velocity per volume) of spiral galaxies has been measured directly for
the Virgo cluster by \citet{Papastergis11}.  The full spectrum of
(projected) line widths ranges from 20~\kms, where corrections for
turbulence dominate measurement uncertainty, to more than 400~\kms.
\footnote{The fastest reported galaxy ``disk'' is that of UGC 12591,
an S0/Sa galaxy rotating at 500~\kms and having a total mass within
$R_{25}$ equal to $1.4 \times 10^{12} \Msol$ ($H_0=70$ \kms Mpc$^{-1}$)
\citep{Giovanelli86}.  
Considering its X-ray emission and surprisingly low baryon mass
fraction \citep[3-5\%;][]{Dai2012}, this galaxy is likely a massive
spheroid that has accreted its, now rotating, neutral gas after
assembling most of its stellar mass.  The late merger/accretion
event that formed the disk of UGC 12591 may have also turned on
a massive outflow to drive its gas halo out to very large radii.
As such, UGC 12591 should clearly not be compared against normal
spiral galaxies.}


For a spherically symmetric system, the total mass enclosed within
a radius $R$ can be written in solar units as: 
\be M(R) = 2.33 \times 10^5 R \Vobs^2 / \sin^2(i) \ \ \Msol
\label{eq:mass}
\ee
where $R$ is the radius along the major axis in kpc and $\Vobs$ is the
observed rotation velocity in \kms. Galaxy masses are thus best
measured for systems with accurate distances; galaxies in unvirialized
clusters or close enough ($cz \lsim 5000$ \kms) to experience
substantial deviations from the Hubble flow clearly suffer (linearly)
from significant distance estimate errors.

Note that while the total enclosed mass, $M(R)$, is corrected for
projection through \Eq{mass}, the complete deprojection of a rotation
curve into a radial mass profile requires a tilted-ring analysis of
the light distribution to account for the combined effects of a bulge,
bar and disk (and sometimes a stellar halo) and isophotal warps
\Fig{chemintilt}.  Other physical effects discussed below in
\se{inner} make this endeavor, especially in the inner parts of a
galaxy, a rather challenging and uncertain one.  To probe the total
galaxy potential in its outskirts would require other tracers such as
planetary nebulae, globular clusters and satellites.


\subsection{Inner Parts}\label{sec:inner}

In addition to errors introduced by deprojection effects and distance
uncertainty, 
fundamental physical complications thwart the direct interpretation of
galaxy rotation curves via \Eq{mass} in their inner parts.  These
include deviations from circular orbits due to bar-like perturbations,
differential dust opacity in the bulge and inner disk, density profile
variations due to a triaxial halo, and more.  We address these briefly
below.

Because a large fraction of gas-rich galaxies have non-axisymmetric
inner parts, non-circular velocities are often observed within the
corotation radius of gas-rich galaxies
\citep{Lindblad96,Weiner01,Courteau03,Valenzuela07,Spekkens07,Sellwood10}.  
Once identified, correcting for non-circular motions is a daunting task.
To relate those motions with a photometric bar requires detailed,
model-dependent fluid dynamical simulations (see \se{fluid}).  
Beyond the self-consistent treatment of the bar in these fluid-dynamical
models, the correlation between the observed photometric bar strength
(or length) and the amplitude of bar-like non-circular motions has not
been widely explored \citep{Spekkens07,Sellwood10,Kuzio12}; the latter
requires high-quality 2D velocity fields and an empirical means to
measure bar-like flows.  Furthermore, the contribution of a bar or
oval distortion to non-circular motions depends on the angle between
the bar axis and the major axis.  The amplitude of a rotation curve
will be biased high/low if a bar close to the minor/major axis is
neglected \citep{Spekkens07,Sellwood10}.

\begin{figure*}[t]
\centering
\includegraphics[width=0.4\textwidth]{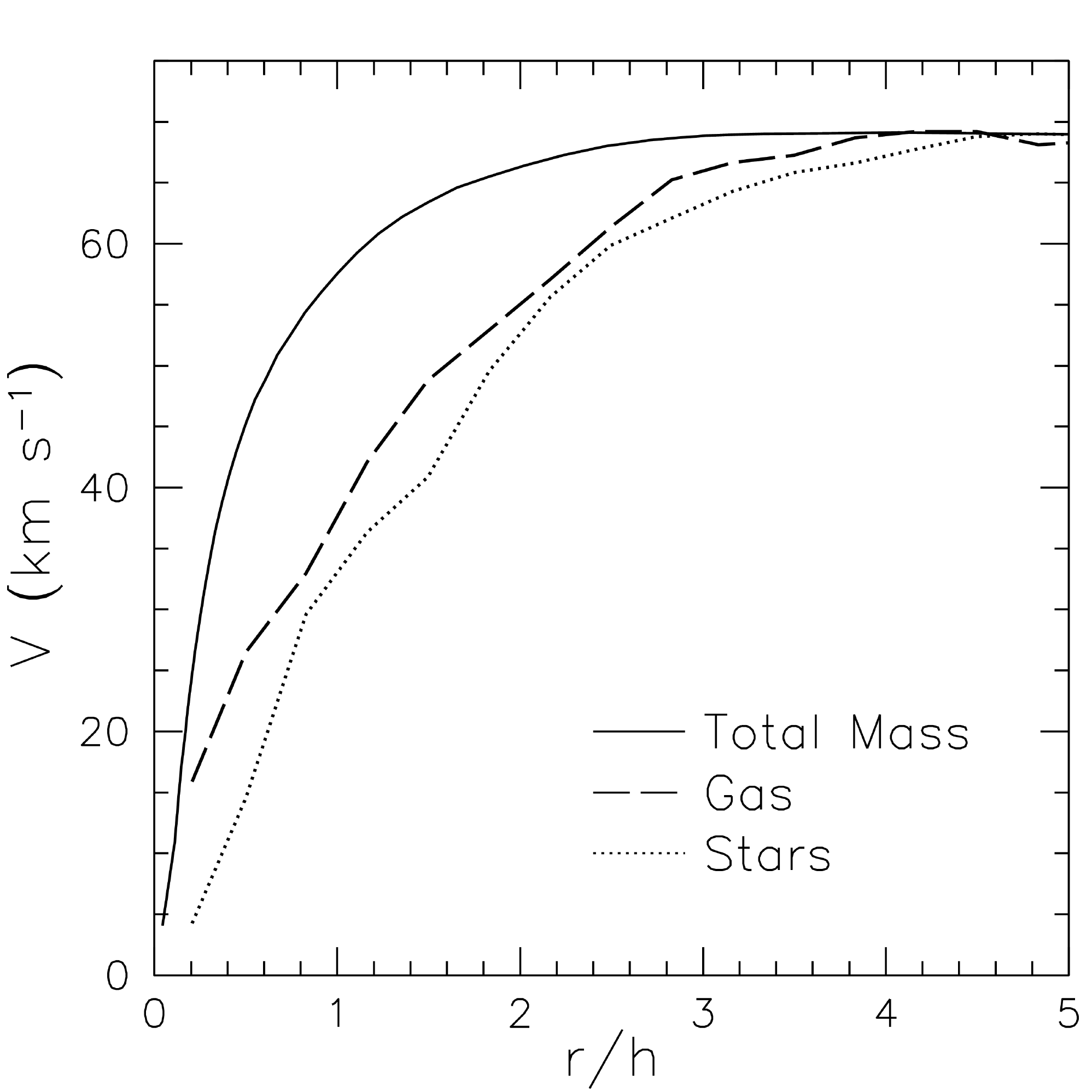}
\caption{Simulated velocity curves for a dwarf galaxy from
 \citet{Valenzuela07}.  The solid curve shows the spherical
 averaged circular velocity (GM$_{\rm tot}/r$)$^{1/2}$.
 The long-dashed curve is for the azimuthally-averaged rotation
 velocity of gas particles.  The stellar rotation velocity is
 shown by the dotted curve.  Curves for the gas and stars
 are substantially below the circular velocity for $r < 3h$, 
 where $h$ is the disk scale length.  
}
\label{fig:Valenzuela}
\end{figure*}

As an example for the phenomenological description of the influence
of a bar on measured rotation curves, we show in \Fig{Valenzuela}
a diagram adapted from \cite{Valenzuela07}.  These authors used
N-body simulations to show that non-circular motions, combined
with gas pressure support and projection effects, can result
in a large underestimation of the circular velocity in the
central $\sim1$~kpc region of a gas-rich dwarf galaxy.
While those highlighted effects are stronger in barred systems, they
are also present in axisymmetric disks.  Their simulations show that
$V_{\rm obs} \simeq V_{\rm circ}$ only beyond 3 disk scale lengths.

The differential opacity in spiral disks \citep{Bosma92,Giovanelli02}
also cautions for a careful interpretation of optical rotation curves.
The effects of dust extinction on velocity curves, which tend to mimic
solid-body rotation, can be significant at high tilt ($>80^\circ$) and
for wavelengths bluer than \ha\ \citep{Courteau92,Bosma92} within one
disk scale length.  This ``tapering'' effect is also luminosity
dependent, being stronger for the most luminous systems
\citep{Giovanelli02}.  This problem is mostly overcome by measuring
rotation curves at near-infrared wavelengths (e.g. through the
Pa~$\alpha$ and Br~$\gamma$ hydrogen recombination lines) where
extinction is minimised\footnote{Clearly, the issue of extinction
is less conspicuous in nearly face-on systems \citep{Andersen06}.}.
Radio observations nearly alleviate this concern, though lower
resolution at \hi\ may affect the rise of the rotation curve in the
central regions \citep{Bosma78,vdbosch01}, and the HI optical depth
and related self-absorption effects are not negligible in more
inclined galaxies \citep{Rupen91}.
The high-spatial resolution kinematics of galaxies' inner regions from
high-$J$ CO line spectroscopy using the Atacama Large Millimeter Array
will soon lessen these issues, though similar concerns as above for
more inclined galaxies apply.
Effects due to triaxiality and flattening of the disk have been
discussed in \citet{Dutton05} and \citet[][Fig. 2.13]{Binney08}.

For distant galaxies, the inner rise of the rotation curve is also
critically damped by both resolution effects and enhanced central
galaxy activity which contributes more dust per unit area at early
times.  Both dust extinction and resolution yield observed rotation
curves that are shallower than the true velocity profile
\citep{Forster06}.

Due to the complex and somewhat uncertain modeling involved in
correcting for non-circular motions and internal extinction, masses
for nearby galaxies are often extracted beyond corotation or,
equivalently, beyond two to three disk scale lengths
\citep{Kranz03,Valenzuela07}.

\subsection{Mass Modeling}\label{sec:modeling}

An ultimate goal of galaxy mass studies is the decomposition of a mass
profile into its main components -- the bulge, disk and dark matter
halo -- {\sl at all radii}.  Unlike gas-poor systems (\se{gaspoor})
whose spectral features are faint beyond one effective radius, mass
modeling decompositions can be attempted for spiral galaxies since
dynamical tracers are conspicuous from the center to the optical edge.
Pioneering mass models have been derived by \citet{Casertano83},
\citet{Wevers84}, \citet{Carignan85}, \citet{vanAlbada85} and others.

Mass modeling is possible because the gas particles and stars are
sensitive to the full potential contributed by the baryons and dark matter. 
If the matter distribution is axially symmetric and in centrifugal
equilibrium, then the total circular velocity is given by,
\be
  V_{\rm circ} \simeq \sqrt{V_{\rm{gas}}^2 + V_*^2 + V_{\rm{halo}}^2},  
\label{eq:vsqr}
\ee
at each radius $R$ in the plane of the galaxy, where $V_{\rm{gas}}$,
$V_*$, and $V_{\rm{halo}}$ are the observed rotation curves of the
gas, stars, and halo components respectively.  The latter accounts
for the baryons and presumed dark matter particles in the halo.
$V_{\rm{halo}}$ is usually inferred once the gas and stellar component
have been subtracted from the observed overall rotation curve.

The velocity curves for the stars and gas found in the galaxy bulge
and disk can be obtained by inverting their respective light
(emission) profiles into mass profiles using suitable potentials
and stellar $M/L_*$ ratios.  For instance, the stellar rotation curve
is obtained by multiplying the light profile, ideally in a band where
dust extinction effects are minimised, with an optimized $M/L_*$
consistent with stellar population models (\se{stellpop}).

The neutral atomic gas, M$_{\hi}$, can be estimated from the total
\hi\ flux density, $S_{21}$, measured from the 21-cm line in absorption
and emission:
\be
{\rm M}_{\hi} = 2.33 \times 10^5 S_{21} (D/{\textrm{Mpc}})^2 \Msol
\ee
where $D$ is the physical distance to the source in Mpc and S$_{21}$
is the integrated flux density of the source in Jy \kms.
The total gas mass which accounts for helium and other metals is given by: 
\be
{\rm M}_{\rm gas} = 1.33 {\rm M}_{\hi}
\ee
for an optically thin gas. 
An estimate of the molecular gas mass is more challenging since
H$_{2}$, the most abundant molecule in the Universe by far, has no
permanent electric dipole moment and thus cannot emit in the state in
which it is typically found. 
Consequently, the second most abundant molecule, the CO molecule which
has an electric dipole and is often optically thick, can be used since
it is collisionally exited by H$_2$.  However, the conversion from CO
intensity to H$_2$ mass, via some ``$X_{\rm CO}$'' factor, is
notoriously uncertain \citep[e.g.][]{Bolatto08}.  Still, Braine \etal
(1993) have estimated the mass contribution of molecular gas to $\sim
20$\% of the \hi\ mass, and this fraction decreases in later-type
systems.  For these reasons, we do not consider the molecular mass
content further.

Detailed mass models of spiral galaxies, including the equations for
the density profiles and associated rotation curves for the bulge,
disk and halo profiles can be found in, e.g., \citet{Dutton05},
\citet{deBlok08}, \citet{T512} and Section~11.1 of \citet{Mo10}.
Disks are often modeled as idealized infinitesimally thin, radially
exponential, collections of dust, gas and stars with surface density
distributions \citep{Freeman70}:
\be \Sigma(R) = \Sigma_0 \exp (-R/R_{\rm d}), \ee
where $R_{\rm d}$ is the scale length of the specific disk component.
The total mass of the disk is $M_{\rm d} = 2\pi \Sigma_0 R_{\rm d}^2$.  
\citet{Dutton11b} show that disk scale lengths of the gas are
on average 1.5 times greater than disk scale lengths measured in
the R-band light.  The case of a thicker disk \citep{Casertano83}
adds only a small effect to the overall rotation curve \citep{Mo10}.
Real disks, however, often show spiral arm features, truncations,
anti-truncations, and other deviations from a pure exponential
surface density distribution \citep[see e.g.][]{vdKFreeman2011}
that are best modeled through a free form reconstruction of the
stellar mass by inversion of the light profile as discussed
above after \Eq{vsqr}.


The halo profile is obtained as the difference (in quadrature) between
the observed rotation curve and the inferred baryonic components
(\eq{vsqr}).  The halo is typically modeled as a pseudo-isothermal
profile \citep{Burkert95}, a cosmologically-motivated dark profile
\citep[][hereafter NFW]{NFW96,NFW97}, or an Einasto fitting function
\citep{Einasto65,Einasto69}.  The halo profile can be conveniently
parameterized via the following function \citep{Kravtsov98}:
\be
\rho_{\rm halo}(r)= \frac{\rho_0}
{(r/r_{\rm s})^{\alpha}(1+r/r_{\rm s})^{3-\alpha}},
\label{eq:KravEq}
\ee 
where $\rho_0$ is a central density, $r_{\rm s}$ is a scale
radius, and $\alpha$ is a shape index.  This density profile has an
inner logarithmic slope of $-\alpha$.  For $\alpha=1$ this reduces to
the NFW profile, and at the scale radius, $r_{\rm s}$, the slope of
the density profile is $-2$ (isothermal).  At large radii, the
logarithmic slope is $-3$.

\subsubsection{Mass Modeling Limitations}\label{sec:adiabat}
The greatest source of uncertainty in mass modeling is the assessment
of realistic stellar $M/L_*$ ratios, followed by ill-constrained
covariances amongst halo parameters as well as between halo and disk
parameters, as we discuss below.

The computation of mass models through \Eq{vsqr} usually involves four
fundamental parameters: one for the stellar $M/L_*$, and three for the
halo component (as in \eq{KravEq}).  In the language of NFW, those
three quantities are the dark matter halo shape index, $\alpha$, a
velocity normalization, $V_{200}$, and a concentration, $c$.

\begin{figure*}[t]
\centering
\includegraphics[width=0.8\textwidth]{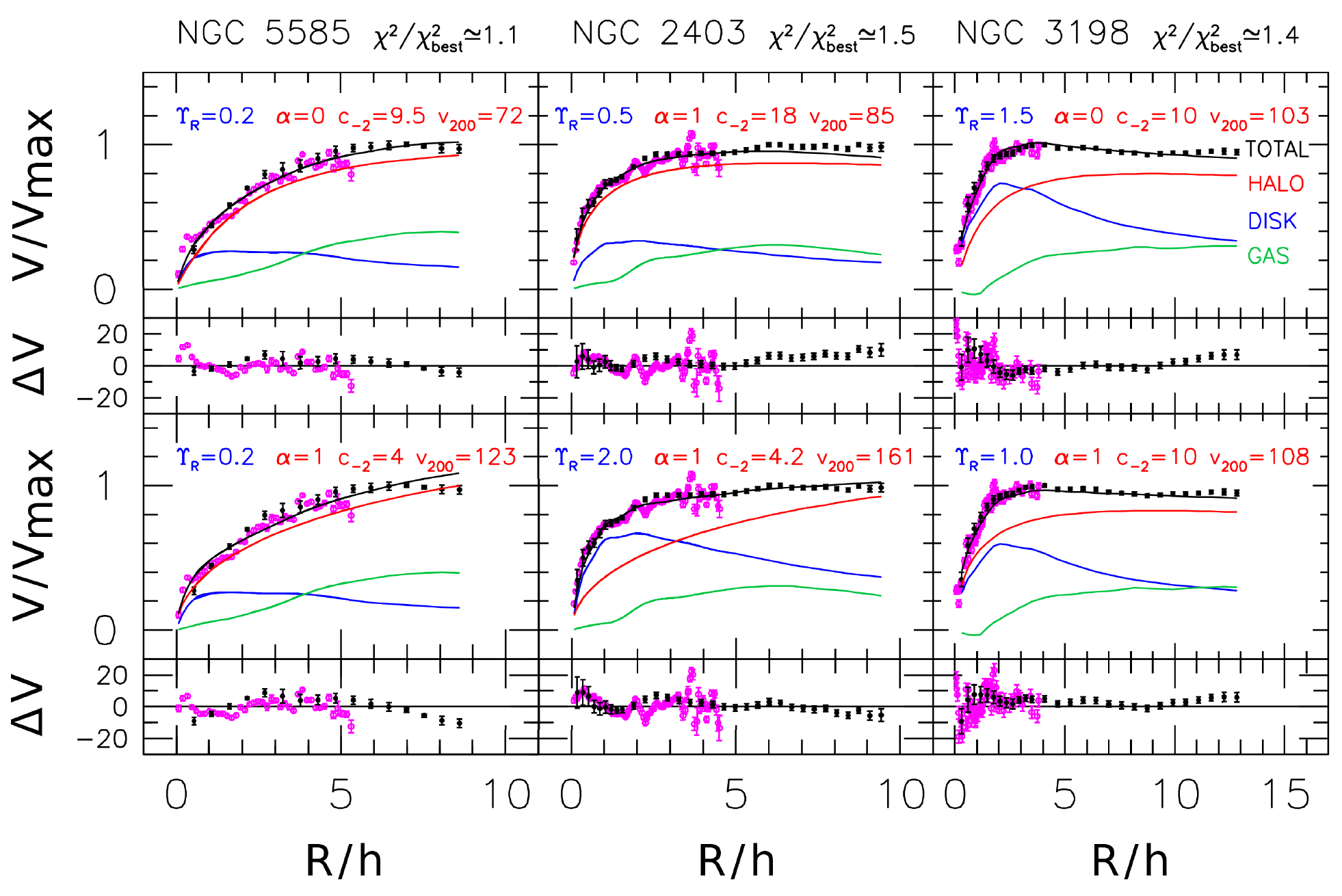}
\caption{Mass models for three spiral galaxies with a range of peak
  circular velocity.  The X-axis is in units of the optical disk scale
  length, $h$. The velocity on the Y-axis is normalized by the maximum
  observed orbital velocity.  There are four free fitting parameters:
  $\Upsilon_{\rm R}$, is the stellar $M/L_*$ (here in the R-band) and
  $\alpha$, $c$, and $V_{200}$ describe the NFW dark matter halo
  \citep[see][for details]{Dutton05}.  Variations in the fitting
  parameter can result in the same data-model residuals, as gauged by
  the model residuals $dV$ (shown in the lower windows) and the 
  overall $\chi^2$-square statistic.  The model decompositions in
  each column (two per galaxy) have the same overall $\chi^2$-square
  statistic, shown at the top. Mass models are thus non-unique.
  See text for details.}
\label{fig:Dutton}
\end{figure*}

\Fig{Dutton} highlights many of the challenges inherent to mass
modeling such as the intrinsic degeneracy of current mass model
solutions due to strong covariances between the disk and halo model
parameters \citep[see also][]{vanAlbada85}. \Fig{Dutton} shows
examples of modeled rotation curves for galaxies in three different
mass range. The purple and black points are \ha\ and \hi\ velocity
data from \citet{BlaisOuellette00}.  The fitted components are shown
for the gas (green), disk (blue) and halo models (red).  In each
panel, one model parameter is held fixed while the others can adjust
to achieve a best fit solution (by minimizing the global data-model
$\chi^2$ statistic).   In all cases, the model decompositions (two
per galaxy) in a given column have the same overall $\chi^2$ statistic.
Variations in the fitting parameters can result in the same data-model
residuals ($dV$, shown in the smaller horizontal windows) thus yielding
non-unique solutions.  The two right-side panels highlight the
well-known disk-halo degeneracy between the stellar disk $M/L_*$,
here expressed as $\Upsilon_{\rm R}$ in the R-band, and the halo
inner slope $\alpha$ (a cusp has $\alpha=1$, a core has $\alpha=0$;
see \Eq{KravEq}).  The two central panels show another facet of the
disk-halo degeneracy assuming that all dark matter halos are cuspy.
The range of acceptable parameters is large.  Finally, the left
panels show that model degeneracies exist even amongst halo parameters,
assuming a common best-fitting stellar $M/L_*$ ratio (in agreement
with stellar population models).  Thus, in order to achieve realistic
mass models, both accurate stellar $M/L_*$ {\sl and} well-constrained
cosmological models are needed.  Stellar $M/L_*$ are only accurate
to factors of two (\se{stellpop}) and the current range of allowed
halo parameters, $c$, $V_{200}$ and $\alpha$ is still too broad to
provide tight (unique) mass model solutions \citep[e.g.][]{Maccio08}.
The inner shape parameter, $\alpha$, is especially difficult
to constrain observationally due to the added complication that
only a small number of rotation curve points constrain this value;
see \citep{Dutton05} and \citep{deblok10} for reviews. 

\begin{figure*}[t]
\centering
\includegraphics[width=0.5\textwidth]{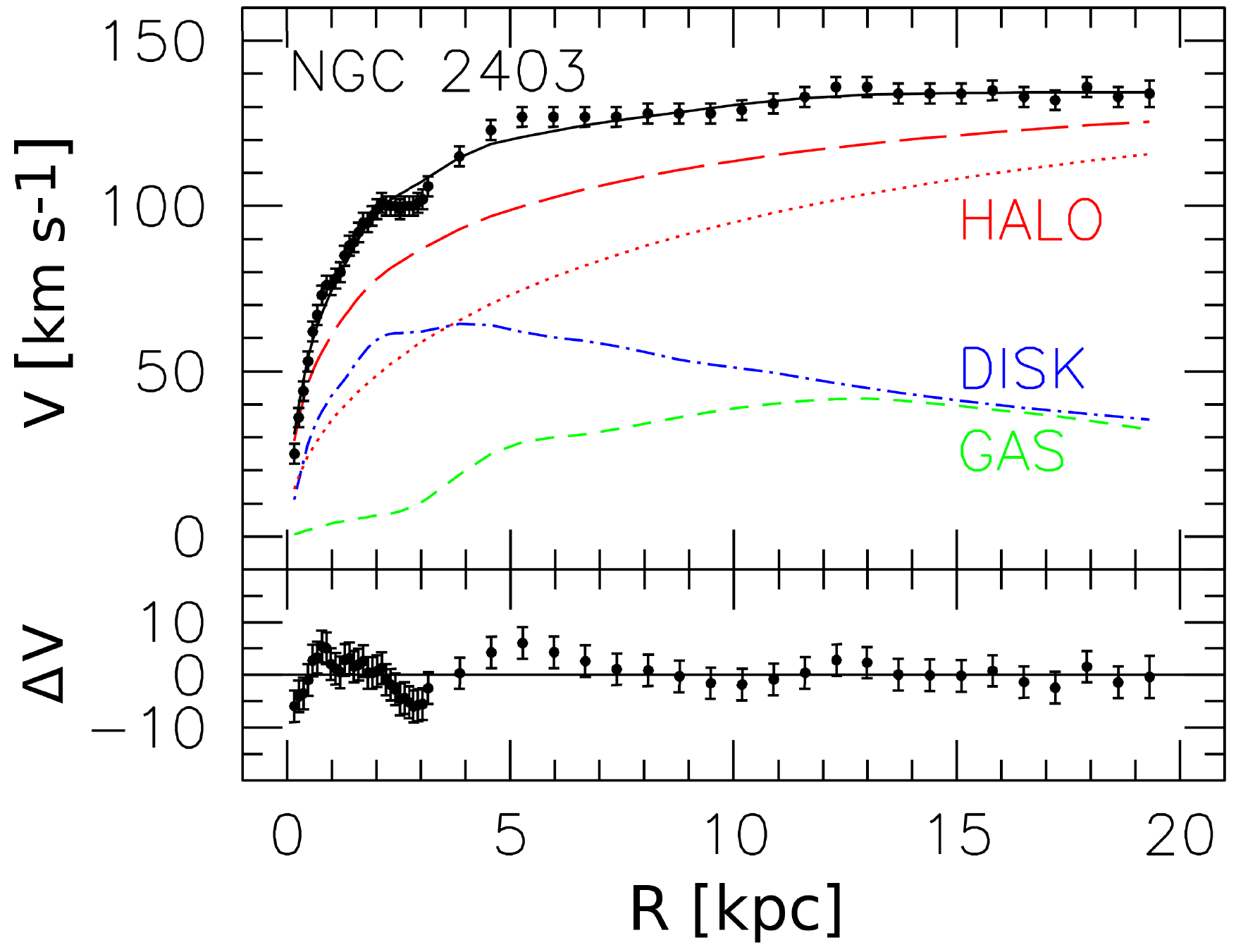}
\caption{Mass model for the bright spiral galaxy NGC 2403 with
  an adiabatically contracted halo model (long dashed red line)
  or not (initial model shown with red dotted line).
  The disk and gas profiles correspond to the contracted final model.
  The data-fit residuals, ${\Delta}V$, are shown in the lower part
  of the figure. 
  The ratio of dark to baryonic mass within the optical disk
  of a galaxy can increase by almost 40\% within $V_{2.2}$ if
  AC is invoked.
  The observed velocity data (black points) are from \citet{BlaisOuellette00}. 
  See text for details about the modeling technique.}
\label{fig:N2403}
\end{figure*}

An additional complication to the mass modeling exercise is whether
the {\sl initial} distribution of dark matter particles is affected by
the gradual cooling of the baryons as a galaxy forms.  If the
potential variations due to the dissipating baryons occur slowly
compared to the orbital period of a dark matter particle in circular
orbit, then the end state of the system is independent of the path
taken \citep{Blumenthal86,Gnedin04,Abadi10}.  Thus, contraction of the
dark matter occurs when baryons collapse and come to the central
region; if the gravitational potential increases, matter naturally
follows.  This ``adiabatic'' contraction of the halo due to the
cooling baryons can be modeled as \be r_{\rm f} V_{\rm f}(r_{\rm f}) =
r_{\rm i} V_{\rm i}(r_{\rm i}), \ee where $V_{\rm i}(r_{\rm i})$ and
$V_{\rm f}(r_{\rm f})$ are the initial and final rotation curves,
respectively, and $r V(r)$ is the adiabatic invariant \citep{Mo10}
\footnote{These authors advocate using $r V(r)$ as the adiabatic
  invariant, instead of the usual $r M(r)$ invariant, since disks are
  not spherical.  Note also that the algorithm for compression due to
  adiabatic infall in a spherical halo model may take a different form
  when random motions are accounted for \citep{Sellwood05}.}.  The
example of a modeled rotation curve with and without adiabatic
contraction (hereafter AC) is shown in \Fig{N2403}.  This model has a
fixed stellar $M/L_*=1$ and halo $\alpha=1$.  The ratio of dark to
baryonic mass within the optical disk of a galaxy can increase by
almost 40\% if AC is invoked.  In more extreme cases, AC can transform
an initial $\rho\propto r^{-1}$ NFW-type halo into a $\rho \propto
r^{-2}$ isothermal halo.

While AC is undoubtedly at play in all forming galaxies, it has been
known for many years that baryonic effects such as supernova feedback
and dynamical fraction can, in principle, result in reduced halo
contraction or even halo expansion \citep[e.g.][]{NEF96,El-Zant01}.
Only recently have these effects been demonstrated in fully
cosmological simulations of galaxy formation 
\citep{Johansson09,Abadi10,Duffy10,Maccio12,Martizzi12,Governato12}.
Despite recent progress, cosmological simulations have not yet
provided unique predictions for the response of dark matter halos
to galaxy formation. A wide range of possibilities (from adiabatic
contraction to expansion, from cuspy to cored halos) should thus
be accounted for when attempting mass models of galaxies.

\bigskip

\subsection{Other Galaxy Mass Constraints} 

Besides mass modeling of individual galaxies, with all the limitations
that this entails as we saw above, potentially tighter constraints for
the mass of baryonic and dark matter in disk galaxies may be achieved
through dynamical or statistical arguments. As a way of introduction,
we first present various methods to constrain the stellar $M/L_*$,
independently of stellar population models.

\subsubsection{Maximal and Sub-maximal Disks}\label{sec:maxdisks}

The hypothesis that the bulge and disk could contribute ``maximally''
to the rotation curve was introduced by \citet{Carignan85} and 
\citet{vanAlbada85} in order to overcome the intrinsic
uncertainties of stellar $M/L_*$ ratios.  A maximal disk obeys
$${\cal{F}} \equiv V_{\rm disk}(R_{\rm max})/V_{\rm tot}(R_{\rm max}) > 0.85,$$
where $V_{\rm disk}$ is the inferred velocity of the disk (stars and
gas), $V_{\rm tot}$ is the total observed velocity, and $R_{\rm max}$
is the radius at which $V_{\rm disk}$ reaches its peak value
\citep{Sackett97}.  For a pure exponential disk, this is $V_{2.2}$
(see \se{masses}).  In other words, for a maximal disk galaxy, $V_{\rm
  disk}^2(R_{\rm max})/V_{\rm tot}^2(R_{\rm max}) \gtrsim 0.72$ and
the disk contributes at least 72\% of the total rotational support at
$R_{\rm max}$.  Note that this is just an arbitrary convention
following \citet{Sackett97}.  A lesser contribution of the disk to the
overall rotation curve at $R_{\rm max}$ is deemed ``sub-maximal''.
Since a galaxy with a sub-maximal disk with a significant bar or bulge
component can still be baryon-dominated within $R_{\rm max}$, we
define a galaxy to be maximal at $R_{\rm max}$ if ${\cal{F}} > 0.85$,
where $V_{\rm disk}$ is the inferred velocity of the baryons (bulge
and disk).  Even in the presence of a maximal disk, rare are the
galaxies that do not require a halo component to match fully extended
rotation curves.  For dwarf galaxies, maximal disks often involve
stellar $M/L_*$ ratios that are physically implausible
\citep{Swaters11}.


The central panel of \Fig{Dutton} provides a good example of a
sub-maximal (top) versus a maximal (bottom) disk fit.  Both fits are
equally valid numerically \citep[see also][]{Kassin06,Noordermeer07}.
Thus, without further constraints and especially extended \hi\
rotation curves, the choice of a (sub-)maximal disk solution remains
ill-constrained.  Fortunately, arguments about the
dynamical structure of disks and the existence of velocity-luminosity
($VL$, aka ``Tully-Fisher'') relation of spiral galaxies allow for new
constraints to be implemented as we discuss in \se{VLres} (see also
lucid discussions on this topic by vdKF11).  In general, those other
techniques listed below point to galaxy disks whose stellar masses are
significantly below the so-called maximal value.

\subsubsection{Velocity Dispersion Measurements}\label{sec:veldisp}

For a self-gravitating, radially exponential disk with vertical
profile of the form $\rho(R,z)=\rho(R,0) {\rm sech}^2(z/z_0)$,
\cite{vdk88} and \cite{Bottema93} showed that the peak circular
velocity of the stellar disk, $V_{\rm{disk}}$, measured at
$R=2.2R_{\rm d}$ can be related to the vertical velocity dispersion,
$V_z$, and the intrinsic thickness (or scale height) of the disk,
$z_0$, via:
\be
V_{\rm disk}(R_{\rm max}) = c_{\rm max} \,
{\left\langle V_z^2 \right\rangle}^{1/2}_{R=0}  
\,\sqrt{\frac{R_{\rm d}}{z_0}}.  
\label{eq:Freeman}
\ee
%
where $c_{\rm max} \simeq 0.88 (1-0.28 z_0/R_{\rm d})$
\citep{Bershady11}.  A more detailed discussion of \eq{Freeman}
is presented in vdKF11 (see their Section 3.2.4).


The power of \Eq{Freeman} is that the disk $M/L_*$, derived via
$V_{\rm disk}(R_{\rm max})$, can be determined independently of the
dark matter halo.  However, the measurements involved are difficult
since the scale height, $z_0$, and the vertical component of the
velocity dispersion, $V_z$, cannot be measured simultaneously.  Thus
this method is statistical in nature, at least for nearly edge-on
\citep{Kregel05} and nearly face-on \citep{Andersen06,Andersen13}
systems. Indeed, for face-on systems, $V_z$ can be measured but $z_0$
must be inferred; and vice versa for edge-ons.  The tightness of the
statistical correlations used to infer $z_0$ and $V_z$ for face-on and
edge-on galaxies, respectively, is reviewed by \citet{Bershady10b}.
Furthermore, \cite{Bershady11} showed that the edge-on sample of
\cite{Kregel05} and the face-on sample (so-called ``DiskMass Survey'')
of \cite{Bershady10a} yield similar distributions of $V_{\rm disk}$
versus $V_z$.  Systematic errors in $M/L_*$ estimates based on
\Eq{Freeman} are thus relatively small.

\citet{Kregel05} painstakingly determined the intrinsic stellar disk
kinematics through \Eq{Freeman} for 15 intermediate and late-type
edge-on spiral galaxies using a dynamical modelling technique and
assuming that $\sigma_z/\sigma_r=0.6\pm0.1$ (based on various
plausible arguments).  For 12 of their 15 spirals, they find on
average ${\cal{F}} = 0.53 \pm 0.04$.
They also find that the contribution of the disk to $V_{2.2}$ is
independent of barredness, in agreement with the ``Tully-Fisher''
($VL$) analysis of barred galaxies\footnote{The fact that barred
and unbarred galaxies share the same Tully-Fisher relation \citep{Courteau03}
reflects that the angular momentum transferred from the bar to the halo
is relatively small and easily absorbed by the halo. Thus, bars of all
strengths belong to the same Tully-Fisher relation, as found in
\citet{Sheth12}.}
by \cite{Courteau03} and \cite{Sheth12} and the $N$-body simulations
of \citet{Valenzuela03}.

More recently, \citet{Bershady10a} applied the velocity dispersion
method on a sample of 46 nearly face-on (inclinations $\simeq 30$
degrees) galaxies. This survey uses integral-field spectroscopy to
measure stellar and gas kinematics using the custom-built SparsePak
and PPAK instruments.  For the high-surface-brightness galaxy UGC 463,
\citet{Westfall11} find the galaxy to be sub-maximal at 2.2 disk
scale lengths with
${\cal{F}} =0.61^{+0.07}_{-0.09}({\rm ran})^{+0.12}_{-0.18}({\rm sys})$.
The ratio ${\cal{F}}$ could be much smaller for lower surface
brightness systems.  In fact, \citet{Bershady11} confirm for
30 DiskMass systems covering a range of structural properties
that the fraction $\cal{F}$ ranges from 0.25 to 0.65 and increases
with luminosity, rotation speed, and redder color.
The DiskMass project does not include a dark matter component
in their analysis since their data in the plane of the disk
is largely baryon dominated (even in areas where dark matter
dominates the enclosed mass).
The impact of the dark halo is to make the disks less maximal
\citep[as advocated by][]{Bottema93} at the 20\% level.

\subsubsection{Scaling Relations Residuals}\label{sec:VLres}
\citet{CourteauRix99} suggested that sub-maximal disks provide a
solution to the surface brightness independence of the $VL$ relation
\citep[see also][]{Zwaan95}.  Courteau \& Rix found that,
{\it on average}, high-surface brightness spiral galaxies have ${\cal{F}}
\lsim 0.6 \pm 0.1$, as recently verified by 
\citet[][see \se{veldisp}]{Bershady11}.  Their
argument relies on the assumption that the scatter in the $VL$
relation and the size-luminosity relation is largely controlled by the
disk scale length, that spiral galaxies have self-similar $M/L$
profiles (different than having constant $M/L$ ratios), and that dark
matter halos are adiabatically contracted.  \citet{Dutton07} revisited
the Courteau \& Rix method using a more detailed account of baryonic
physics to find that ${\cal{F}} \lsim 0.72 \pm 0.05$ if AC
is ignored or compensated for by other mechanisms which may result from
non-spherical, clumpy gas accretion, coupled with dynamical friction
transfer of energy from the gas to the dark matter. \citet{Dutton07}
reproduced the \citet{CourteauRix99} result if AC is invoked. Either way,
most spiral disks obey the sub-maximal disk constraints.

\subsubsection{Fluid Dynamical Modeling}\label{sec:fluid}

Dynamical friction between a stellar bar and a dark matter halo is
believed to slow down the pattern speed of the bar, and therefore fast
bars should imply maximal disks \citep{Weinberg85}.  
A compilation of 17 barred galaxies analysed via the ``Tremaine-Weinberg''
method was presented by \cite{Corsini11}; the overall impression
from these analyses favors maximal disks.  Likewise, slow bars
might imply maximal halos \citep[e.g., for UGC 628,][]{Chemin09b}.  
However, some of these claims have been challenged on account
of numerical artifacts yielding e.g. over-efficient bar slow down
\citep{Valenzuela03}.  
Indeed, \cite{Athanassoula03} and \cite{Athanassoula13} have argued
that the decrease in the pattern speed does not depend only on the
mass of the dark matter halo, but also also on other galaxy properties
such as gas fraction, or halo shape. Thus they argue that, on its own,
the pattern speed decrease cannot set constraint to the halo mass.

Based on detailed fluid dynamic modeling, a maximal disk solution
was found by \citet{Englmaier99} for the Milky Way and by \citet{Weiner01}
for the NGC 4123.  Both galaxies are barred \citep[see][for other
examples]{Sellwood10}.  Conversely, the hydrodynamic gas simulations
used by \citet{Kranz03} to model the spiral arm structure of five
grand design non-barred galaxies yield a wide range of
${\cal{F}}$, from closely maximal to 0.6.  Galaxy disks appear
to be maximal if $V_{\rm max} > 200$~\kms, sub-maximal otherwise.
Although detailed comparisons of observed galaxy velocity fields
with hydrodynamic gas simulations are challenging, future galaxy
masses review ought to include more analyses of this kind.

\subsubsection{Gravitational Lensing}

In the rare cases where a distant galaxy or quasar is lensed by a
foreground galaxy, gravitational lensing can be used to place
constraints on the projected ellipticity and mass within the Einstein
radius. For disk-dominated lenses, this extra information coupled with
spatially resolved kinematics can be used to break the disk-halo
degeneracy \citep{Dutton11a}. Until recently, only a handful of spiral
galaxy lenses were known. The best studied being B1600+434
\citep{Jaunsen97,Koopmans98,Maller00} and the ``Einstein Cross''
2237+0305 \citep{Huchra85,Trott02,Trott10}. However, since these
lenses are bulge-dominated, they are not ideally suited to constrain
disk masses.

A number of recent searches for spiral lens galaxies
\citep{Feron09,Sygnet10,Treu11}
have uncovered several new disk-dominated spiral lenses.
A joint strong lensing and dynamics analysis of the disk-dominated
gravitational lens SDSSJ2141-0001, discovered as part of the SLACS
survey \citep{Bolton08}, yields a best fit
${\cal{F}}=0.87^{+0.05}_{-0.09}$ \citep{Barnabe12}, where ${\cal{F}}$ 
includes both disk and bulge. Since the bulge contributes $\sim 30\%$
of the stellar mass within 2.2 disk scale lengths, the {\it disk} is
actually consistent (at the 1$\sigma$ level) with being sub-maximal.

\subsubsection{Two-body Interactions and the Mass of the Local Group}
\label{sec:kahn}

The relative motions of two orbiting bodies, and assumptions about
their angular momentum and total energy, may also be used to infer
the total mass of that system.

Looking at our own Local Group (LG) of galaxies, the Milky Way and
M31 galaxies display largely unperturbed disks suggesting that they
are likely on their first passage since having formed.  Based on that
observation, \citet{Kahn59} were able to compute the relative motion
of the Milky Way and M31 as a two-body problem assuming purely radial
infall (zero angular momentum).  This method, referred to as the
``Kahn-Woltjer timing argument'' \citep[TA; see also][]{Binney08},
led \citeauthor{Kahn59} to measure a total mass for the LG in excess
of the reduced mass of M31 and the Milky Way by a factor greater
than six, thus calling for a sizeable amount of hitherto undetected
intergalactic matter.  \cite{Sandage86IX}, using a similar argument
for the deceleration of nearby galaxies caused by the LG, found
a lower value for the maximum mass for the LG, $M_{\rm LG} = 5 \times 10^{12} \Msol$,
with a best-fit value of $4\times 10^{11} \Msol$.
\citet{CourteauvdB99} also used various mass estimators to compute
$M_{\rm LG} = (2.3 \pm 0.6) \times 10^{12} \Msol$ within 1.2 Mpc.
More recently, Partridge et~al. (2013) 
revised the TA for the LG by accounting for a Dark Energy
component and finding a total 
$M_{LG} = (6.19 \pm 0.56[{\rm obs}] \pm 0.99[{\rm sys}]) \times 10^{12} \Msol$.
The systematic error is obtained by testing the TA model against
LG-like objects selected from a cosmological simulation.
The effect of dark energy is to make the value of $M_{LG}$
roughly 12\% larger than similar TA mass estimates that neglect it.

A major limitation of the TA is its reliance on single-galaxy
interactions with the Milky Way and assumptions about virialization
of the LG \citep{Phelps13}.  The latter implemented a numerical
action method, originally developed by \cite{Peebles89},
which takes into account the peculiar motions of a large
subset of LG galaxies whilst eliminating the mass
degeneracy in the two-body TA.  Their method yields estimates
of $2.5 - 5.0 \times 10^{12} \Msol$ for the Milky Way and
$1.0 - 5.0 \times 10^{12} \Msol$ for M31.
If the putative transverse velocity of M31 \citep{vanderMarel12}
is taken into account, the lower bound for the mass of the Milky
Way drops from $2.5 \times 10^{12} \Msol$ to 
$1.5 \times 10^{12} \Msol$ in the 95\% confidence region. 
The transverse velocity of M31 remains somewhat tentative owing
to the uncertainty in our own orbit around the Milky Way center.  

While the above figures seem high, the action method effectively
measures a maximal possible mass for each bound system. Smaller
mass estimates will thus be obtained on smaller scales.
For instance, independent estimates for the dynamical mass
of the Milky Way within 100 kpc range from
$M_{\rm MW} = (.4-1.4) \times 10^{12} \Msol$
(see \se{milkyway}, Table 1); likewise for M31 with
$M_{\rm M31} = (0.8-1.1) \pm 10^{12} \Msol$ within the virial
radius $R_{200}=189-213$ kpc, depending on the modeled dark
matter distribution \citep{T512}.

Since this review is mostly concerned with mass estimators
for individual galaxies, we do not investigate the mass of
the LG further.  The TA should however be noted as a method
to estimate the mass of the Local Group and/or that of the
Milky Way or M31 if one of the other masses is known
independently.  For more details on modern applications
of the TA, see \citet{Phelps13} and references therein.
We review independent mass estimates of the Milky Way in
\se{milkyway}.
 

\newpage
\subsection{Future Prospects}\label{sec:gasrichsum}

Mass modeling methods will require new constraints to break the
disk-halo degeneracies as well as those internal to the halo model.
Those may come from other mass estimators (not considered here due to
space limits) based on e.g. external tracers such as planetary
nebulae, globular clusters, and satellites \citep[e.g.][]{Yegorova11}.
Nonetheless, various independent methods outlined in \se{gasrich}
present evidence for the prevalence of sub-maximal disks in most
spiral galaxies with:
\be
V_{\rm disk}/V_{2.2} \simeq 0.6 \pm 0.1. 
\ee

However, galaxies do come with a range in disk-to-halo masses, with
the most massive ($V_{\rm max} > 200$~\kms), high surface brightness
spirals being close to having maximal disks.  Almost all other
galaxies, 90\% of them being of dwarf type, are likely sub-maximal.

The dark matter near the virial radius likely has circular velocity at
slightly lower speeds than the inner baryons.  By comparing with the
rotation velocities within the optical disk, $V_{\rm opt}$, one finds
from satellite kinematics and weak lensing that on average for
late-type galaxies $V_{\rm opt}/V_{200}\simeq 1.2$
\citep{Dutton10,reyes12}.

Looking ahead, very large kinematical surveys of spiral galaxies will
enable a global characterization of rotation curve shapes for
thousands of galaxies; e.g. MaNGA (``Mapping Nearby Galaxies at APO'';
PI: Kevin Bundy) is a project to collect IFU velocity maps at Apache
Point Observatory from 2014-2020 for $\sim$10,000 northern galaxies
with stellar masses above 10${^9}\Msol$ over a full range of gas
contents, environments, and orientations.  It will cover the range
360-1000~nm at a resolution of about 2500, with an emphasis on
spectrophotometric calibration at all wavelengths.  On shorter
timescales, the Calar Alto Legacy Integral Field Area Survey (CALIFA)
will already provide 2D PPAK IFU maps for $\sim$600 nearby galaxies
\citep{CALIFA}.  However, a short-coming of the large IFU surveys in
progress and planned is their low spectral resolution.  This will not
enable measurements of disk velocity dispersions except in the inner
regions where disentangling effects of the bulge kinematics will make
interpretation difficult without more sophisticated analysis methods.
One such method is the ``Jeans Anisotropic Multi-Gaussian Expansion''
described in detail in \se{aximod}, although it will need to be
augmented to consider the effects of dust extinction.  CALIFA, and
particularly MaNGA, will still truly refine our definition of velocity
fields and the stellar populations in galaxies.  Large samples of
interferometric data will also be needed to extract velocity fields
homogeneously for comparisons with optical velocity fields.  This will
be addressed by upcoming \hi\ surveys with SKA pathfinders, notably
WALLABY on ASKAP and its northern counterpart on WSRT.  These projects
may span $\sim$5 years or more.

Despite the extensive new galaxy dynamics data bases, the rigorous
separation of baryonic and dark matter profiles in galaxies will
require accurate stellar $M/L_*$'s from stellar population models
and/or dynamical measurements (e.g. Eq.~\ref{eq:Freeman}) {\it as well as}
precise constraints from $\Lambda$CDM structure formation models to
resolve the many degeneracies between the luminous and dark matter
components.


\newpage

\section{Dark Matter and Mass Models of the Milky Way}\label{sec:milkyway}
\subsection{Introduction}

As we saw in \se{gasrich}, Galactic rotation curves are one piece of
the multifaceted dark matter puzzle.  Around the time of
\cite{Rubin70}, cosmologists began to understand that relics from the
early Universe, in the form of subatomic particles, could contribute
significantly to the present-day mass density of the Universe.  Today,
the leading dark matter candidate is a stable, electrically neutral,
supersymmetric particle with a mass between $1\,{\rm GeV}$ and
$1\,{\rm TeV}$.  This weakly interacting massive particle or WIMP,
would have been non-relativistic during the formation of large scale
structure and hence, represents an example of cold dark matter (CDM;
see, for example, \cite{Bertone05}).  WIMPs naturally form halos with
roughly the right structure to explain flat rotation curves
\citep{Blumenthal84,Dubinski91,NFW96}.  Moreover, the halo mass
function that is predicted by the standard CDM theory of structure
formation is consistent with the observed hierarchy of virialized
systems from dwarf galaxies to clusters
\citep{Press74,BBKS86,Tinker08}.

In the mid-80's, several researchers pointed out that WIMPs might be
detected in the laboratory \citep{Drukier84,Goodman85}.  At present over a
dozen groups have deployed or are building terrestrial dark matter
detectors \citep{Bertone05,Feng10}.  These experiments have the
potential to probe the local density and velocity
dispersion of dark matter at the position of the Earth.  By the same
token, the interpretation of these experiments, and in particular, the
constraints inferred on the mass and scattering cross section of dark
matter candidates, depend on astrophysical estimates of the local dark
matter distribution function.  Thus, there is a direct link between
mass models of the Milky Way and dark matter detection in the
laboratory.

Apart from the dark matter question, the Milky Way presents an
opportunity to observe a ``typical'' (barred SBc) spiral galaxy
from a unique vantage point.  Thus, observations of the Milky Way
hold a special place in our attempt to understand the formation
and structure of spiral galaxies.  The remainder of this section
focuses on what we know about the distribution of baryons
and dark matter in the Galaxy.

Studies of the Milky Way are invariably challenged by our position
within it and by our frame of reference, which orbits about the
Galactic center.  In particular, our distance to the Galactic center
and the circular speed at the position of the Sun remain uncertain at
the 5\% level
\citep{Ghez08,Gillessen09,Reid09,Bovy09,Brunthaler11,Schonrich12}.
These uncertainties enter our interpretation of various observations
and our determination of the mass of the Galaxy.  On the other hand,
the Milky Way offers a unique opportunity to probe the distribution of
both visible and dark matter relatively close to the Galactic center.
By contrast, for external disk galaxies, rotation curves provide
evidence for dark matter only in the outermost regions where dark
matter dominates.  Estimates of the density of dark matter in the
inner regions of large spiral galaxies require assumptions about the
shape of the dark halo density profile.

The Milky Way is also unique in that there exist extensive catalogs of
halo stars, globular clusters, and satellite galaxies with reliable
Galactocentric distances and velocities.  These tracers provide
important constraints on the Galactic potential.  Moreover, numerous
stellar streams, such as the Sagittarius and Monoceros streams, may
provide another handle on the Galactic potential.

\subsection{Multicomponent Models for the Milky Way}
 
The literature is replete with models of the Milky Way.  An important
early example can be found in \citet[][hereafter BSS]{Bahcall82}.
Their model comprised a double-exponential disk, a (deprojected) de
Vaucouleurs stellar halo, a cuspy bulge, and a dark matter halo.  The
parameters that described the disk and stellar halo were taken from
the earlier work by \cite{Bahcall80}, which, in turn, were based on
star counts.  The BSS model for the dark halo assumed a constant
density core and a $r^{-2.7}$ power-law fall-off at large radii.
Notably, BSS showed that the different components of the Galaxy could
``conspire'' to produce a flat rotation curve for a wide range of
model parameters, particularly the disk mass-to-light ratio and the
structural parameters of the halo \citep[see also][]{Bahcall85,
vanAlbada85}.  \cite{Blumenthal84} argued that adiabatic
contraction, the respose of dark matter to the baryonic component as
it condenses and forms the disk and bulge, could ``explain'' the
apparent conspiracy that leads to flat rotation curves.  Nevertheless,
the disk-halo {\it conspiracy} in the context of a $\Lambda$CDM
universe remains an outstanding problem in galactic astronomy while
the disk-halo {\it degeneracy} continues to plague attempts to pin
down the structural parameters of dark matter halos \citep{Dutton05}.

\cite{Sellwood85} took the BSS model a step further by realizing it
as an N-body distribution and numerically evolving it forward in time.
He found that the BSS model was stable against bar formation though
it did develop a two-armed spiral.  Recall that a self-gravitating
disk is generally unstable to the formation of a bar while a disk
of particles on circular orbits in a background potential is stable.
Disk galaxies lie somewhere between these
extremes with the gravitational force felt by the disk particles
coming from both the disk itself and the other components. 
In the BSS model, the bulge plays the key role in stabilizing the
disk\footnote{\cite{Ostriker73} suggested that a dark matter halo
also could stabilize the disk against bar formation, but more modern
simulations, including a live halo whose resonances are adequately
described, have shown that the halo has a more complex role.
During the bar formation phase, a more massive halo
slows down the bar formation, but in the later, secular evolution
phases, the halo actually helps the bar grow stronger by absorbing
a considerable part of the angular momentum emitted by the bar
region \citep{Athanassoula02,Athanassoula03}. See also \se{fluid}.}.

Another influential model, especially for its focus on the Galactic
bulge, was devised by \cite{Kent92}.  \cite{Kent91} constructed a
luminosity model for the bulge based on the 2.4~$\mu m$ map of the
Galactic plane from the Spacelab Infrared Telescope. \cite{Kent92}
combined this model with velocity dispersion data to determine the
mass-to-light ratio for the Galactic bulge.  He then constructed
disk-bulge-halo mass models designed to fit the rotation curve.  The
results hinted at the existence of a supermassive black hole in the
Galactic center. (For an earlier discussion of the existence of a
central supermassive black hole, see \cite{Lacy82}).  Moreover, Kent's
model requires that one allow for non-circular motions in the gas, as
in \cite{Gerhard86}.  The Milky Way is now known to be a barred spiral
galaxy (\cite{Blitz91,Binney97}) and mass models that incorporate a bar
include \cite{Fux97, Fux99} and \cite{Englmaier06}.

\cite{Dehnen98} constructed a suite of disk-bulge-halo Galactic mass
models.  The observational constraints for their work included the
circular speed curve, the velocity dispersion toward the bulge, the
Oort constants, the local velocity dispersion tensor, and the force
and surface density in the solar neighborhood.  \cite{Dehnen98}
surveyed the 10-dimensional parameter space of models using a
restricted maximum likelihood analysis in that they considered 25
examples wherein some parameters were held fixed while the remaining
parameters are allowed to vary so as to minimize the likelihood
function.

\citet[][hereafter WPD]{Widrow08} constructed dynamical models for the
Galaxy using observational constraints similar to those considered by
\cite{Dehnen98}.  WPD deployed Bayesian inference and a Markov Chain
Monte Carlo (hereafter MCMC) algorithm to construct the full
probability distribution function (PDF) over the model parameter
space.  The PDF is found to include regions of parameter space in
which the model is highly unstable to the formation of a strong bar,
so much that the models are almost certainly unphysical.  In other
regions of parameter space, the models are found to be mildly unstable
to the formation of a weak bar and therefore may well represent an
axisymmetric, idealized approximation to the Milky
Way. \cite{Binney10a}, \cite{Binney11} and \cite{McMillan12} have
continued to develop observationally-motivated models for the Milky
Way along similar lines.

As mentioned above, both \cite{Dehnen98} and WPD use the Galactic
circular speed curve as a model constraint.  As with external
galaxies, observations of neutral hydrogen provide a measure of the
Galactic rotation curve, which translates to a circular speed curve
provided the gas follows circular orbits.  Inside the solar
circle, \hi~ observations are usually presented in terms of the
so-called terminal velocity, $v_{\rm term}$, which is defined as the
peak velocity along a line of sight at Galactic coordinates $b=0$ and
$|l|<\pi/2$.  If one assumes axisymmetry, then the \hi ~ emission
corresponding to the peak velocity originates from the galactocentric
radius $R=R_0\sin{l}$ where $R_0$ is our distance from the Galactic
center.  Thus $v_c(R) = v_{\rm term} + v_c(R_0)\sin {l}$ where $v_c$
is the circular speed center.  \cite{Malhotra95} for example, has
determined the terminal velocity to the HI measurements of
\cite{Weaver73, Bania84,Kerr86} and her measurements were used in both
the \cite{Dehnen98} and WPD analyses.

The radial velocity of an object at Galactic coordinates
$\left (l,\,b\right )$ relative to the local standard of rest,
$v_{\rm LSR}$ is related to the circurlar velocity via the expression
\be
v_{\rm LSR} = \left\{\frac{R_0}{R} v_c(R) - v_c(R_0)\right\}\cos{b}\sin{l}~.
\ee
\cite{Brand93} consider a sample of HII regions/reflection nebulae
with distances and radial velocities are available and use this method
to infer the rotation of the Galaxy out to $17\,{\rm kpc}$.  Unlike
the terminal velocity measurement, $R$ must here be inferred from
observations of the heliocentric distance $D$ through the relation $R
= \left(D^2\cos^2 b + R_0^2 - 2R_0D\cos{b}\sin{l}\right)^{1/2}$.
\citet{Dehnen98} and \citet{Widrow05} present a statistical method for
accomplishing this and incorporate the \cite{Brand93} data into their
Galactic model constraints.  Note that both terminal velocity and
outer rotation curve methods require $R_0$ and $v_c(R_0)$, both of
which are uncertain at the $5\%$ level.  A proper statistical
analysis using these methods therefore requires a marginalization over
$R_0$ and $v_c(R_0)$, subject to prior probabilities for these
parameters.

\cite{Xue08} derived a Galactic rotation curve out to radius of
60~kpc.  Their rotation curve is based on observations of blue
horizontal branch (BHB) stars from the Sloan Digital Sky Survey (SDSS)
which provided line-of-sight velocity distributions at different
galactocentric radii.  To construct the rotation curve, they compared
these observations with mock observations of simulated Milky Way-like
galaxies.  Further investigations of halos stars are discussed below.

\begin{figure}[t]
\centering
\includegraphics[width=0.45\textwidth]{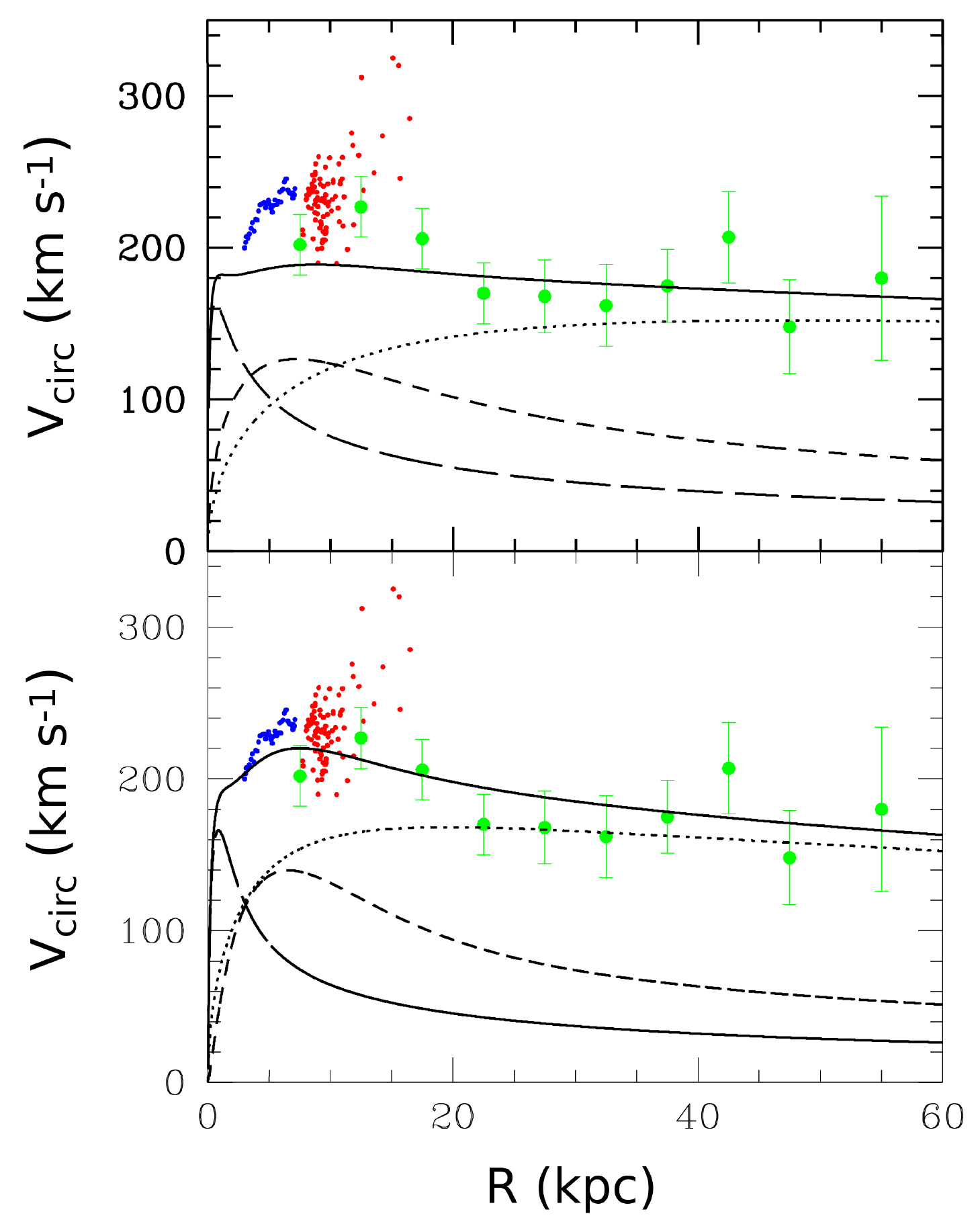}
\caption{Rotation curve, as determined from the analysis of BHB stars
  by \cite{Xue08} (green), the terminal velocity method by
  \cite{Malhotra95} (blue), and the analysis of Hii regions/reclection
  nebulae by \cite{Brand93} (red).  The upper panel shows the model
  predictions from \cite{Xue11} for the total rotation curve (solid)
  and the contributions from the disk (dashed), bulge (long-dashed),
  and halo (dotted).  The lower panel shows the WPD model.}
\label{fig:XueWPD}
\end{figure}

\Fig{XueWPD} shows the inner rotation curve as inferred from
\hi\ observations by \cite{Malhotra95}, the outer rotation curve
inferred from observations of \hii\ regions by \cite{Brand93}, and
the \cite{Xue08} rotation curve.  The upper panel also shows the
circular speed curve for one of two mass models from \cite{Xue08}.
This model assumes a Hernquist bulge, a {\it sphericalized}
exponential disk, and an NFW halo.  The lower panel shows one of
the more stable examples from the WPD Bayesian analysis.  Note
that the \cite{Xue08} model was not constrained by \cite{Malhotra95} or
\cite{Brand93}'s data while the WPD model was constructed independent
of the \cite{Xue08} results.

Along rather different lines, \cite{Klypin02} constructed mass models
for both the Milky Way and M31 that were motivated by
disk formation theory in the standard $\Lambda$CDM cosmology.  To be
precise, they assume that the proto-galaxy has a NFW halo with a
concentration parameter in agreement with pure dark matter
simulations.  The present-day halo is derived by adiabatically
contracting the early-time halo.

\begin{table}
\centering
\centerline{Table 1: Derived values for various models}
\label{tab:tableMW}
\begin{tabular}{lcccccc}
 & & & & \\
model & $M_d$ & $h$ & $M_b$ & $M_{100}$ & $V_{\rm disk}^2/V_{\rm tot}^2$
 & $\rho_{DM,\odot}$\\
 & & & & \\ [-7pt]
\hline
 & & & & \\ [-5pt]
BSS & 5.6 & 3.5 & 1.1 & 144 & 0.53 & 0.009 \\
Kent (1992) -- low $M_d$ &  3.7 & 2.8 & 1.2 & 125 & 0.4 & 0.013\\
Kent (1992) -- high $M_d$ & 5.5 & 2.8 & 1.2 & 116 & 0.67 & 0.0097\\
Dehnen \& Binney (1998) & 0.36-1.5 & 2-3.4 & 3.3-5.1 & 40-75 & 0.33-0.83 & 0.009-0.012\\
Klypin et~al. (2002) & 4 & 3.5 & 0.8 & 55 & 0.37 & 0.007\\
Widrow et~al. (2008: WPD) & $4.1$ & $2.8$ & $0.96$ & $40^{+22}_{-19}$ & $0.5$ & $0.008\pm 0.0014$ \\
Smith et~al. (2007) & $5$ & $4$ & $1.5$ & $55$ & 0.4 & $0.011$ \\
Xue et~al. (2008)  & $5$ & $4$ & $1.5$ & $49$ & 0.33 & $0.006$\\ 
Gnedin et~al. (2010) & $5$ & $3$ & $0.5$ & $89$ & 0.33 & $0.014$\\ 
McMillan et~al. (2012) & $5.7$ & 2.9 & $0.9$ & 83 & 0.63 & $0.0104$ \\
Moni Bidin et~al. (2012) & & & & & & $0 \pm 0.001$ \\
Bovy \& Tremaine (2012) && &&& & $0.008 \pm 0.003$ \\
 & & & & \\ [-7pt]
\hline
\end{tabular}
\caption{Selected mass models of the Milky Way as presented in the text.
  The disk mass, $M_d$, bulge mass, $M_b$, and total mass within
  $100\,{\rm kpc}$, $M_{100}$, are given in units of $10^{10}\,M_\odot$.
  The disk scale length, $h$, is given in units of kpc.  Note that
  \cite{Smith07} assume a Miyamoto-Nagai model \citep{miyamoto75} for
  the disk where the peak in the contribution of the disk to the circular
  speed curve occurs at $R_{\rm max} \simeq 1.4h$.
  \cite{Xue08} assume a ``spherical'' disk potential
  where $R_{\rm max} \simeq 1.9h$.  All other models assume an exponential
  disk where $R_{\rm max} \simeq 2.2h$.  Column 6 gives the ratio
  $V_{\rm disk}^2/V_{\rm tot}^2$ evaluated at $R_{\rm max}$
  where $V_{\rm tot}$ and $V_{\rm disk}$ are the total observed
  velocity and contribution from the baryonic disk, respectively.
  The final column gives the density of dark matter in the solar
  neighborhood in units of $M_\odot {\rm pc}^{-3}$.}
\end{table}

In \Table{tableMW}, we present various structural parameters for the
models described in this and the next section.  We include, in
addition to the disk, bulge and halo masses, the local dark matter
density, the key quantity for terrestrial detection experiments.  Note
that the assumed canonical value for these experiments is $0.0079
M_\odot \,{\rm pc^{-3}} = 0.3\,{\rm GeV}\,{\rm cm}^{-3}$
\citep{Lewin96,Bertone05}.  The local density of dark matter is of
particular interest for direct detection experiments.  A more complete
discussion of this parameter can be found in \cite{Catena10,Catena11}
and WPD.

As we saw in \se{maxdisks}, a working definition for the maximal disk
hypothesis is that
$$ V^2_{\rm disk}(R_{\rm max})/V^2_{\rm tot}(R_{\rm max}) \gtrsim 0.72.$$
With the exception of a few of the \cite{Dehnen98} models and possibly
the heavy disk model of \cite{Kent92}, all models shown in
\Table{tableMW} find (or assume) a submaximal disk for the Milky Way.
Roughly speaking, systems with smaller values of $V_{\rm
  disk}^2/V_{\rm tot}^2$ will be more stable against bar formation,
though this ratio alone is an inadequate predictor of whether a bar
will form (see, e.g. \cite{DeBattista00} and WPD).

\subsection{Further Observational Constraints on the Milky Way Potential}

\subsubsection{Circular Speed at the Sun's Position in the Galaxy}

As mentioned above, our position within the Milky Way offers unique
observational opportunities.  Recently, \cite{Reid09} (see also
\cite{Brunthaler11} reported VLBI measurements of trigonometric
parallaxes and proper motions for 18 masers located in several of the
Galaxy's spiral arms.  These measurements yielded several structural
parameters of Galaxy, most notably, the circular rotation speed at the
position of the Sun.  Their value $V_c = 254\pm 16\,{\rm km\,s^{-1}}$,
is 15\% higher than the standard IAU $220\,{\rm km\,s^{-1}}$.
(\cite{Brunthaler11} analysed the same VLBI data but with updated values for the
solar motion and found $V_t = 239\pm 7\,{\rm km\,s^{-1}}$.)
By contrast, \cite{Koposov10} found $V_c = 221\pm 18\,{\rm km\,s^{-1}}$
by fitting an orbit to the GD-1 stellar stream.  Naively, a 15\%
change in circular speed implies a 33\% change in mass, though the
implications for the bulge, disk, and halo masses would require an
analysis that incorporates other constraints such as the ones carried
out by \cite{Dehnen98} or WPD.  Moreover, \cite{Bovy09} reanalysed the
maser data and found a somewhat lower value with $V_c=236 \pm 11$.
Their Bayesian analysis differed from that of \cite{Reid09} by
assuming a more general orbital distribution for the masers.  

Recently, \cite{Schonrich12} described various ways to determine $V_c$
based on as stellar kinematics.  The idea is to model the streaming
motion of stars using the full phase space information available from
the SDSS/SEGUE survey.  The results are consistent with those
described above and again suggest that $V_c$ is $\sim 10\%$ higher
than the standard IAU value.

\subsubsection{Local Escape Speed}

\cite{Smith07} use a sample of high-velocity stars from the RAdial
Velocity Experiment (RAVE; \citeauthor{Steinmetz06}) to infer that the
local escape speed of the Galaxy is $544^{+64}_{-46}\,{\rm
  km\,s^{-1}}$ (90\% confidence interval).  The fact that the escape
speed is significantly higher than $\sqrt{2} V_c$ provides compelling
evidence that the Galaxy is embedded in a dark matter halo that
extends well-beyond the solar circle.  \cite{Smith07} go on to combine their
estimate for the escape speed with other kinematic constraints to
derive (model-dependent) estimates for the virial radius, virial mass,
and concentration of the Galactic dark halo.  In particular, they
consider a model for the Galaxy comprising a Miyamoto-Nagai (1975)
disk, a Hernquist (1990) bulge, and an NFW halo.  The halo potential
can be written
\be
\Phi_{\rm NFW}(r) = \frac{GM_{\rm vir}}{\ln{\left (1+c\right )} - 
c/\left (1+c\right )} \frac{1}{r}\ln{\left (1 + \frac{r}{r_s}\right )}
\ee
where $r_s$ is the scale radius (see Equation 11), $M_{\rm vir}\equiv
M(r=r_{\rm vir})$ is the virial mass, and $c = r_{\rm vir}/r_s$ is the
concentration parameter.  The virial radius $r_{\rm vir}$ is defined
such that the mean density within $r_{\rm vir}$ is $\delta_{\rm
  th}\rho_{\rm cr}$ where $\rho_{\rm cr}$ is the background density
(\cite{NFW96}).  The NFW potential has two free parameters.  Thus,
by combining the escape speed constraint
\be
\Phi_{\rm total} = -\frac{v_{\rm esc}^2}{2}
\ee
with the constraint on the local circular speed, one can infer the
parameters of the dark halo.  \cite{Smith07} assume $M_d=5\times
10^{10}\,{\rm M}_\odot$, $M_b = 1.5\times 10^{10}\,{\rm M}_\odot$,
$\delta_{\rm th} = 340$, and $V_c = 220\,{\rm km\,s^{-1}}$ to obtain
$M_{\rm vir} = 0.85^{+0.55}_{-0.29}\times 10^{12}\,{\rm M}_\odot$.

\subsubsection{Kinematic Tracers}

Yet another route to contraining the mass distribution of the Galaxy,
especially beyond the solar circle, is provided by
kinematic tracers such as halo stars, globular clusters, and satellite
galaxies.  Line-of-sight velocities are naturally more easily measured
than proper motions.  For tracers in the outer parts of the Galaxy,
these line-of-sight velocities are primarily radial (with respect to
the Galactic center) whereas line-of-sight velocities of tracers in
other galaxies will be a mixture of radial and tangential velocities.
Thus, the analysis of Galactic tracers takes a different form to that
for external galaxies.

Early work on dynamical tracers used simple mass estimators such as
the following example due to \cite{LyndenBell83}.
\be
M_{\rm est} = \left (\frac{2}{\langle e^2\rangle_s}\right ) \left
  (\frac{1}{GN} \sum_{i=1}^N v_{ri}^2 r_i\right ) 
\ee
where the sum is over the $N$ objects (satellites, globular clusters,
etc.)  of the sample.  The quantity $\langle e^2\rangle_s$ is the mean
orbital eccentricity ($\langle e^2\rangle_s=1/2$ for an isotropic
distribution) and encapsulates our ignorance of the orbit
distribution.  The above expression is suitable for a system of
objects orbiting a point mass $M$.  Recently, \cite{Watkins10} carried
out a detailed systematic study of mass estimators for both the Milky
Way and Andromeda galaxies based on satellite kinematics.  Their
results illustrate just how sensitive mass estimates are to
assumptions about the velocity structure of the satellite population.
In particular, the mass estimates at 100 kpc range from $1.8\times
10^{11}\,{\rm M_\odot}$ to $2.3\times 10^{12}\,M_\odot$ depending on
the degree of radial velocity anisotropy.  A detailed discussion of
mass estimators as they might be applied to Gaia data can be found in
\cite{An12}

With enough kinematic tracers, one can determine a velocity dispersion
profile.  \cite{Battaglia05} for example, derive a radial velocity
dispersion profile for the Galaxy from a sample of 240 halo objects
(mainly globular clusters and halo stars).  The dispersion is roughly
constant at $120\,{\rm km\,s^{-1}}$ within $30\,{\rm kpc}$ and then
decreases to about $50\,{\rm km\,s^{-1}}$ at a Galactocentric distance
of $120\,{\rm kpc}$.  These results are consistent with those from a
sample of 2400 BHB stars by \cite{Xue08}, though
that sample only extended to a Galactocentric radius of $60\,{\rm
  kpc}$.  As discussed above, \cite{Xue08} use the kinematics of BHB stars 
to derive a rotation curve by comparing their data with mock
observations of simulated galaxies.  Alternatively, one can model the 
dispersion profile directly.

The decline in the dispersion profile beyond $30\,{\rm kpc}$ places
constraints on the density profile of the dark halo.  In particular,
an isothermal halo, where the density is $\propto r^{-2}$ is
inconsistent with the data since the dispersion profile in this case
is constant with radius.  \cite{Battaglia05} compare their dispersion
profile with model profiles derived from the Jeans equation and find
that the data are consistent an NFW halo that has a relatively high
concentration ($c=18$) and $M_{\rm vir}\simeq 0.8\times 10^{12}\,{\rm
  M}_\odot$.

\cite{Gnedin10} also used halos stars to infer a rotation curve for
the Milky Way.  Their analysis was based on the spherical Jeans
equation with the assumption that both the density and velocity
dispersion profiles for the halo stars are described by power-law
functions of radius.  They find that the velocity dispersion declines
very gradually out to $80\,{\rm kpc}$ and thus they infer a shallow
slope for the dispersion profile.  For this reason, they infer a
larger mass for the virial radius and mass of the Galactic halo than
\cite{Battaglia05}.  Best-fit values for their three-component
Galactic model (motivated, to a large extent, by the \cite{Klypin02}
cosmological model) are given in \Table{tableMW}.

In principle, one can model the phase space DF for a set of tracers
thereby taking full advantage of the data.  This idea was discussed by
\cite{Little87}, who also introduced the use of Bayesian statistics to
the problem.  One begins by calculating the likelihood function (the
probability of the data (e.g., radial velocities and distances for a
population of kinematic tracers) given a model for the tracer DF and
gravitational potential.  Bayes's theorem then allows one to calculate
the corresponding probability of the model given the data.  In
general, one since on is interested in the potential, one marginalizes
over those model parameters that describe the DF.  More sophisticated
models for the Milky Way potential were considered by
\cite{Kochanek96} and \cite{Wilkinson99} who modelled the kinematics
of the satellite galaxy population.  Similarly, \cite{Deason12}
modelled the distribution function of BHB stars to obtain constraints
on the Galactic potential from 15-40 kpc.

\subsubsection{Vertical Force and Surface Density in the Solar
  Neighborhood}

The vertical structure of the Galactic potential provides a
potentially powerful probe of the amount of dark matter in the solar
neighborhood.  The vertical force is approximately proportional to the
surface density, which, in turn, can be compared to the total surface
density from stars the interstellar medium.  The classical means of
determining the local vertical force requires a sample of stars with
known vertical distances and velocities.  The first detailed analysis
of this type was carried out by \cite{Oort32} who built upon earlier
and important work by \cite{Kapteyn22} and \cite{Jeans22}.  Attempts
to understand the local vertical structure of the Galaxy from stellar
kinematics have come to be known as the Oort problem.

Of course, in order to infer the local density of dark matter from the
vertical structure of the Galaxy near the Sun, one requires a detailed
model of the local distribution of visible matter.  A detailed model
of the visible components in the solar neighborhood can be found in
\cite{Flynn06}.  Note that the total estimated density in
visible matter (their Table 2) is $0.091\,M_\odot\,{\rm pc}^{-3}$, which is a factor
of $6-10$ times greater than the predicted local dark matter density.
\cite{Flynn06} also found the local mass-to-light ratios in the $B$,
$V$, and $I$ bands to be $(M^*/L)_B=1.4\pm0.2$, $(M^*/L)_V=1.5\pm0.2$,
and $(M^*/L)_I=1.2\pm0.2$, in good agreement with population synthesis
predictions that use typical solar neighborhood IMFs.

In general, the kinematics of the stars in the disk of the galaxy can
be described by a phase space distribution function (DF), $f({\bf
  r},\,{\bf v},\,t)$, which obeys the collisionless Boltzmann equation
\be \frac{\partial f}{\partial t} + {\bf v}\cdot \frac{\partial
  f}{\partial {\bf r}} - \frac{\partial f}{\partial {\bf v}}
\frac{\partial \Phi}{\partial {\bf r}} = 0~.  \ee Typically, one
assumes that the stars in the sample are in dynamical equilibrium
($\partial f/\partial t = 0$) and that vertical motions decouple from
motions in the disk plane.  With these assumptions, $f({\bf r},\,{\bf
  v},\,t)\propto F\left (z,\,v_z\right )$ and \be v_z \frac{\partial
  F}{\partial z} - \frac{\partial F}{\partial v_z} \frac{\partial
  \Phi}{\partial z} = 0~.  \ee Finally, Jeans theorem implies $F\left
  (z,\,v_z\right ) = F(E_z)$ where $E_z = v_z^2/2 + \Phi(R_0,z) -
\Phi(R_0,0)$.

Formally, each star in the sample may be viewed as a $\delta$-function
in $z$ and $v_z$ and therefore the data provide an estimate of $F$,
which, in principle, may be used to infer $\Phi$.  In general
restrictive assumptions are required to extract $\Phi$.  For example,
the distribution may be assumed to be isothermal so that $F\propto
\exp{\left (-E_z/\sigma_z^2\right )}$ where $\sigma_z$ is the vertical
velocity dispersion.  One then finds $\Phi(z) = \ln{\left
    (\nu(z)/\nu(0)\right )}$ where $\nu(z)=\int Fdv_z$ is the density
run for the tracers.  Variations of this method have been used by
\cite{Oort32}, \cite{Bahcall84}, and \cite{Holmberg00}.

\cite{Kuijken89} proposed an alternative method which does not rely on
the assumption of isothermality.  The DF is related to the density and
potential through an Abel transform: \be F(E_z) =
\frac{1}{2^{1/2}\pi}\int_{E_z}^\infty \frac{d\nu}{d\Phi}\frac{d\Phi}
{\left (\Phi - E_z \right )^{1/2}} \ee In order to use this equation,
Kuijken \& Gilmore make an ansatz for the form of $\Phi$.  For each
choice of potential, they calculate $F$ and then compare the velocity
distribution of the model to that of the data.  The best-fit over the
space of potentials leads to an estimate for $\Phi$.  In principle,
the local density may be determined from $\Phi$ through the Poisson
equation.  The difficulty is that one requires second derivatives of
the potential.  An alternative is to determine the local surface
density, which is (approximately) proportional to the force.  The dark
matter distribution can then be assumed to be approximately constant
close to the Galactic plane (i.e., dark matter is distributed in a
halo and not a disk (but see \cite{Read08} for an alternative
viewpoint.)

Recently, \cite{MoniBidin12b} argued that there is little
or no dark matter in the solar neighborhood.  Their analysis is based
on an analysis of some 400 red giant branch stars found in the Two Micron
All Sky Survey \citep{Skrutskie06}.  The stars in their sample lie
in the direction of the South Galactic pole between 1.5 and 4.0 kpc
from the Galactic midplane and are presumably mainly from the thick disk.
Their analysis is based on the Poisson and Jeans
equations under the assumptions that the stars in the solar
neighborhood are in equilibrium and that the Galaxy has azimuthal
symmetry.  With these assumptions, the integrated surface density within
a distance $z$ of the Galactic midplane is
\be
\Sigma(z) = -\frac{1}{4\pi G}\int_{-z}^{z}
\frac{1}{R}\frac{\partial}{\partial R}\left (RF_R\right )dz
- 2\left [F_z(z)-
F_z\left (0\right )\right ]
\ee
where $F_R$ and $F_z$ are the radial and vertical components of the
force, respectively.  They then use Jeans equation to write these
components in terms of radial and vertical derivatives of the
components of the velocity dispersion tensor and the density.  While
$z$-derivatives are estimated from the data, the data are too sparse
to provide information about the radial derivatives.  Rather, the
authors assume that all second moments of the velocity dispersion
tensor scale as $\exp{\left (-R/h\right )}$ where $h$ is the radial
disk scale length for the surface density.  Furthermore, they assume
that the rotation curve is locally flat for all $z$.

With these assumptions, \cite{MoniBidin12b} found that $\Sigma(z=1.5
{\rm kpc})\simeq~ 55 M_\odot {\rm pc}^{-2}$ and is very nearly flat
between $1.5$ and $4.0$ kpc.  In fact, their $\Sigma(z)$ curve is
well-fit by the curve for visible mass.  By contrast, the standard
disk-halo model for the solar neighborhood has $\Sigma\left (z=1.5
{\rm kpc}\right )\simeq 75\,M_\odot\,{\rm pc}^{-2}$ and rises to $\sim
100\,M_\odot\,{\rm pc}^{-2}$ by $z=4\,{\rm kpc}$.  \cite{MoniBidin12b}
therefore conclude that there is little room for dark matter in the
solar neighborhood.

These results have been called into question by \cite{Bovy12} who
point out that while the circular speed ($=\sqrt{-RF_R}$) may indeed
be locally flat near the midplane, the mean velocity exhibits a
significant lag with respect to the local standard of rest due to
asymmetric drift.  When this effect is taken into account, the data
imply a local dark-matter density of $0.008\pm 0.002 M_\odot \,{\rm
  pc}^{-3}$, a result that is consistent with the values shown in
\Table{tableMW}. As well \cite{Garbari12} have analysed kinematic
data for 2000 K dwarfs using a method that is also based on the
Poisson-Jeans equations and found a relatively high value for the
local dark matter density ($\rho_{DM,\odot} = 0.022^{+0.015}_{-0.013}
M_\odot \,{\rm pc}^{-3}$.

Virtually all approaches to the Oort problem rely on the assumption
that the stars in the solar neighborhood are in dynamical equilibrium.
\cite{Widrow12} found evidence that the disk near the Sun has been
perturbed.  The results are based on their analysis of solar
neighborhood stars from SDSS \citep[Data Release 8;][]{aihara11} and
the SEGUE spectroscopic survey \citep{Yanny09}. The evidence comes in
the form of an asymmetry between the number density north and south of
the Galactic midplane. The asymmetry function (difference between the
number density to the North minus the number density to the South
divided by the average) has the appearance of a wavelike perturbation.
In addition, there appears to a gradual trend in the bulk velocity
across the Galactic midplane.  This result has also been observed in
the RAVE survey \citep{Williams13}.  The perturbations are fairly
small (10\% or less in the density) and therefore the uncertainty in
the surface density due to this effect is likely less than current
observational uncertainties.  Nevertheless, as the observational
situation improves, it may become important to account for departures
from equilibrium.  In any case, the \citet{Widrow12} result may well
represent a new window into the interaction between the disk and halo.
In particular, such perturbations may have arisen from a passing
satellite or dark matter subhalo \cite{Widrow12,Gomez13}.

A second issue concerns the assumption that $E_z$ is an integral of
motion, at least to a good approximation. \cite{Statler89} has
stressed that, for $|z| > 1\,{\rm kpc}$, the approximation breaks down
and has proposed the use of St\"ackel potentials, which admit three
exact integrals of motion.  Unfortunately, the St\"ackel potentials
are a fairly restricted set and, to date, no realistic
disk-bulge-halo model has yet used them.

\subsection{Future Prospects}

The future for mass modelling of the Milky Way is undoubtedly
promising with the anticipated explosion of data from observational
programs such as Gaia \citep{Perryman01, Wilkinson05} and LSST
\citep{Ivezic08}.  Gaia will yield distances and proper motions for
about one billion Milky Way stars to its faint limit of V$\sim$20 mag
and also provide radial velocities for about 150 million stars as
faint as V$\sim$16 mag. LSST has expected uncertainties in parallax
and proper motions that are well matched to those expected for the
faintest Gaia stars and will provide meaningful measurements down to
r$\sim$24 mag.  We can therefore expect direct parallax distances with
accuracies better than 10\% for turn-off stars to $\sim$1 kpc and for
bright RGB stars to $\sim$10 kpc.  Parallax distance accuracies of 10
\kms\ will be achieved for transverse velocities out to 10 kpc;
similar accuracies for larger distances will require other
(photometric) distance estimates to be folded in.  Astronomers will
then have an unprecedented description of the phase space distribution
function for the stellar component of the Galaxy.  Major achievable
science goals with these data include the discovery of stellar streams
from tidally disrupted satellites and new constraints on the local
distribution of dark matter. Current Milky Way models may lack the
richness and sophistication worthy of the data; several groups have
indeed begun laying the groundwork for accurate Milky Way mass models
in the era of Gaia and LSST \citep[][and the many articles in these
proceedings]{Binney10b,Sharma11,An12,McMillan13,Magorrian13}.


\newpage

\section{Dynamical Masses of Gas-Poor Galaxies}\label{sec:gaspoor}
\subsection{Introduction}

As we go from gas-rich spiral systems to early-type galaxies (ETGs)
or dwarf spheroidals (dSphs), common practice is to abandon the
systematic use of the extended neutral gas component of the former
as a tracer to determine the mass distribution.  However, many ETGs
have a significant amount of gas, either ionised
\citep[e.g.][]{Bertola+1984,Fisher1997,Sarzi+2006}, sometimes
molecular \cite[e.g.][]{Sage+2007,Young+2011} or even neutral
\citep{Knapp+85,Morganti+2006,diSeregoAlighieri+2007}.  In such cases,
it is possible to conduct parallel approaches to constrain the overall
mass profiles of the galaxies. This has been exploited in the context
of, say, the search for supermassive black holes
\citep[see e.g.][]{Neumayer10}, the kinematics of the central regions
\citep[][and references therein]{corsini99,VegaBeltran+2001,Pizzella+2004,Sarzi+2006} or large-scale kinematics \citep{Franx+94,weijmans08}.

However, the scarcity and complexity of observed gaseous distributions
and kinematics and the associated difficulty of properly modelling the
dissipative content of galaxies with multi-component morphologies
\citep[though see e.g. ][]{Weiner01} has led to further reliance on
stellar dynamics: the interpretation of the large- (and small-) scale
rotation curves revealed by the emitting gaseous content has thus
generally been overtaken by state-of-the-art modelling of one of the
existing dissipationless tracer, e.g.  old stars, globular clusters,
planetary nebulae. The associated side-products, i.e., constraints on
the orbital structure of the galaxy under scrutiny, have the advantage
of representing a rich source of information to establish its overall
formation and evolution history.

In the following sub-sections, we will therefore restrict ourselves
to briefly introducing the basic ingredients needed for the
\emph{kinematic modelling} of dissipationless systems, i.e., the
determination of the total mass distribution, thus yielding the dark
matter (DM) distribution after subtraction of the visible component.
Determining the mass distribution requires extending beyond the simple
use of the first velocity moment, the mean velocity $V$, as the
centered second velocity moment, the stellar velocity dispersion
$\sigma$, becomes a non-negligible actor.  Orbital shapes also depart
from the commonly-assumed circularity.  Kinematic modelling is
significantly more accurate through the measurement of the detailed
shape of the velocity distribution, which is directly related to the
orbital shapes, and thus allowing a better understanding of the
formation of the galaxies under study.

We first provide some insights on the dependence of mass estimators
based on the measurement of the line-of-sight (LOS) velocity
dispersion on details of the probed aperture.  We then describe the
standard techniques of kinematic modelling, based upon either the
Jeans equations of local dynamical equilibrium or the six-dimensional
distribution functions and we highlight the recent improvements to
these methods.  We finally illustrate the power of these methods with
recent analyses of observed gas-poor galaxies, often obtaining useful
constraints on the compatibility of the DM profiles with those in
$\Lambda$CDM halos.  In several cases, one can also obtain useful
constraints on the DM normalization, concentration, inner slope, as
well as the orbital \emph{velocity anisotropy} (hereafter anisotropy)
in the inner and outer regions.  This review does not specifically
address the mass modelling of central supermassive black holes (see
references in, e.g.  \citealp{Kormendy13}), however most of the
techniques that we discuss here also apply to that problem.  The
reader is also referred to \citet{Gerhard13} for another recent review
on dark matter profiles determinations based on multiple tracers.

\subsection{Simple Mass Estimators}
\label{sec:simple}

Before engaging in the complexity of the mass modelling described
in \se{methods}, we review simple mass estimators that have been
proposed, all based on the scalar virial theorem (sVT).

The sVT states that for an isolated system in steady state
\be
2K+W=0\,,
\label{eq:virth}
\ee where the total kinetic energy $K=\frac{1}{2}M\langle v^2\rangle$,
$M$ is the total galaxy mass and $\langle v^2\rangle$ is the mean
square velocity of its stars, integrated over the entire galaxy, while
$W$ is the total potential energy, which depends on the distribution
of the stars and the possible dark matter.  This energy budget derives
from a time average and depends on the isolation of the dynamical
system (to ensure that the tracer is not affected by a neighboring
system).

For a non-rotating spherical galaxy, the mean square velocity
of the stars is related to the observed LOS velocity dispersion:
$\langle v^2\rangle =3\langle \sigma_{\rm LOS}^2\rangle$ (where
both averages extend to infinity).  Assuming finite mass,
\Eq{virth} gives
\be 
M = c\, \frac{r_r \,\langle \sigma_{\rm LOS}^2\rangle}{G} \,,
\label{eq:vlos}
\ee 
where $r_r$ is a characteristic radius of the galaxy, while
$c = 6 M^2/\{r_r\,\int_0^\infty [M(r)/r]^2\,{\rm d}r\}$, in
the self-consistent case. When the $\langle \sigma_{\rm LOS}^2\rangle$
integral is extended over the {\em entire} galaxy, \Eq{vlos} is
completely independent on the radial variation of the stellar anisotropy
(\citealp{Binney08}, \S~4.8.3). In the general case, the coefficient $c$
depends uniquely on the total ($\rho$) and tracer ($\nu$, i.e. stellar
in galaxies) density profiles of the system.

The practical application of that formula has been the source of
some confusion to be addressed before engaging into tentative
interpretations.
Firstly, the physical radius $r_r$ is measured by the angular radius
that it subtends; therefore, it depends directly on the assumed distance
$D$ of the object. Any uncertainty on $D$ thus translates linearly into $M$. 
Secondly, there is an inherent uncertainty associated with $r_r$'s
measurement: while it can be strictly defined as, for example, the
radius at which half of the galaxy's total light is encompassed, the
notion of total light itself is ill-defined (it often depends on a
subjective extrapolation); the nature of the data also plays a role
(e.g.  bandpass and signal-to-noise effects).  It is thus common to
retrieve specific radii values differing by factors of two or more
for the same well-studied systems \citep[e.g.][]{Kormendy09,Chen10sdss}.
Both issues should be carefully addressed, especially when considering
samples of galaxies for which distances and aperture radii emerge
from heterogeneous sources and/or methods.  Thirdly, when working
at a finite radius $r_r$, one must add a non-negligible surface
term into the virial theorem \citep{TW86}.

The application of \Eq{vlos} as a robust mass estimator is further
compounded by its limited applicability to real stellar systems:
one can rarely observe the stellar $\sigma_{\rm los}$ for the entire
galaxy, due to the rapid surface brightness drop with galactocentric
radii.  Therefore, the effect of using a finite aperture must be
considered \citep{Michard1980, Bailey1981, Tonry83}. In early works,
the coefficient $c$ was only determined for specific galaxy surface
brightness profiles \citep{Poveda1958,Spitzer69}. Using Jeans models,
it was realized that the coefficient $c$ however depends significantly
on the shape of galaxy surface brightness profile
and on the dynamical structure of the galaxy \citep[that is,
its anisotropy, see e.g. ][]{Prugniel1997,Bertin2002}.

Moreover, galaxies contain unknown amounts of dark matter, making
the total mass $M$ an uncertain and ill-defined quantity. For these
reasons, following \citet{TBB04}, we rewrite \Eq{vlos} as
\be
M(r_M) = c\,\frac{r_r \,\sigma_{\rm ap}^2\left(R_{\rm \sigma}\right)}{G} \ ,
\label{eq:mass_r}
\ee
where the $\sigma_{\rm los}$ and mass integrals are restricted to finite
radii, and where
\be
\sigma_{\rm ap}^2({R_\sigma})=\frac{\int_0^{R_\sigma} \Sigma(R) \,
\sigma_{\rm LOS}^2(R)\, R\, {\rm d} R}{\int_0^{R_\sigma} \Sigma(R)\, R\, {\rm d} R}
\label{eq:c1}
\ee
is the squared \emph{aperture velocity dispersion} averaged
over a cylindrical aperture on the sky of projected radius ${R_\sigma}$.

\citet{Cappellari+06} calibrated \Eq{mass_r} using the observed
surface distributions and integral-field kinematics within typically
the \emph{effective radius} containing half the projected luminosity,
$R_{\rm e}$, for a sample of ETGs, in combination with Schwarzschild's
axisymmetric dynamical models. They found that the enclosed mass within
the effective radius ($r_r=R_{\rm e}$) of ETGs can be robustly recovered
using a best-fitting coefficient\footnote{Their expression for
$(M/L)(r_M=R_{\rm e})$ was converted to enclosed mass assuming
$(M/L)(r_M=R_{\rm e})\approx(M/L)(r_M=r_{1/2})$.} $c\approx2.5$,
which varies little from galaxy to galaxy, with $r_M=r_{1/2}^{\rm light}$
(the radius of a sphere enclosing half of the galaxy light) and with
$r_r=R_\sigma=R_{\rm e}$ (the radius of a cylinder enclosing half of
the galaxy light).
Using $r_M=r_{1/2}^{\rm light}$, $r_r=R_{\rm e}$ and $R_\sigma\to\infty$,
\citet{Wolf+10} analytically derived $c\simeq 4.0$ for systems with
$\sigma_{\rm LOS}(R) \approx \rm Cst$ and proved that $c$ depends very
little on anisotropy (as expected given their infinite aperture for
$\sigma_{\rm ap}$).\footnote{\citet{Churazov+10} suggested a generalization
of the approach of \citet{Wolf+10} by computing the mass at the radius
where mass is least dependent on anisotropy, assuming the 3 cases of
isotropic, radial and circular orbits.}

\begin{table}
\begin{center}
\caption{Structural constant $c= G M(r_M)/[r_r \sigma_{\rm ap}^2(R_\sigma)]$}
\label{tab:ctable}
\tabcolsep=1.5pt
\begin{tabular}{lcccc}
& & & & \\ [-7pt]
& Spitzer & Cappellari  & Wolf & $3\,R_{\rm e}$ \\
& & & & \\ [-11pt] \hline
& & & & \\ [-7pt]
$r_M$ & $\infty$ & $r_{1/2}^{\rm light}$ & $r_{1/2}^{\rm light}$  & $3\,R_{\rm e}$ \\
$r_r$ & $r_{1/2}$ & $R_{\rm e}$ & $R_{\rm e}$ & $R_{\rm e}$ \\
$R_\sigma$ & $\infty$ & $R_{\rm e}$ & $r_{\rm vir}$ & $3\,R_{\rm e}$ \\
& & & & \\ [-9pt]
\hline
& & & & \\ [-7pt]
\emph{Predicted}& \ \ \ \ \ \,7.5  & 2.5\,\ \    & 4.0\,\ \  & --- \\
Hernquist       & \ \ \ \ \ \ \ 7.46      & 3.31      & 4.84 & 5.74 \\
$n$=2.0 S\'ersic  & \ \ \ \ \ \ \ 7.23      & 3.63      & 4.85 & 7.22 \\
$n$=4.0 S\'ersic  & \ \ \ \ \ \ \ 6.59      & 2.96      & 4.44 & 5.36 \\
$n$=5.5 S\'ersic & \ \ \ \ \ \ \ 5.91     & 2.49    & 3.96 & 4.37 \\
$n$=2.0 S\'ersic + Einasto DM & 112      & 3.74 & 4.96 & 8.38\\
$n$=4.0 S\'ersic + Einasto DM & 103      & 3.20 & 4.33 & 6.70 \\
$n$=5.5 S\'ersic + Einasto DM & \ \,94 & 2.76  & 3.95  & 5.70 \\
& & & & \\ [-9pt]
\hline
\end{tabular}
\end{center}
\end{table}

\Table{ctable} lists the values of $c$ for some popular models of elliptical
galaxies for comparison with different predictions (assuming isotropic
velocities; anisotropic velocities are discussed later).  The values
of $c$ are computed by inserting the LOS velocity dispersion of
Eq.~(\ref{eq:Pwkernel}) into Eq.~(\ref{eq:c1}), yielding
\be
c = 3\,\frac{M(r_M)\,M_{\rm p}(R_\sigma)/ (4 \pi r_r)}
{\int_0^\infty r\, \nu\, M\, {\rm d}r - \int_{R_\sigma}^\infty \left(r^2-R_\sigma^2\right)^{3/2}\,\nu
 \, M\, {\rm d}r/r^2}
\label{eq:c}
\ee
where $\nu(r)$ and $M_p(R)$ are the stellar mass density at $r$ and
stellar mass enclosed in the cylinder of radius $R$, and where the
denominator is obtained, for isotropic orbits,
by \citet{ML05a,ML06a}.\footnote{The aperture velocity dispersions
generally involve a triple integral. However, for simple anisotropic
velocity models, one-third times the denominator of \Eq{c} becomes
$\int_0^{R_\sigma}\! R \,{\rm d}R\,\int_R^\infty
K_\beta(r/R,r_a/R)\,\nu(r)\,M(r)\,{\rm d}r/r$, where $K_\beta$ is
a dimensionless kernel given in \citet{ML05b}.}
Note that our models are idealized as they do not include realistic
kinematics, galaxy rotation or multiple photometric components as in
the real stellar systems on which the estimators were originally
calibrated. The first three models (from \citealp{hernquist90} and
\citealp{Sersic68}) assume no DM, while the last three include an
$m$=6 Einasto DM model (which \citealp{Navarro+04} first found to
fit best the halos in $\Lambda$CDM pure DM simulations), with radius
of density slope $-2$ equal to one tenth the quasi-virial radius,
$r_{200}$,\footnote{At $r_{200}$, the mean mass density is defined
to be 200 times the critical density of the Universe.} 
within which the DM accounts for 90\% of the total.

For the \citeauthor{Cappellari+06} and especially the
\citeauthor{Wolf+10} estimators, the mass within $r_{1/2}$ depends
little on the DM (the \citeauthor{Spitzer69} relation, originally
formulated for single-component polytropes, with
$r_M=r_{1/2}^{\rm light}$, matches that of \citeauthor{Wolf+10},
except for the Hernquist model, where $c=4.96$).  Inclusion of
velocity anisotropy, with
$\beta=r/2/(r+a_\beta)$ \citep{ML05b}, where $a_\beta = 2\,R_{\rm
  e}$ as found for ellipticals formed by mergers by \citet{Dekel+05},
makes no difference when $R_\sigma\to\infty$ (as theoretically
expected) and decreases $c$ by typically only 4\% for finite
$R_\sigma$.

In general, galaxies are not spherical and rotate. For this reason,
Eq.~(\ref{eq:mass_r}) is not rigorously correct. However, for an aperture
that extends to $1R_{\rm e}$, Eq.~(\ref{eq:mass_r}) was empirically found
to still provide a reliable enclosed-mass estimator
\citep{Cappellari+06,Cappellari+12models}. In this case, $\sigma_{\rm e}$
is measured from a single `effective' spectrum within an aperture, centered
on the galaxy, enclosing half of the galaxy light. The spectrum can be
obtained from integral-field observations for nearby galaxies, or from
a single aperture for high-redshift ones. The resulting $\sigma_{\rm e}^2$
provides a good approximation to the luminosity-weighted velocity second
moment $\overline{v_{\rm LOS}^2}\approx\overline{V^2 + \sigma^2}$ inside
the given aperture, where $V$ is the observed mean stellar velocity at
a given location and $\sigma$ is the corresponding dispersion. For this
reason, $\sigma_{\rm e}$ automatically includes contributions from both
rotation and velocity dispersion and is only improperly called $\sigma$.
The inclusion of rotation is essential for the reliability of the mass
estimator.

The \citet{Spitzer69} and \citet{Wolf+10} formulae require kinematic
measurements out to the virial radius\footnote{The virial radius is
defined to be that where the radial streaming motions are small,
typically 4/3 of $r_{200}$.}. Such data can currently be obtained only
for dSph's or globular clusters using individual stellar velocities.
When those data are available, the latter formula is weakly sensitive
to the differences in the input models. However, the \citet{Cappellari+06}
formula should be used instead when only central kinematics
(within $\sim\,1\,R_{\rm e}$) is available, as is currently the
case for most ETGs. The data in \Table{ctable} suggest that most of the
difference in the coefficients $c$ of \citeauthor{Cappellari+06} and
\citeauthor{Wolf+10} is attributable to their use of different apertures
to measure kinematics.  Incidentally, both formulae provide formally
correct results for a self-consistent S\'ersic model with $m=5.5$,
where the difference in $c$ is entirely explained by the different
apertures.

\Table{ctable} also illustrates that aperture averaged velocity
dispersions out to $3\,R_{\rm e}$ are insufficient to measure the
DM fraction at that radius (which is determined more accurately
using the radial profile $\sigma_{\rm LOS}(R)$ out to $3\,R_{\rm e}$).

Below, we consider the more refined methods for determining the
distribution of total mass 
of ETGs lacking a spatially extended gas tracer.

\subsection{Methods based on Dynamical Modelling}
\label{sec:methods}

The mass distribution in a gas-poor (and luminous) galaxy is expected
to be generally dominated by baryons (mostly stars) in the inner parts
and DM-dominated in the outskirts.  The exact location and shape of
the transition between these two regimes has been the subject of an
active debate, with conclusions that seem to depend on the type (and
mass) of the sampled galaxies.

The density profiles of DM halos in dissipationless $\Lambda$CDM
simulations (hereafter $\Lambda$CDM halos) seem to converge
\citep{Navarro+04} to the ``Einasto'' model
\citep[e.g.][]{Einasto+89}, which is mathematically ``prettier'' than
the traditional \citeauthor{NFW96} (\citeyear{NFW96}, hereafter NFW)
model as its central density and total mass are both, unlike NFW,
finite.  These fits are now established from $\simeq 10^{-3}$
\citep{Navarro+10} to 2 or 3 \citep{Prada+06} virial radii.

The inclusion of dissipative gas in cosmological simulations has the
effect of concentrating the baryons in the centers of their
structures, where they dominate the gravitational potential and drive
the DM component deeper inside.  On the scales of ETGs, this effect,
commonly referred to as \emph{adiabatic contraction}
\citep[]{Blumenthal86,Gnedin04} alters the DM density profile towards
the singular isothermal ($\rho \propto 1/r^2$) model.  This result is,
however, expected to be very sensitive to the details of the baryonic
feedback processes, and orbit diffusion by the quickly varying
potential could be an important agent in flattening of the DM halo
cusp \citep[e.g.][]{PontzenGovernato11}.  See more discussion on
this issue in \se{adiabat}.

For all galaxy types, the dissipative nature of baryons leads them to
accumulate in galaxy centers; indeed, if the baryons were negligible
everywhere, the Einasto (or NFW) models found in $\Lambda$CDM halos
would lead to much lower local stellar $M/L$ and aperture velocity
dispersion than observed \citep{ML05a}.  The dominance of baryons in
the center and of DM in the envelopes of ETGs has been confirmed by
X-ray measurements \citep{Humphrey+06,Humphrey2010} and dynamical
modelling \citep[e.g.][]{Cappellari+06, Thomas+11}; see
\se{resultsETG} below.

In the central region (within $\sim 1-2\,R_{\rm e}$) of an ETG, we
rely on a tracer that may generate the majority of the {\em local}
potential, making the stars a nearly {\em self-consistent} component
of the galaxy\footnote{This should never be assumed but rather
  demonstrated.}.  This would contrast with the galaxy's outskirts,
where the potential would be completely dominated by invisible matter,
and our visible tracers are merely a set of orbiting entities.

The holy grail of dynamicists is the \emph{distribution function}
(hereafter DF), that is the density in phase space (the union of
position and velocity spaces) \emph{of the observed tracer}
(luminosity for unresolved ETGs, numbers of stars for resolved
dSphs). Its evolution is set by the collisionless Boltzmann equation
(hereafter CBE), which states the incompressibility of the system 
in 6-dimensional phase space, or in simpler terms that the DF, $f$,
is conserved along trajectories
(${\rm d} f/{\rm d}t = 0$). In vector notation, the CBE reads:
\[
\qquad \qquad \qquad{\partial f \over\partial t} + {\bf v} \cdot
\nabla f
- \nabla \Phi \cdot {\partial f \over \partial{\bf v}} = 0 \ ,
\qquad \qquad ({\rm CBE})
\]
where $\Phi$ is the gravitational potential.  In the last term, 
$- \nabla \Phi$ is the force per unit mass acting on stars (and
other bodies).
The \emph{total density} $\rho$ is uniquely determined from $\Phi$
through the Poisson equation:
\[
\qquad \qquad \qquad \qquad
\nabla^2\Phi = 4\pi G\rho \ .
\qquad \qquad \qquad \quad\ \ ({\rm Poisson})
\]

Solving the CBE coupled with the Poisson equation is a challenging task,
especially since $f$ is a function of at least 6 variables (3 positions,
3 velocities, ignoring any time dependence), and one can rarely access
the tracers representing the total density $\rho$.  A way out of this
conundrum is to consider \emph{local variables}, to eliminate the direct
dependence of $f$ with respect to velocities, as detailed below.

\subsubsection{Jeans Analysis}
\label{sec:Jeans}

The traditional simpler approach to mass modeling involves writing
the first velocity moments of the CBE, yielding the
\emph{Jeans equations}\footnote{The Jeans equations
are also called ``equations of stellar hydrodynamics'' or
``hydrostatic equations''.} that specify the local dynamical equilibrium
\be
{\partial \overline{\bf v} \over \partial t}
+ (\overline{\bf v} \cdot \nabla) \overline{\bf v} = -  \nabla \Phi
- {1\over \nu}\,\nabla \cdot \left (\nu\, \mbox{\boldmath$\sigma^2$} \right )
\ ,
\label{eq:jeans4vec}
\ee
where
$\nu = \int f \,{\rm d}^3 {\bf v}$ is the space density of the tracer,
$\mbox{\boldmath$\sigma^2$}$ is the tracer's dispersion tensor,
whose elements are $\sigma_{ij}^2 = \overline{v_i\,v_j} - \overline{v_i}\,\overline{v_j}$,
where $\nu\,\overline{v_i}= \int v_i \,f\,{\rm d}^3 {\bf v}$ and
$\nu\,\overline {v_i v_j} =  \int v_i v_j \,f\,{\rm d}^3 {\bf v}$.
The product $\nu\,\mbox{\boldmath$\sigma^2$}$ represents the
\emph{anisotropic dynamical pressure tensor} of the tracer.

The CBE and the Jeans equations (Eq.~\ref{eq:jeans4vec}) apply to all systems,
even out of dynamical equilibrium, as long as the tracers behave like test
particles in the gravitational potential, hence do not interact (otherwise
the right-hand-side of the CBE would be non-zero).
In other words, the two-body relaxation time must be much longer
than the age of the Universe, as is the case for ETGs, dEs (except
in their nuclei) and dSphs.  In particular (as mentioned above),
in both the CBE and the Jeans equations, there is no requirement that
the observed tracer density, $\nu$, be proportional to the total mass
density, $\rho$.

With the simplifying assumptions of stationarity (ignoring any direct time
dependence, i.e. removing the first term on the left-hand side of the CBE),
these \emph{stationary Jeans equations} specify the local dynamical
equilibrium:
\be
\nu\,(\overline{\bf v} \cdot \nabla) \overline{\bf v}
+ \nabla \cdot \left (\nu \,\mbox{\boldmath$\sigma^2$} \right )
= -\nu\,\nabla \Phi
\ .
\label{eq:jeans2}
\ee

Using the stationary Jeans equations (Eq.~\ref{eq:jeans2}),
one can relate the orbital properties, contained in the streaming (1st) and
pressure (2nd) terms with the mass distribution contained in the potential
(right-hand-side), through Poisson's equation.
Such a \emph{Jeans analysis} is fairly simple,
as it circumvents the difficult problem of recovering the DF,
by only considering its first few moments, which
more directly relate to real astronomical observable quantities (depending
on spatial coordinates).  However, one is still left with a degeneracy
between mass and the anisotropy of the pressure tensor, as we will discuss
in the following subsections. Moreover, using moments does not guarantee
that the DF is positive or null everywhere \citep{NB84}.

\subsubsection{Spherical Modelling}

The small departures from circular symmetry of many astrophysical systems
observed in projection, such as globular clusters, dSphs and the rounder
early-type galaxies as well as groups and clusters of galaxies, has
encouraged dynamicists to often assume spherical symmetry in their kinematic
modelling.  The \emph{stationary non-streaming spherical Jeans equation} can
then be simply written
\be
\frac{{\rm d} \left(\nu\,\sigma_r^2\right)}{{\rm d} r}
  + 2\,\frac{\beta}{r}\,
\nu\,\sigma_r^2
= - \nu(r)
\,
\frac{v_c^2}{r} \ ,
\label{eq:jeanssph}
\ee 
where
$\nu \sigma_r^2$
is the \emph{radial dynamical pressure} (hereafter \emph{radial pressure}),
$v_c^2(r) = GM(r)/r = r {\rm d}\Phi/dr$
is the squared circular velocity at radius $r$, while
$M(r)$ is the \emph{total} mass profile, and where
\be
\beta(r) = 1 - \frac{\sigma_\theta^2+\sigma_\phi^2}{2\,\sigma_r^2} = 1 -
\frac{\sigma_\theta^2}{\sigma_r^2}
\label{eq:beta}
\ee
is the tracer's anisotropy profile
with $\sigma_r \equiv \sigma_{rr}$, etc., $\sigma_\theta =
\sigma_\phi$, by spherical symmetry,
and with $\beta = 1$, 0, $\to -\infty$ for radial, isotropic and circular
orbits, respectively.  The stationary non-streaming
spherical Jeans equation provides an excellent estimate of the mass profile,
given all other 3D quantities, in slowly-evolving triaxial systems such as
$\Lambda$CDM halos \citep*{TBW97} and for the stars in ETGs
\citep[e.g.  formed by mergers of gas-rich spirals in dissipative
$N$-body simulations;][]{Mamon+06}.

As one is left with two unknown quantities, the radial profiles of
mass and anisotropy, linked by a single equation, one must contend
with a nefarious \emph{mass-anisotropy degeneracy} (MAD). The simplest
and most popular approach to circumvent the MAD is to assume simply
parameterized forms for both the mass and anisotropy profiles.
One can then express the product of the observable quantities: 
the \emph{surface density} profile, $\Sigma(R)$, and the \emph{line
of sight square velocity dispersion} profile, $\sigma_{\rm LOS}^2(R)$,
versus \emph{projected radius} $R$ through the
\emph{anisotropic kinematic projection equation} \citep{BM82}
expressing the observed quantity
\be
\Sigma(R)\, \sigma_{\rm LOS}^2(R)=
2\int_R^\infty \!\!\left (1\!-\!\beta\,{R^2\over r^2} \right )
\nu\,\sigma_r^2\,{r\,{\rm d}r\over \sqrt{r^2-R^2}} \ .
\label{eq:anisproj}
\ee

One can insert the radial pressure from the spherical stationary
Jeans \Eq{jeanssph} into \Eq{anisproj} to determine the line-of-sight
(LOS) velocity dispersions through a double integration over
$\nu\,M\,{\rm d}r$.
\citet[][Appendix]{ML05b}
have simplified the problem by writing the projected
pressure as a single integral
\be
\Sigma(R)\,\sigma_{\rm LOS}^2(R)
= 2\,G\int_R^\infty \!\!\!K_\beta\!\left({r\over R},{r_\beta\over
  R}\right)\nu(r)\,M(r)\,{{\rm d}r\over r} \,,
\label{eq:Pwkernel}
\ee
where they determined simple analytical expressions for
the dimensionless kernel $K_\beta$ for several
popular analytical formulations of $\beta(r;r_\beta)$ (\citealp{Tremaine+94}
previously derived $K_\beta(r,R)=\sqrt{1-R^2/r^2}$ for $\beta=0$).
The number density $\nu$ is obtained by Abel inversion
$$\nu(r) = -(1/\pi) \int_r^\infty ({\rm d}\Sigma_{\rm tot} /
  {\rm d}R)\,{\rm d}
  R/\sqrt{R^2-r^2}.$$
When both $\rho(r)$ and $\nu(r)$ are expanded as sums of spherical
Gaussian functions \citep{Bendinelli91}, \Eq{Pwkernel} can be applied
to the individual Gaussians, which can have different $\beta$ values.
This leads to an expression involving a single quadrature for nearly
general $\beta(r)$ profiles \citep{Cappellari08}.

The next step in complexity is the non-parametric
\emph{mass inversion}, where $\beta(r)$ is assumed,
involving first the
anisotropic kinematic deprojection by inverting \Eq{anisproj}
\citep{MB10,Wolf+10} and then \emph{directly} obtaining the mass
profile by inserting the derived radial pressure into the Jeans
\Eq{jeanssph}.  For simple $\beta(r)$ models, the mass profile can
be written as a single integral \citep{MB10}.  Interestingly, for
systems with roughly constant $\sigma_{\rm LOS}(R)$ (as is the case
for most galaxies), the mass profile at the half-light radius
$r_{1/2} \simeq 1.3\,R_{\rm e}$ is almost independent
of the assumed $\beta(r)$, as analytically derived by \citet{Wolf+10}.

Alternatively, a mass profile is assumed and
one directly determines the anisotropy profile through the non-parametric
\emph{anisotropy inversion}, first derived by \citet{BM82}, with other
algorithms by \citet{Tonry83}, \citet{BBCK89}, \citet{DM92}, and
especially \citet{SS90}.

None of these approaches can lift the MAD. One promising alternative
approach is to consider the variation with projected radius of the LOS
velocity dispersion \emph{and kurtosis} \citep{Lokas02,LM03}.
This method has been successfully tested \citep{SLM04} on $\Lambda$CDM
halos viewed in projection, despite the fact that these halos are triaxial
\citep[][and references therein]{JS02}, with anisotropy that increases with
radius \citep[][and references therein]{ML05b}, substructures and streaming
motions.  Unfortunately, the LOS projection of the 4th order Jeans equation,
required in the dispersion-kurtosis method, is only possible when
$\beta = \rm Cst$, whereas ETGs formed by major mergers show rapidly
increasing $\beta(r)$ \citep{Dekel+05}.  Nevertheless, \citet{RF13a}
recently generalized this approach for systems where the 4th order
anisotropy is a function of the usual 2nd order one, as is indeed
seen in $\Lambda$CDM halos \citep{Wojtak+08}.
             
\subsubsection{Axisymmetric Modelling}\label{sec:aximod}
\begin{figure*}
\centering
\includegraphics[width=0.8\textwidth]{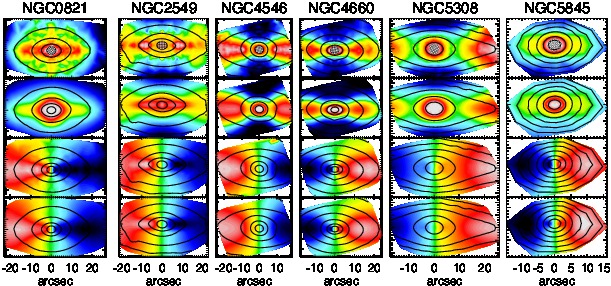}
\caption{Data-model comparison of six fast rotator galaxies (previously
classified as either E's and S0's) using the ``Jeans Anisotropic MGE'' (JAM)
method \citep[from][]{Cappellari08}. From \emph{top} to \emph{bottom}:
bi-symmetrized observations of
$V_{\rm rms}\equiv\sqrt{V^2+\sigma^2}$, model
of same, bi-symmetrized observations of $V$, model of $V$.  The
\emph{contours} show the isophotes. The models generally agree with
the original non-symmetrized data within the statistical errors.}
\label{fig:jam}
\end{figure*}

Still, the large majority of the galaxies in the Universe are to first
order axisymmetric (except for spiral arms and bars) and possess disks
even for ETGs \citep{McDonald11,krajnovic11}. This includes fast-rotators
\citep{emsellem07,cappellari07,emsellem11} and spirals.

If we rewrite the stationary CBE in cylindrical coordinates  $(R,z,\phi)$
and assume {\em axial symmetry} and steady state, we obtain two non-trivial
Jeans equations (\citealt{Jeans22}; \citealt{Binney08}, Eq. 4.222b,c) that
are functions of four unknowns, $\sigma_R^2$, $\sigma_z^2$,
$\overline{v_\phi^2}$, and $\overline{v_R v_z}$ and do not
uniquely specify a solution.
By assuming that the velocity-ellipsoid is aligned with the cylindrical
coordinates, we further simplify such equations as:
\begin{eqnarray}
    \frac{\nu\sigma_R^2-\nu\overline{v_\phi^2}}{R}
    + \frac{\partial(\nu\sigma_R^2)}{\partial R}
    & = & -\nu\frac{\partial\Phi}{\partial R}
    \label{eq:jeans_cyl_R}\\
    \frac{\partial(\nu\sigma_z^2)}{\partial z}
    & = & -\nu\frac{\partial\Phi}{\partial z} \ .
    \label{eq:jeans_cyl_z}
\end{eqnarray}
The two Eqs.~(\ref{eq:jeans_cyl_R}) and (\ref{eq:jeans_cyl_z}) now depend only
on $\sigma_R^2$, $\sigma_z^2$, $\overline{v_\phi^2}$, but one must still
specify at least one function of $(R,z)$ for a unique solution.

The generality of such equations can be maintained by writing a direct
dependence between the two dispersions in the meridional plane via
a function $b$ such that
$\sigma_R^2=b\,\sigma_z^2$, with the boundary condition $\nu\sigma_z^2=0$
as $z\rightarrow\infty$.  This yields \citep[e.g.][]{Cappellari08}
\begin{eqnarray}
    \nu\sigma_z^2(R,z)
    & = & \int_z^\infty \nu\frac{\partial\Phi}{\partial z}\mathrm{d}z
    \label{eq:jeans_sol_z}\\
    \nu\overline{v_\phi^2}(R,z) & = &
    b\left[
    R \frac{\partial(\nu\sigma_z^2)}{\partial R}
    + \nu\sigma_z^2 \right]
    + R \nu\frac{\partial\Phi}{\partial R}
    \label{eq:jeans_sol_R}\,.
\end{eqnarray}
For a given {\em observed} surface brightness and {\em assumed} total mass
distribution, when Eqs.~(\ref{eq:jeans_sol_z}) and (\ref{eq:jeans_sol_R})
are projected onto the plane of the sky and integrated along the LOS,
they produce a {\em unique} prediction for the observed 2nd moment
$\overline{v_{\rm LOS}^2}$, as a function of
$b$ and the inclination $i$.
The 2nd moment $\overline{v_{\rm LOS}^2}$ is empirically well approximated
by $v_{\rm rms}^2 \equiv V^2 + \sigma^2$ (the squares of the centroid of
a Gaussian fit to the LOS velocity profile and of its dispersion), which
is easily observed in galaxies.
This implies that $V$ and $\sigma$ do not provide separately
any extra information on the galaxy mass that is not already contained
in their quadratic sum. It also implies that, when galaxy rotation $V$
is significant, one cannot neglect its contribution to the galaxy mass
determination, and that one needs to know the inclination of the
galaxy accurately.
The dependence of the mass distribution on $V_{\rm rms}$ alone can be
physically understood: for a given dynamical model, any star along a
given orbit can have its sense of rotation reversed without altering
$\overline{v_{\rm LOS}^2}$ (or the mass).

To predict galaxy rotations from the Jeans equations one must make an
{\em extra} assumption on how the 2nd moments around the symmetry axis
$\overline{v_\phi^2}$ divides into ordered and random motions:
$\overline{v_\phi^2}=\overline{v_\phi}^2 + \sigma_{\phi}^2$. 
The simplest assumption to define this division is to adopt an
\emph{oblate velocity ellipsoid} (OVE), namely assume
$\sigma_\phi=\sigma_R>\sigma_z$ \citep{Cappellari08}.
This OVE model has a streaming velocity uniquely defined by
\be \label{eq:v_phi2}
\overline{v_\phi}^2 = \overline{v_\phi^2} - b\, \sigma_z^2 \,,
\ee
with $\overline{v_\phi^2}$ and $\sigma_z^2$ as given in
Eqs.~(\ref{eq:jeans_sol_z}) and (\ref{eq:jeans_sol_R}).

With $b=1$, Eqs.~(\ref{eq:jeans_sol_z}) and (\ref{eq:jeans_sol_R})
define a circular velocity-ellipsoid in the $(v_R,v_z)$ plane: this
is the historical \emph{semi-isotropic} assumption that implies
$\sigma_R=\sigma_z$ (and $\overline{v_R v_z}=0$), and is sufficient
to `close' the set of equations to provide a unique solution for the
remaining variables $\sigma_z^2$ and $\overline{v_\phi^2}$
\citep{nagai76,satoh80,binney90,vanderMarel90,emsellem94}. If we
consider \Eq{v_phi2}, we retrieve the special case of the classic
{\em isotropic rotator} \citep{binney78}.

When the total mass and surface brightness are described via the
\emph{Multi-Gaussian Expansion} (MGE) method of \citet{emsellem94},
the potential and Jeans Equations can be expressed in a simple form
and $\overline{v_{\rm LOS}^2}$ only requires a single quadrature both
for the semi-isotropic case \citep{emsellem94, emsellem99}. This is
also true for the general case $\sigma_R\neq\sigma_z\neq\sigma_\phi$,
and for all the six projected second moments, including radial
velocities and proper motions, as demonstrated by
\citet{Cappellari08}.  This flexibility can be more practically
witnessed when using an implementation of the ``Jeans Anisotropic MGE''
(JAM) modelling method \citep[see][]{Cappellari08} augmented by
the possibility to probe the parameter space within a Bayesian
framework (e.g.  \se{IMF}). A key feature of the OVE rotator with
constant $b = (\sigma_R/\sigma_z)^2$ is that it maintains the
simplicity of the isotropic (or semi-isotropic) rotator, but contrary
to the latter, it provides a remarkably good description of the
observations. In fact, these suggest that both fast-rotator ETGs
\citep{cappellari07,Thomas09} and disk galaxies have a dynamical
structure roughly characterized by a flattening of the velocity
ellipsoid in the $z$ direction parallel to the galaxy symmetry axis
\citep{gerssen97,gerssen00,shapiro03,Noordermeer08}.  Indeed, once an
accurate description of the surface brightness of the galaxies is
provided via the MGE, the OVE rotator with constant $b$ accurately
predicts (\Fig{jam}) both the first ($V$) and 2nd ($V_{\rm
  rms}=\sqrt{V_{\rm rot}^2+\sigma^2}$) moment of the LOS velocity as
inferred with state-of-the-art integral-field observations of the
stellar kinematics of large samples of fast-rotator ETGs
\citep{Cappellari08,Scott09,Cappellari+12models}. The success of the
cylindrical oriented approximation may be related to the disk-like
nature of the majority of the galaxies in the Universe, where this
particular alignment of the velocity ellipsoid appears natural
\citep{Richstone84}.

Real galaxies need not have accurately cylindrically oriented velocity
ellipsoids. In fact, theoretical arguments and numerical experiments
suggest the velocity ellipsoid cannot be perfectly cylindrically oriented
\citep[e.g.][]{dehnen93}. However, comparison with realistic $N$-body
simulations of galaxies indicate that the cylindrically-oriented velocity
ellipsoid approximation can be used to reliably measures the mean values
of the internal anisotropy or to recover mean $M/L$ even in realistic
situations where the anisotropy is not constant \citep{Lablanche12}.

\subsection{Distribution Function Analysis}\label{sec:basis}

Although the Jeans analysis is simple and fast, it has two disadvantages:
firstly, the 2nd LOS velocity moment does not describe the full information
of projected phase space (hereafter, PPS) $(\alpha,\delta,v_{\rm LOS})$,
where $(\alpha,\delta)$ are the equatorial sky coordinates and $v_{\rm LOS}$
is the LOS velocity), and even the inclusion of the higher order moments
\citep[see e.g. ][]{Magorrian94,Magorrian95} is less informative than using
the full PPS; Secondly, for spherically modeled galaxies, the solutions of
the Jeans analysis
depend on the required radial binning of the velocity moments.  Moreover,
the variation of the velocity moments with projected radius is often noisy,
requiring smoothing of the data.
We now describe a more general family of mass modelling methods, which
solves for the gravitational potential and the DF, by fitting the PPS
distribution predicted for combinations of gravitational potential and
DF to the observed PPS distribution.

\subsubsection{Spherical Distribution Function Modelling}\label{sec:spherical}

In spherical symmetry, the PPS is simply $(R,v_{\rm LOS})$, where $R$ is the
projected radius, and the DF projects onto PPS as a triple integral \citep{DM92}:
\be
g(R,v_{\rm LOS})\!=\!2\!\int_R^\infty\!\!\!\! {r\,{\rm d}r\over \sqrt{r^2\!-\!R^2}}
\int_{-\infty}^{+\infty} \!\!\!\!\!\!{\rm d}v_\perp \!
\int_{-\infty}^{+\infty}\!\!\!\!\!\!f(r,{\bf v})\, {\rm d}v_\phi\,.
\label{eq:gDM}
\ee
So, with the knowledge of the DF shape, one can fit its parameters to match
the PPS.

In spherical systems with isotropic non-streaming velocities, the DF
is a function of energy only, {\it i.e.\/} $f=f(E)$, while in anisotropic
non-streaming ({\it e.g.\/}, non-rotating) spherical systems it is a
function of energy and the modulus of the angular momentum.
The PPS distribution in isotropic systems is \citep{Strigari+10}
\be
g(R,v_{\rm LOS}) = 4\pi\int_R^\infty\!\!\!\! {r {\rm d} r\over
  \sqrt{r^2-R^2}}\int_{\Phi(r)+v_{\rm LOS}^2/2}^0 \!\!\!\!\!\!\!f(E)\,{\rm
  d}E \ .
\label{eq:giso}
\ee
where the DF is given by the Eddington formula \citep{Eddington16}.

However, $\Lambda$CDM halos are anisotropic \citep[e.g.][]{CKK00,AG08}.
\citet{Wojtak+08} have recently shown that $\Lambda$CDM halos have DFs
that are separable in energy and angular momentum, with
$f(E,L) =
f_E(E)\,L^{2\,(\beta_\infty-\beta_0)}\,\left(1+L^2/L_0^2\right)^{-\beta_0}$,
with $\beta_0 = \beta(0)$ and $\beta_\infty
= \lim_{r\to\infty} \beta$, and where $L_0$ is a free parameter related
to the ``anisotropy radius'' where $\beta(r) = (\beta_0+\beta_\infty)/2$.
Unfortunately, the energy part of the DF is non-analytical, though
\citeauthor{Wojtak+08} show how it can be efficiently evaluated
numerically (they also provide an analytical approximation for $f_E(E)$).
This $\Lambda$CDM halo-based DF can then be applied to fit the distribution
of objects in PPS using \Eq{gDM}, as shown by \citet{WLMG09}.

For quasi-spherical galaxies, where, in contrast to clusters and $\Lambda$CDM
halos, dissipation ought to play an important role, it is not yet
clear that the DF is separable in energy and angular momentum as
\citet{Wojtak+08} have found for $\Lambda$CDM halos.
Moreover, the triple integral in \Eq{gDM} makes the $\Lambda$CDM halo
DF method computationally intensive.  The simplest and popular alternative
is to fit the PPS assuming a Gaussian distribution for the LOS velocities,
and the radial profiles of mass and anisotropy 
\citep{Battaglia+08,Strigari+08_Nat,Wolf+10}. Unfortunately, this method
provides very weak constraints on the anisotropy
\citep{Walker+09}.\footnote{\citet{Walker+09} did not fit the PPS
but $\sigma_{\rm LOS}(R)$.}.
One can assume instead a Gaussian shape for the 3D velocity
distribution, as in the MAMPOSSt method of \citet*{MBB13},
again adopting radial profiles of mass and anisotropy, and
fitting the predicted distribution of particles in PPS.
This operation only involves a single integral.

Both the $\Lambda$CDM halo DF and MAMPOSSt methods have been
successfully tested on $\Lambda$CDM halos viewed in projection
\citep[][respectively]{WLMG09,MBB13}.  They both yield useful
constraints on both the mass and anisotropy profiles: with $\sim500$
tracers, the mass $M_{200}$ within the (quasi-virial) radius $r_{200}$
and the outer anisotropy are recovered with $\sim 30\%$ and $\sim
20\%$ relative accuracy, while the scale radius of the DM is obtained
to within a factor 1.5.  The bias in the recovered $M_{200}$
correlates with the ratio of LOS velocity dispersion measured within
the virial sphere, estimated along the LOS to that measured in 3D
(corrected by $\sqrt{3}$) so that the limiting factor for accurate
mass measurements is the triaxiality of $\Lambda$CDM halos
\citep{MBB13}.

\subsubsection{Towards Flattened Systems}

The majority of the gas-poor ETGs have important elongation in the
plane of the sky.  A number of methods have been developed and tested
in an attempt to retrieve the full DF for flattened systems. In the
case of the semi-isotropic approximation, \citet{HunterQian93},
expanding upon the Eddington formula for spherical systems,
demonstrated that a direct inversion of the mass density $\rho$ can be
obtained analytically, which can then be applied to galaxies with
complex morphologies \citep[see e.g. ][]{emsellem99}. Besides the fact
that this involves analytic extrapolations of functions into the
complex plane, flattened systems do not seem consistent with the
semi-isotropic hypothesis. It may therefore be worth re-examining
this technique with the OVE assumption in mind.

Other techniques based on, for instance, the expansion of the DF into
a set of basis functions \citep{Dejonghe89}, have enjoyed some success
\citep{Kuijken95,Gerhard+98,emsellem99}. When the potential is of the
St\"ackel form, the DF can be readily expressed
\citep{deZeeuw85,Arnold+94}, and the orbital structure is then a
simple function of basic building blocks corresponding to explicit
integrals of motion \citep[see applications in e.g. ][]{Hunter+92,
  Statler2001}.  For these specific cases, the general solution of the
Jeans Equations providing the moments in terms of standard integrals
has even been worked out by \citet{vandeVen+03}.  Unfortunately, the
difficulty of choosing a relevant set of basis functions for the DF,
or to design models that fit specific galaxies from the very centre to
the outer parts, has led modelers to consider other more natural
methods that treat a galaxy as the sum of well-chosen orbits: this is
the subject of the following section.

\subsubsection{General Orbit-based Modelling}

A popular approach to non-spherical potentials (as well as spherical
ones) is that of \emph{orbit modelling}
\citep{schwarzschild79,RT84}. In the axisymmetric case, one considers
orbits of given $E$, $L_z$ and $I_3$ (a non-classic integral of
motion) in a given potential (i.e., the DF is made of delta functions
in $E$, $L_z$ and $I_3$).  One searches for a linear combination of
these orbits that minimizes the residuals between predicted and true
observables, enforcing positive weights.  These weights are obtained
either by averaging the observables over an orbit
(\citeauthor{schwarzschild79}) or by continuously updating them
(\citealp{Syer96}; NMAGIC code of \citealp{deLorenzi07};
\citealp{Dehnen09}; \citealp{Long10}). Such a technique can also be
generalized to triaxial systems \citep{vanDenBosch08}. Although more
challenging to implement than Jeans analyses, orbit-based and
particle-based methods constitute the state-of-the-art methods of
kinematic modelling.

Due to its generality, Schwarzschild's method is more robust to the
biases that may affect some of the other methods. It can also handle
observable quantities more effectively, such as higher order
Gauss-Hermite moments, while Jeans analyses are mostly concerned with
the first two exact velocity moments.  The robust measurement of the
latter can be challenging given the complex LOS velocity
distributions.  Schwarzschild's method has been used extensively to
measure masses of supermassive black holes in galaxies
\citep[e.g.][]{vanDerMarel98,cretton99ngc4342,verolme02,gebhardt03,valluri04},
to measure $M/L$ or DM profiles
\citep[e.g.][]{Cappellari+06,thomas07,Weijmans+09} or to study the
orbital anisotropy
\citep[e.g.][]{cappellari07,vanDenBosch08,Thomas09}.

The generality of Schwarzschild's method is linked with the presence
of degeneracies in the recovered parameters, and with the general need
to regularize the sampling of the PPS by adding minimization
constraints.  Indeed, the unknown three-dimensional shape and using
plane-of-sky velocities makes the dimension of the observable
projected phase space too low. In fact, observations can at best
provide a 3D quantity, namely the LOSVD at every projected location on
the sky plane. This observable has the same dimension as the DF which,
for Jeans's theorem, generally depends on the three isolating
integrals of motion. The dimensionality equivalence between the
observables and the DF explains why one can uniquely recover the DF
from the data, when all other model parameters are known
\citep{thomasj04,krajnovic+05,vandeVen08,Morganti12}. However it is
unlikely to robustly constrain additional quantities from the same 3D
data \citep{valluri04}, namely the 3D total mass distribution and the
angles at which it is observed.  Important degeneracies are indeed
found when trying to measure the galaxy shape
\citep{krajnovic+05,vanDenBosch+09} or mass distributions
(\citealt{deLorenzi+09}; see also \Fig{dmfracs3379} below) with very
general approaches.

\subsection{Results}
\label{sec:resultsETG}

As emphasized above in \se{spherical}, the CBE can be applied to any
tracer, as long as the system is sufficiently isolated.  The choice of
tracers typically involves the (old) stellar population, the globular
cluster or planetary nebula systems, satellite galaxies, or X-ray
emission from hot gas when present.\footnote{See also the use of
low-mass X-ray binaries as mass/dynamical tracers in dSph galaxies
by \citet{Dehnen+06}.}
Leaving the analysis of resolved dwarf spheroidal galaxies for
\se{resultsdSph}, nearly all kinematic studies of ETGs have focused
on the bright end of the luminosity function, as dwarf ellipticals
are often too difficult to study.\footnote{The more rapidly declining
surface brightness profiles of dwarf ellipticals relative to their
giant counterparts makes the spectroscopic measurements at several
$R_{\rm e}$ especially challenging.}

\subsubsection{Integrated Stellar Light: the Inner Regions and the IMF}
\label{sec:stellpopimf}

The integrated stellar light component is the prime choice of tracer
when considering regions within $\simeq 1-2\,R_{\rm e}$ (see
\Fig{jam}), and relatively luminous galaxies.  Most of the
observations then generally extend beyond the first two velocity
moments \citep{vanDerMarel94b}, which often helps to break the MAD
\citep{Gerhard+98,Napolitano+11}.  For galaxies that are intrinsically
flattened or with complex morphologies, it is critical to make use of
the two-dimensional kinematic maps provided by, for instance,
integral-field spectroscopy as shown by \citet{krajnovic+05} and
\citet{cappellari05}.  Progress can also be made by comparing
dynamical and stellar population estimates to infer DM fractions
and more generally the mass distribution in the central regions of
galaxies \citep{gerhard01,Cappellari+06,Napolitano2010,Thomas+11,Wegner12}.
Among others, these works confirm that the total $M/L$ in the inner
region of ETGs does not agree with the one predicted using stellar
population models with a universal IMF.
This can be interpreted as evidence for a variation in the dark matter
fraction in the galaxies central regions, if the IMF is universal, or
that the IMF is not universal and likely a function of total mass.

\begin{figure*}[ht]
\centering
\includegraphics[width=0.85\textwidth]{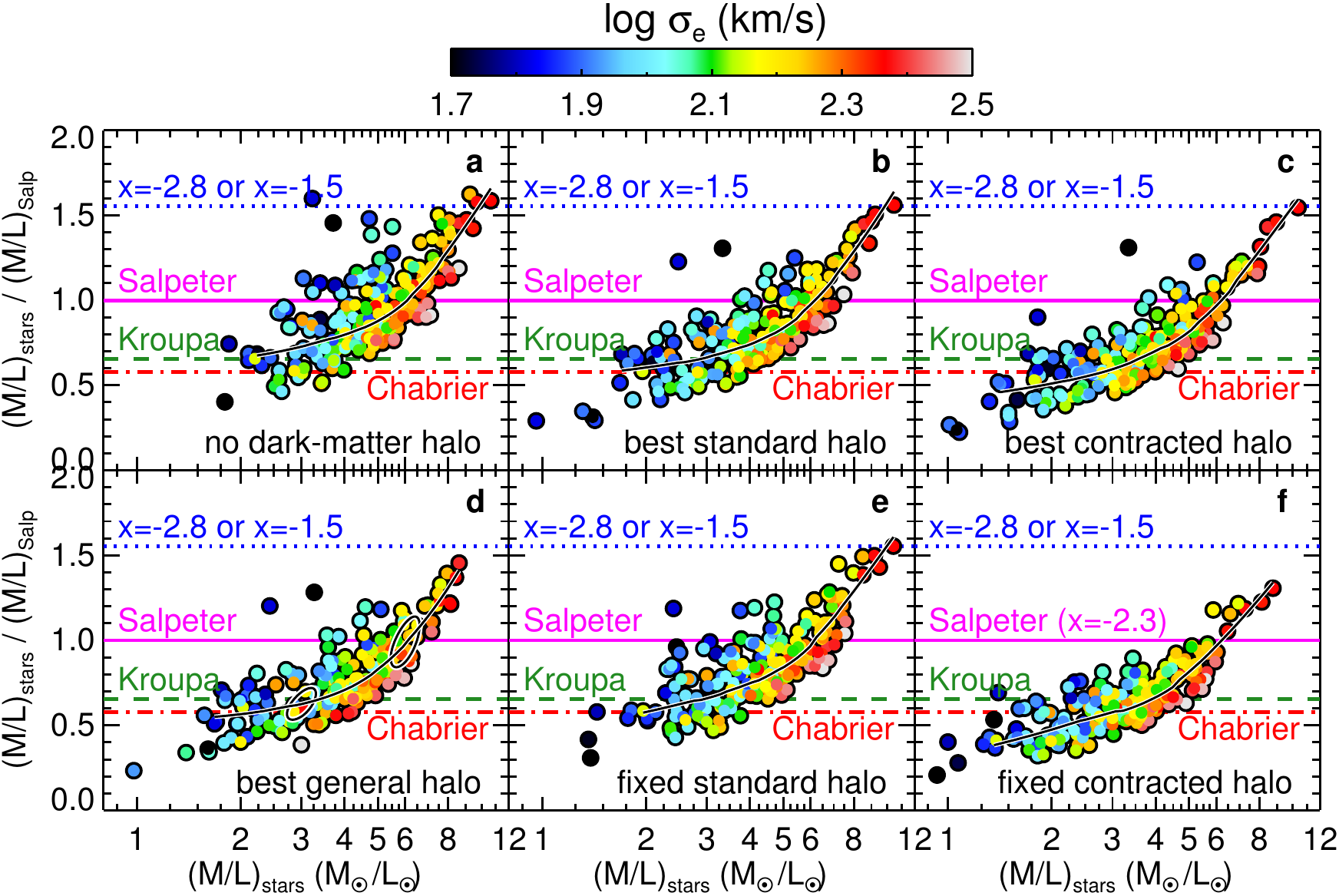}
\caption{Systematic variation of the stellar IMF in ETGs. The six
  panels show the ratio between the $(M/L)_{\rm stars}$ of the stellar
  component, determined using dynamical models, and the $(M/L)_{\rm
  Salp}$ of the stellar population, measured via stellar population
  models with a Salpeter IMF, as a function of $(M/L)_{\rm
  stars}$. The black solid line is a {\em LOESS} non-parametric
  regression to the data.  Colours indicate the galaxies' stellar
  velocity dispersion, $\sigma_{\rm e}$, which is related to the
  galaxy mass. The horizontal lines indicate the expected values for
  the ratio if the galaxy had (i) a Chabrier IMF (red dash-dotted
  line); (ii) a Kroupa IMF (green dashed line); (iii) a Salpeter IMF
  ($x=-2.3$, solid magenta line) and two additional power law IMFs
  with (iv) $x=-2.8$ and (v) $x=-1.5$ respectively (blue dotted line).
  Different panels correspond to different assumptions for the dark
  matter halos employed in the dynamical models as written in the
  black titles. A curved relation is clearly visible in all panels.
  From \citet{Cappellari+12}.}
\label{fig:jam_imf}
\end{figure*}

The most extensive set of detailed dynamical models to date,
accurately reproducing both the galaxy photometry and the
integral-field stellar kinematics, was constructed for the 260 ETGs
of the \atlas\ survey \citep{Cappellari+12models}. This study used
axisymmetric anisotropic models based on the Jeans equations (JAM in
\se{aximod}) and includes a rather general dark halo, where both its
slope and normalization are varied to reproduce the data within a
Bayesian framework. The halo inner logarithmic slope is allowed to
vary from the values predicted by
halo contraction models \citep{Abadi10} to the nearly constant
density profiles expected from halo expansion models
\citep{PontzenGovernato11}.  The median fraction of dark matter
inferred from the models, within a sphere of radius $r=R_{\rm e}$,
is just 10-20\%.  \citet{Cappellari+12models} find this dark matter
fraction to be consistent with predictions for the same galaxies inferred
by linking NFW halos to the real galaxies and assuming halo masses
via the halo abundance matching technique
\citep[e.g.][]{Behroozi2010}. Using satellites as tracers of the
gravitational potential, the NFW model of \citet{WM13}
extrapolates\footnote{The projected radii of the satellite
galaxies analyzed by \citet{WM13} begin at $5\,R_{\rm e}$.}
to a similar DM fraction within $R_{\rm e}$ for red galaxies
of masses $1.6 - 5\times 10^{11} M_\odot$ (see \Fig{dmfracs3379} below),
but to much larger dark matter fractions for lower and higher mass
red galaxies.


The study of \citet{Cappellari+12} finds that dark matter cannot
explain the systematic increase in the total $M/L$ with the galaxies
velocity dispersion $\sigma_e$.  This implies a systematic variation
of the stellar IMF with $\sigma_e$, with the mass normalization
changing by a factor up to 2--3, or from \citet{chabrier03} or
\citet{kroupa01} to heavier than \citet{Salpeter55} over the full
galaxy population (\Fig{jam_imf}; \citealt{Cappellari+12}).
The Salpeter or heavier IMF for the most massive ETGs is consistent with
recent findings from the analysis of IMF sensitive spectral features
\citep{vanDokkum10,Spiniello12,Conroy12} and with strong lensing
results \citep[see, e.g.,][and discussion in
\se{strong}]{Treu10,Auger10,DuttonMMS13},
under the assumption of cosmologically motivated dark matter halos.
This result also smoothly bridges the gap between the IMF inferred
for massive ETGs and the lighter Chabrier/Kroupa inferred for spiral
galaxies \citep[][see \se{gasrich}]{Bell01}.

Alternatives to a non-universal IMF do exist, to explain the dynamical
or lensing results, but they require that (i) either all current stellar 
population models (\se{stellpop}) systematically and severely under
predict the $M/L$ for the galaxies with the largest $\sigma$, which
are characterized by the largest metallicities, or (ii) the dark matter
accurately follows the stellar distribution, contrary to what all
simulations predict.  Moreover the IMF trends inferred from spectral
absorption features would need to be explained by a conspiracy of
chemical abundance variation with galaxy $\sigma$. There is currently
no evidence for any of these effects, but further investigations
in these directions are still important.

Several recent studies have provided kinematic measurements of the
integrated stellar light of ETGs beyond $3-4\,R_{\rm e}$, using
long-slit \citep{thomas07,Proctor2009,Coccato+10,Arnold+11} or
two-dimensional spectroscopy \citep{Weijmans+09}.  This, however,
remains a rather challenging task, and as we probe outwards we must
consider more discrete tracers such as globular clusters and planetary
nebulae (see \se{outerregions} below).  The results obtained so far
have provided a picture where ETGs are dominated by the stellar mass
out to $1-3\, R_{\rm e}$, thus playing the counterpart of the
{\em maximum disk} hypothesis in gas-rich systems (see \se{maxdisks})
but for hotter stellar systems, while the outer halos are generally
consistent with $\Lambda$CDM predictions.

\subsubsection{Globular Clusters and Planetary Nebulae: the Outer Regions}
\label{sec:outerregions}
In order to probe the distant radii beyond $\simeq 4\,R_{\rm e}$,
observations from individual globular clusters \citep[e.g.][]{Foster+11}
or planetary nebulae \citep{Douglas+07,Romanowsky2009,Coccato+09,Napolitano+11}
become essential, even though these populations are often scarce.
See \citet{Gerhard13} for a recent review.

Globular clusters (GCs) have been used extensively to probe the mass
distribution of the outer halos of ETGs \citep[see e.g.][and references
therein]{Cote+03,Hwang+08,Lee+10}. There are three drawbacks to adopting
GCs as dynamical tracers. 1) As most GCs are red and very old, they have
orbited many times around their host galaxy, and the most adventurous ones
with the smallest pericentres will have been progressively tidally stripped
by the potential of the host galaxy. Therefore, for a given apocentre, the
GCs with the largest pericentres will have survived, leading to a bias towards
more circular orbits (in comparison with the underlying stellar population).
2) Their dynamics is thought to originate from rather violent physical
processes (early collapse, gas-rich mergers) and may therefore not be
strictly linked with the orbital structure of the old stellar population
\citep{Bournaud+08}.
3) A bimodality in the colour distribution of GCs in bright ETGs is often
observed \citep{Brodie+06}, which may then call for several decoupled
dynamical components in the final modelling.  As for any tracer embedded
in the outermost regions of a galaxy potential, it is sometimes difficult
to assess the steady-state and dynamically relaxed nature of a certain
tracer, and address whether or not the observations still probe the galaxy
potential or lie beyond the boundary with the intra-cluster potential
\citep{Doherty+09}.

One of the most thorough studies of a GC system by \citet{Schuberth+10}
includes nearly 700 GCs for the central Fornax cluster massive early-type
galaxy NGC\,1399.  Using a $\beta=\rm Cst$ Jeans analysis, these authors
showed that the red GC population traces very well the field stellar
population, while the blue one appears to be the superposition of several
sub-populations including accreted or true cluster members.
A similar study of the same galaxy with (4 times fewer) PNe by
\citet{McNeil+10} illustrates the relative merits of using GCs
and PNe as tracers of the gravitational potential.

It is fortunate that planetary nebulae (PNe) do not suffer from the three
drawbacks affecting GCs.  PNe are generally thought to represent the
distribution and dynamics of the galactic stellar halos with high
fidelity \citep[see, however, ][]{Mendez+01,Sambhus+06}.
Moreover, they are easy to observe, especially thanks
to their very strong \oiii\ emission line at 5007\,\AA.
Henceforth, several dynamical studies have targeted PNe around
bright ETGs using multi-slit or slit-less spectroscopy,
or with the dedicated Planetary Nebulae Spectrograph \citep{Douglas+02}.

First analyses \citep{Mendez+01,Romanowsky+03,Douglas+07} based on Jeans
analysis and Schwarzschild modeling suggested that the host galaxies studied
lacked DM, and were consistent with no DM \citep{Romanowsky+03}.  However,
recent results (e.g. \citealp{Das+11}) strongly confirm and quantify the
discrepancy between the observed dynamics and that expected from the sole
stellar light in giant ETGs such as NGC\,4649.  The extracted PNe luminosity
distribution has also served to improve the distance estimated to that
galaxy \citep{Teodorescu+11}.  Although PN-based kinematic modeling is
usually limited by the number of tracers (typically 100 to 200), the PN.S
team has performed an ambitious observational program observing PNe in the
outer regions of a dozen ETGs, with detailed results on a number of
prototypical systems such as NGC\,4374 in the Virgo cluster, reaching
out to $\simeq 5\,R_{\rm e}$ \citep{Napolitano+11}.  This kinematic modeling
usually assumes spherical symmetry, but there now exist several studies using
axisymmetric models: e.g.
NGC 4494 \citep{Napolitano+09}, 
NGC 3379 \citep{deLorenzi+09},
NGC 4697 \citep{Das+11}.

The main limitation of such studies is the often assumed hypothesis
of spherical symmetry for the mass distribution, but these results
can still serve as strong guidelines to constrain the presence of
dark matter in the outer halos of ETGs.

\subsubsection{Other Tracers and Combined Approaches}

At such large radii reached by the GC populations, as in NGC\,1399,
many studies take advantage of the presence of large X-ray
halos around specific ETGs to constrain the corresponding radial mass,
and make direct comparisons \citep{Humphrey+06,Schuberth+10,Das+11,Gerhard13}.
A number of galaxies have been surveyed, mostly massive ETGs as they are
more often embedded within such X-ray halos \citep{Fukazawa+06, Nagino+09}.
The assumed hypothesis of hydrostatic equilibrium for the hot gas may
sometimes hamper the robustness of such conclusions, but these effects
are generally thought to be small. This is convincingly confirmed by
comparing several concomitant tracers
\citep{Churazov+08,Humphrey2010,Humphrey+11} though some discrepancies
have been reported, sometimes suggesting a transition from the galaxy
halo to the cluster intergalactic medium \citet{Schuberth+10}, or
sometimes not yielding firm conclusions on their origins
\citep{Romanowsky2009}.

As also mentioned above, even ETGs have sufficiently abundant (and
well-behaved) gas components that can be used to constrain mass profiles
out to large radii \citep{Franx+94,weijmans08}.

Orbits of individual satellites may further help constraining the potential
around a galaxy \citep{Geehan+06, vanDerMarel+08}.  \citet{Prada+03} and
\citet{Klypin2009} used the SDSS to stack the PPS built from the satellites
of thousands of otherwise fairly isolated galaxies and found it to be
consistent with the predictions of $\Lambda$CDM simulations.
\citet{Conroy+07_dyn} and \citet{More+11} analysed galaxy satellites
based on SDSS data, and derived the variation of virial mass with host
galaxy luminosity, separating red and blue galaxies. They found both
that red host galaxies of given blue luminosity have typically double
the halo mass as their blue counterparts of same stellar mass.
These two studies make assumptions that thwart the derivation of
useful constraints on the anisotropy. Using the $\Lambda$CDM halo
DF model \citep{WLMG09} on a larger SDSS sample, \citet{WM13} are able to
obtain more reliable relations between halo and stellar mass or luminosity,
confirming to first order the results of \citeauthor{Conroy+07_dyn} and
\citeauthor{More+11}. The stellar fractions within the virial radius of red
galaxies with $\log M_{\rm stars} > 10$ exceed 1\%, peaking at 2\% for
$\log M_{\rm stars} \approx 11.9$ and decreasing again to 1\% at larger
masses. However \citeauthor{WM13} also find that red galaxies have more
concentrated halos than blue galaxies of same stellar or halo mass, and
that the orbits of satellites are marginally radial in the central regions,
and increasingly radial with distance to the center.
The reader may consult sections \se{weak} and \se{strong} on lensing
for further details about the galaxy mass profiles at large radii.

As hinted above, a promising path toward robust mass profiles comes
from the simultaneous use of all available tracers, hoping for a
consistent picture to emerge.  Many studies have been conducted
towards this end, mostly targeting ETGs \citep{Foster+11} and more
specifically very massive ones \citep{Woodley+10,Das+11,Arnold+11,Murphy+11}.
In agreement with results stated above, the overall impression from
these studies calls for ETGs as being dominated by baryons within
$1\,R_{\rm e}$, with DM representing about half of the mass within
$2-4\,R_{\rm e}$ and dominating at larger radii.

\subsubsection{The Mass-Anisotropy Degeneracy}

\begin{figure}
\centering
\includegraphics[width=0.8\textwidth]{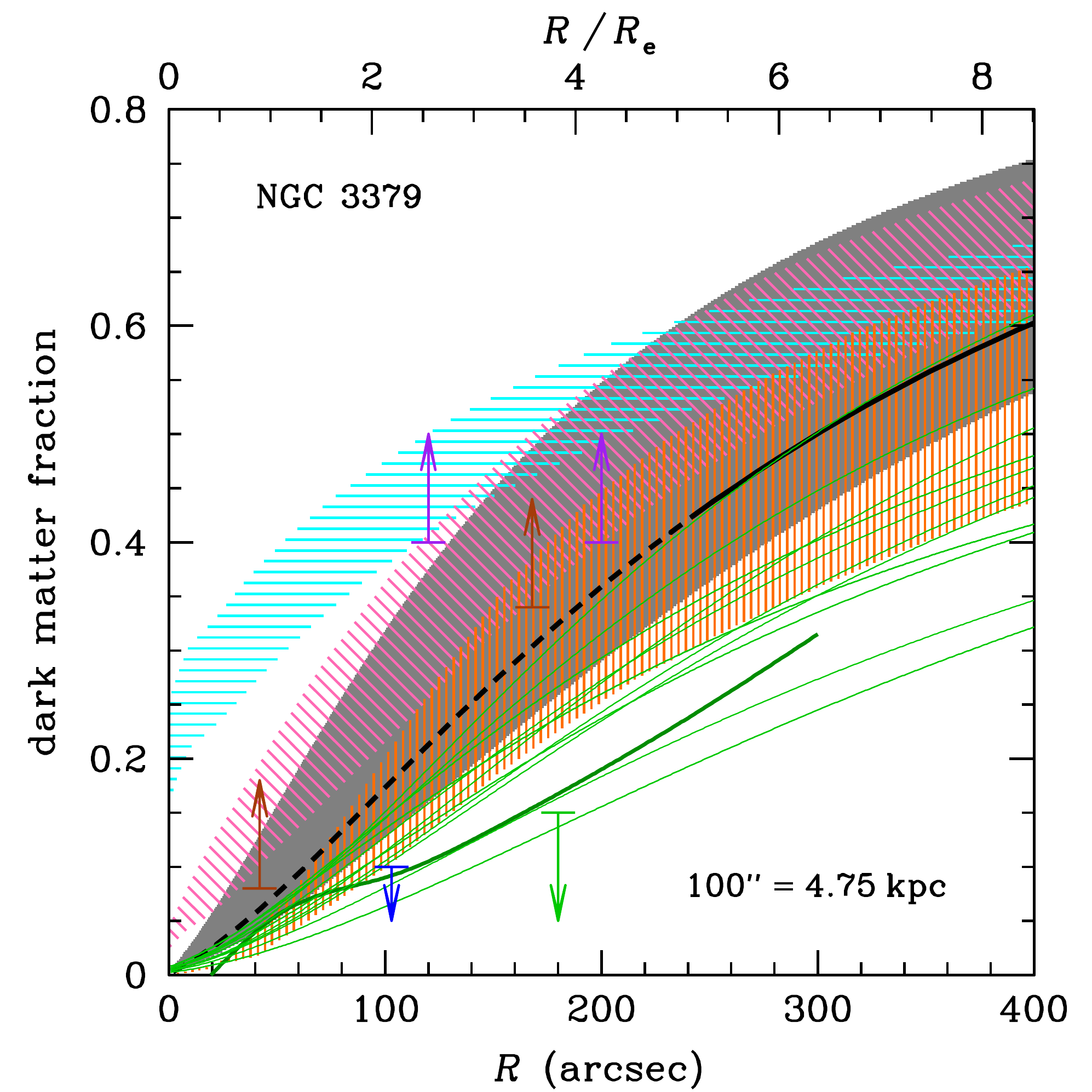}
\caption{Dark matter fraction versus physical radius in NGC~3379
(assuming $R_{\rm e}$=$47\arcsec$ and S\'ersic index $n$=4.74,
following \citealp{Douglas+07}). The \emph{light green upper limit}
and \emph{light green curves} respectively show the Jeans (PNe) and
orbit-modelling solutions of \citet{Romanowsky+03} (stars+PNe), while
the \emph{blue upper limit} is the DF-modelling (stars)
of \citet{KSGB00}. The \emph{medium-thickness dark green curve}
shows the spherical Jeans solution (stars+PNe) with double the number
of PNe \citep{Douglas+07}.  The \emph{orange vertically-shaded region}
gives the limits of NMAGIC orbit-modelling (stars+PNe) \citep{deLorenzi+09}.
The \emph{lower limits} shows the orbit-modelling (stars) by
\citet{Weijmans+09} (\emph{brown}) and the isotropic Jeans analysis
(globular clusters) of \citet{Pierce+06} (\emph{purple}).  The
\emph{cyan horizontal-} and \emph{magenta oblique-shaded regions}
give the predictions (see \citealp{Dekel+05}) from equal-mass
merger SPH+cooling+feedback simulations \citep{CPJS04}, of
respectively gas-poor and gas-rich spirals embedded in dark matter halos.
The \emph{black curve} is from the satellite kinematics study of \citet{WM13},
adopting the mean of their 3rd and 4th stellar mass bins for red hosts
(the stellar mass of NGC~3379 is in between), \emph{dashes} for the
extrapolation within the smallest satellite radii analyzed, while
the \emph{grey shaded region} represents the $1\,\sigma$ confidence
from the Monte Carlo Markov Chain analysis of \citeauthor{WM13}. }
\label{fig:dmfracs3379}
\end{figure}

The results presented in this section appear robust in a statistical sense.
However, on an individual basis, measuring the mass profile in gas-poor
galaxies is intrinsically difficult due to the degeneracy in the stellar
dynamical models.  The MAD is best broken by the joint use of several
tracers, especially if they probe the same region of the potential.
The use of tracers with very different orbital anisotropies can also
be very useful to lift the MAD.

As an illustration of this MAD problem, \Fig{dmfracs3379} shows the wide
variety of solutions for the nearby (apparently) roundish ETG, NGC 3379,
which had been the test-bed for the putative suggestion that ETGs are
devoid of DM \citep{Romanowsky+03,Douglas+07}. One notices highly discrepant
conclusions from various modeling attempts, in particular at the outer limit
of spectroscopic observations ($\sim 200''$ or $\simeq 4\,R_{\rm e}$).
The MAD is clearly present as the higher DM fractions indicate fairly
radial orbits in the outer regions, while the lower ones come with
isotropic orbits \citep[e.g.][]{deLorenzi+09}: this emphasizes the fact that
specific tracers may constrain the mass distribution with uncertainties of
different amplitude and nature.  Moreover, there is a wide range
of theoretical predictions.
Also, some of the orbit solutions of \citet{Romanowsky+03} indicate `normal'
levels of DM at large radii \citep{ML05b}. In fact, \emph{all} recent
observational modelling of NGC~3379 lead to an increased fraction of
DM at increasing large radii.  If NGC~3379, which has quasi-circular
isophotes, were a nearly face-on S0 \citep{CVHL91}, as suggested by
its classification as a fast rotator \citep{emsellem07},
one would expect lower DM fractions \citep{MaBa01}.

\subsubsection{Discrete Star Velocities for Dwarf Spheroidal Galaxies}
\label{sec:resultsdSph}
\begin{figure}
\centering
\includegraphics[width=0.8\textwidth]{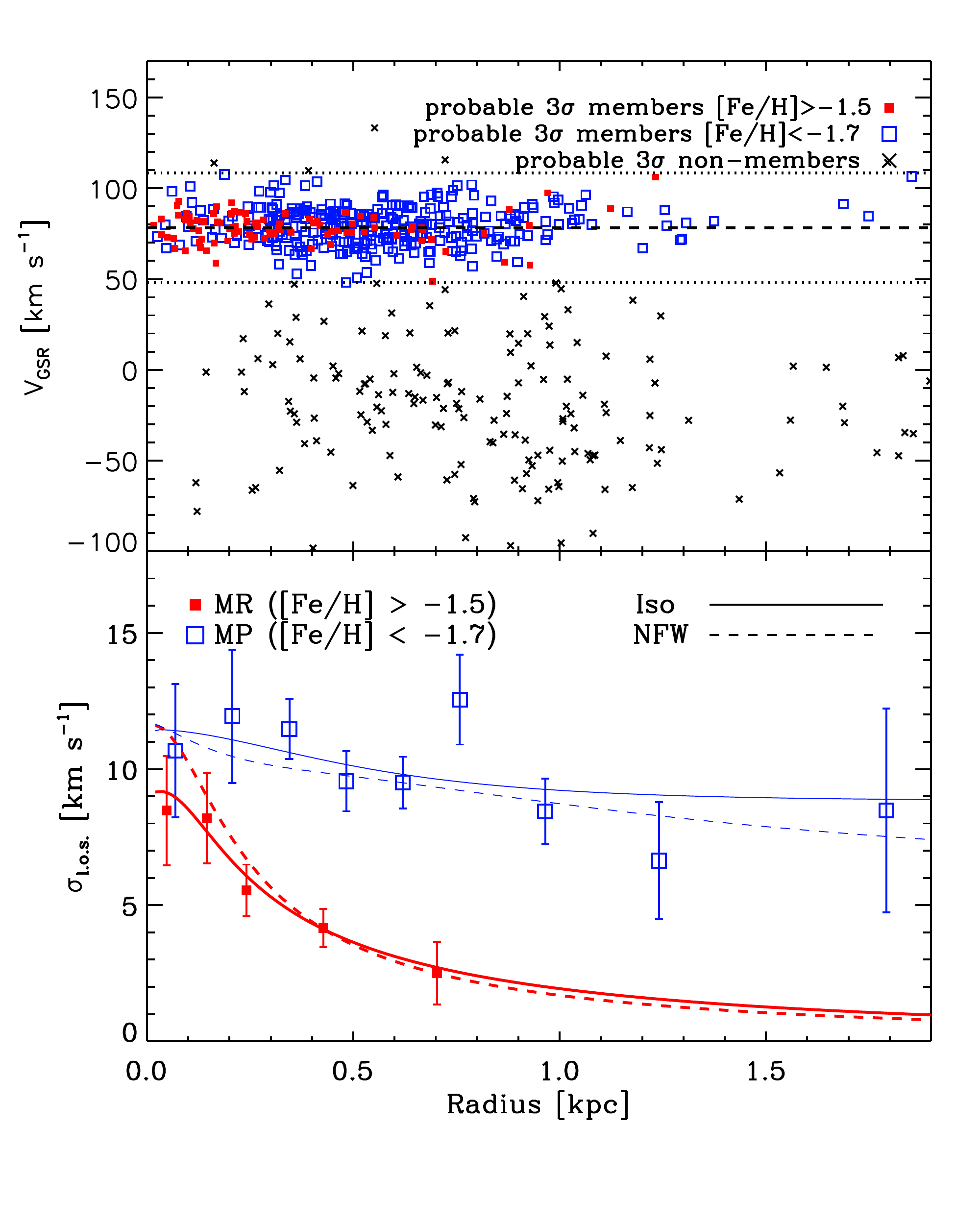}
\caption{VLT/FLAMES velocity measurements for individual stars along
  the line-of-sight to the Sculptor dwarf spheroidal galaxy (dSph).
  \emph{Top panel}:
  Line-of-sight velocities in the Galactic standard of rest versus projected
  radius for probable members to the Sculptor dSph (the \emph{filled} and
  \emph{open squares} show probable members with metallicity [Fe/H]~$ > -1.5$
  and $< -1.7$, respectively) and for probable non-members (\emph{crosses}).
  The region of probable membership is indicated by the two \emph{horizontal
  dotted lines}, while the \emph{dashed line} indicates the systemic velocity
  of the galaxy. 
\emph{Bottom panel}:
  Line-of-sight velocity dispersion profiles for the ``metal-rich"
  ([Fe/H]~$ > -1.5$) and ``metal-poor" stars ([Fe/H]~$< -1.7$), as
  shown by the \emph{filled} and \emph{open squares with error-bars},
  respectively. The \emph{solid} and \emph{dashed lines} show
  the l.o.s. velocity dispersion profiles for the best-fitting
  pseudo-isothermal (cored) and NFW (cusped) dark matter models.
  Figure adapted from results presented in \citet{Battaglia+08}.}
\label{fig:battaglia}
\end{figure}

Relative to giant ETGs, dSphs constitute the other mass extreme.  The
study of such low luminosity objects and very faint galaxies relies mostly
on very nearby (Local Group) galaxies, and largely in the context of
resolved stellar populations \citep[e.g.][]{Gilmore+07}.
Mass modelling is feasible thanks to ambitious observational programs
to measure the stellar kinematics of hundreds and sometimes several
thousands of individual stars\footnote{Similar methods apply to the
study of Milky Way stars; see \se{milkyway}.}
\citep{Tolstoy+04,LMP05,SimonGeha2007,WMO09,Geha+10,Battaglia+2011,Simon+11}.
We refer the reader to the detailed review by \citet{BHB13}.

An example of this method is shown in \Fig{battaglia} for the Sculptor
dwarf spheroidal galaxy, studied in detail by \citet{Battaglia+08}, who
fit the PPS assuming Gaussian LOS velocity distributions (adding
a component for contamination by our Milky Way). By disentangling
the metal-poor and metal-rich stellar sub-populations, the authors
were able to show that, if physically meaningful, these two sub-systems
were compatible with the same potential (best fitted by an isothermal
DM profile) but with different anisotropy, providing some clues about
their origin, as the metal-rich sub-population appears to show a faster
transition to radial orbits than the metal-poor one.
The resulting dynamical mass-to-light ratio $M/L$ reached values
in excess of 150 inside $\sim$2 kpc, demonstrating the dominance
of DM at all radii in such low surface brightness objects.
\citet{WP11} use a similar 2-population analysis to constrain the slopes
of the mass profiles of the Fornax and Sculptor dSph galaxies, ruling out
cusps as steep as $-1$ (NFW) for both, and favoring inner slopes of
$-0.4\pm0.4$ (Fornax) and $-0.5\pm0.5$ (Sculptor).

The DM core in Sculptor has been recently confirmed by \citet{RF13b} using
their new dispersion-kurtosis analysis with general anisotropy \citep{RF13a}.
\citet{AE12c} noted that Fornax has three distinct stellar populations (with
different metallicities),
and with this constraint, \citet{AAE13} have shown that Fornax must indeed
have a core of $1^{+0.8}_{-0.4}\,\rm kpc$ or else an NFW model with an
unlikely very large scale radius.
However, orbit modelling allows a cusp for Sculptor \citep{Breddels+12} and
Fornax \citep{JG12}. Indeed, using orbit modelling, \citet{BH13} conclude
that while Fornax, Sculptor, Carina and Sextans can each accommodate
either a cusp or a core, cores are unlikely when these 4 dSph galaxies
are considered together.  The debate between halo cusps and cores is
thus still ongoing, largely because studies often group together
different stellar populations that share different kinematics and
neglect small but non-negligible rotation, and because non-spherical
modelling increases the space of acceptable solutions.

\citet{Strigari+10} used an isotropic analysis (with \Eq{giso} and
Eddington's formula, and extracted the dispersion and kurtosis profiles
from the former) to show that the classical dSph galaxies have LOS velocity
dispersion, kurtosis and even distributions that are consistent with their
surface density and with subhalos taken from the Aquarius
$\Lambda$CDM simulation of the Milky Way \citep{Springel+08}, 
with dynamical masses between 2 and $15 \times 10^8 M_\odot$.
More generally, it is believed \citep[see][, and references therein]{Mateo98}
that most dSphs have very high virial-theorem $M/L$s. Recently,
\citet{Walker+09} and \citet{Wolf+10} have found that dSph $M/L$s within
the half-light radius increase towards lower masses down to the lowest
mass \emph{ultra-faint dwarfs} (UFDs).
Although extrapolating these systems to their virial radii may be ill-advised,
the data are consistent with all dSphs (including UFDs) having virial masses
above $10^8 M_\odot$ (\citeauthor{Walker+09,Wolf+10}).

However, the detailed modelling of dSphs is challenging because of Milky Way
contamination \citep[e.g.][]{LMP05} and because their likely tidal tails
are expected to lie very close to the LOS \citep{Klimentowski+09b},
which could then lead to an overestimate of their mass \citep{Klimentowski+07}.

Furthermore, when the stellar velocity dispersion reaches extremely low values,
additional ingredients such as the contribution of binary systems must be taken
into account for proper dynamical modelling in particular for ultra-faint dwarf
galaxies \citep[see e.g. ][and references therein]{Martinez+11}. $N$-body
models may be required for an accurate dynamical modelling of these objects.

\newpage
\subsection{Future Prospects}

We have reviewed the basic methods to determine the distribution of total
mass in gas-poor galaxies, whilst addressing a number of intrinsic
degeneracies that may affect current determinations.  We see two main
directions for future applications of the discussed techniques:

One the one hand, for Local Group galaxies, the dynamical degeneracies can
be alleviated by increasing the dimension of the observable space, namely
by observing proper motions together with radial velocities of individual
stars.  At the present, this can be done for nearby star clusters by
including plane-of-sky velocities from stellar proper motions in the
dynamical models \citep[e.g.][]{vandeVen2006,vandenBosch2006,vanderMarel2010}.

The global space astrometry satellite GAIA \citep{Perryman01}
will provide proper motions with unprecedented accuracy. Unfortunately,
the classical dSph galaxies are so distant that the error on proper motions
from GAIA will be of the order of their internal velocity dispersions
\citealp[e.g.][]{BHB13}, so the gain from proper motions with GAIA may
only be significant for the closest dSph galaxies.  However, the future
generation of 30--40m telescopes
should roughly double the GAIA precision on proper motions (with a 5-year
baseline, \citealp{DG10}), and lead to much more accurate mass and orbital
modeling (as first suggested by \citealp{LeonardMerritt89}).

These data will require and exploit the full generality and sophistication
of the models. However, it is likely that such a wealth of data will also
reveal new degeneracies associated with the sub-populations of stars in
galaxies, themselves reflecting their complex formation and evolution history.

Meanwhile, if the increase in computing power grows at the current rate,
one should be able to increasingly resort to $N$-body modelling (or
associated techniques) to determine the distribution of mass in ETGs
and dSphs that are not in perfect dynamical equilibrium and possibly
address such models in some restricted cosmological context.

On the other hand, the same simpler techniques that are being applied today
to relatively small samples of nearby galaxies will be used to study much
larger samples of galaxies with two-dimensional stellar (and gaseous)
kinematics and at increasingly larger redshift. The current state of the
art is defined by the ATLAS$^{\rm 3D}$ \citep{Cappellari+11} and CALIFA
surveys \citep{CALIFA}, which have mapped a few hundred galaxies via
integral-field spectroscopy. Ongoing surveys, such as the SAMI (PI: Scott
Croom) and MaNGA (PI: Kevin Bundy), will extend the sample size by about
two orders of magnitude, using multi-object two-dimensional spectrographs
on dedicated telescopes. Accurate masses, which themselves rely on accurate
distances, will still be a critical ingredient to study galaxy formation
from these larger samples. Finally, the next frontier will involve
constructing dynamical models of galaxies at significant redshift,
to trace the assembly of galaxy masses over time. This will also make
use of multi-object spectrographs, optimized for near-infrared wavelengths,
to effectively reduce the exposure times by orders of magnitude, mounted
on future generation very large telescopes. Within the next ten years
(i.e.\ $\sim 2024$), we may be able to approach the quality of the stellar
kinematics of galaxies obtainable today in the Virgo cluster, up to
the key redshift $z\sim2$, when the Universe was just a quarter of
its current age and much of the galaxy mass was being assembled.


\newpage

\section{Weak Lensing by Galaxies}\label{sec:weak}


\subsection{Introduction}

Most methods to constrain or measure the masses of galaxies are
limited to relatively small radii, where baryons are dominant.
This is, because these methods probe the gravitational potential
through the dynamics of visible tracers. Although it is safe to assume
that galaxies are virialized, uncertainties in the mass estimates
remain, for instance due to anisotropies in the velocity
distributions. Furthermore, these baryon dominated regions are not yet
fully understood, which complicates a direct comparison of the models
of galaxy formation to observational data. We note, however, that
predictions from cosmological numerical simulations keep improving.
Nonetheless, it would be advantageous to have observational
constraints that can be robustly measured from numerical simulations.
The virial mass of the galaxy is an obvious choice, but it is difficult
to measure using dynamical methods. To date, only satellite galaxies
have provided some information using data from large redshift surveys
such as the SDSS \citep[e.g.][]{McKay02,Prada+03}.

In this section we focus on a direct probe of the matter distribution
in the universe, which provides us with a unique opportunity to probe
the outer regions of galaxies. It makes use of the fact that
inhomogeneities in the matter distribution, such as the halos around
galaxies, perturb the paths of photons emitted by distant sources: it
is as if we are viewing these sources with a spatially varying index
of refraction. As a result, the images of the distant galaxies
typically appear slightly distorted (and magnified), an effect that is
known as weak gravitational lensing. The amplitude of the distortion
provides us with a direct measurement of the gravitational tidal
field, which in turn can be used to ``image'' the distribution of dark
matter {\it directly} \citep[e.g.][]{Kaiser93}.
If the distortion is large enough, multiple images of the source can
be observed. This strong lensing provides accurate constraints on the
mass distribution on small scales and its applications are discussed
in \se{strong}. 

The applications of weak lensing are not limited to galaxy-galaxy
lensing, which is the study of the properties of galaxy dark matter
halos. In fact the first detections were made by searching for the
lensing-induced alignments of galaxies behind massive clusters of
galaxies where the lensing signal is larger \citep{Tyson90,Fahlman94}.
In recent years the focus has shifted to the
measurement of the statistical properties of the large-scale
structure: this cosmic shear is a promising probe of dark energy and
has been detected with high significance \citep{Fu08}. This
application is driving much of the developments in improving
measurement techniques, but also in terms of survey
requirements. Consequently, galaxy-galaxy lensing studies benefit as
well, because the data requirements are rather similar: we need to
survey large areas of the sky, preferably in multiple bands in order
to derive photometric redshifts. Such data sets are becoming
available, and significant progress is expected in the coming years as
the analyses of the first multi-color cosmic shear surveys are
completed.

The first attempt to measure the weak lensing signal around
galaxies\footnote{In fact, it was the first attempt ever to measure a
  weak lensing signal.} was made by \cite{Tyson84} using data from
photographic plates with fairly poor image quality. As discussed in
more detail below, the determination of the lensing signal requires
careful measurements of the shapes of faint galaxies which benefit
greatly from good image quality. Consequently, the first detection was
reported over a decade later by \cite{Brainerd96} using deep
ground-based CCD images. Soon after \cite{Hudson98} exploited the
combination of deep HST imaging and photometric redshifts in the
Northern Hubble Deep Field.

An accurate determination of the galaxy-galaxy lensing does not only
require good image quality. As explained below, the signal around an
individual galaxy is too low to be detected. Instead we stack the
signals for a large ensemble of lenses to improve the signal-to-noise
ratio of the measurement. The early studies were all based on small
survey areas, thus yielding small numbers of lens-source pairs.  This
changed with the start of the Sloan Digital Sky Survey (SDSS):
\cite{Fischer00} used only 225 deg$^2$ of commissioning data and
detected a significant galaxy-galaxy lensing signal out to 1~Mpc. The
SDSS data are relatively shallow, but the large survey area provides
the large number of lens-source pairs to measure the lensing signal
with high accuracy. Another important feature of the SDSS is the
availability of redshifts for the lenses (spectroscopic as well as
photometric), which has been used by \cite{McKay01}, \cite{Guzik02} and
\cite{Mandelbaum06}.

In the case of deep observations the reduction in the number of lenses
(due to smaller survey area) is compensated by the increase in the
number of sources, which are also more distant. Hence, even by imaging
tens of square degrees the galaxy-galaxy lensing signal can be
measured accurately \citep{Hoekstra04,Parker07}. Such surveys
typically lack spectroscopic redshift information for the lenses and
use photometric redshifts instead \citep[but see][for an example that
combines SDSS spectroscopy with deeper imaging]{vanUitert11}.

Below we provide a brief introduction to weak galaxy-galaxy lensing
and present a number of highlights, demonstrating the potential of
this technique. However, it is important to stress that this is a field
that is still developing, and many exciting results are expected from
the next generations of surveys.

\subsection{Theory of Weak Lensing}

Due to space limitations, we can only provide the most basic discussion of
weak gravitational lensing. We refer the interested reader to a recent
review by \cite{HJ08} or the thorough introductions by \cite{Bartelmann01}
or \cite{Schneider06}.

Inhomogeneities along the line-of-sight deflect photons originating
from distant galaxies. As these sources are typically small, the
resulting effect is a remapping of $f^{\rm s}$, the surface brightness
distribution of the source:

\begin{equation}
f^{\rm obs}(x_i)=f^{\rm s}({\cal{A}}_{ij}x_{j}),
\end{equation}

\noindent where ${\bf x}$ is the position on the sky and $\cal{A}$ is
the distortion matrix (i.e., the Jacobian of the transformation),
which is specified by the projected surface density of the lens
and the redshifts of the lens and the source.
It is convenient to introduce the deflection potential $\Psi$

\begin{equation}
\Psi({\bf x})=\frac{1}{\pi}\int d^2{\bf x'} \kappa({\bf x'})\ln|{\bf x-x'}|,
\end{equation}

\noindent where the convergence $\kappa$ is the ratio of the projected
surface density $\Sigma({\bf x})$ and the critical surface density
$\Sigma_{\rm crit}$:

\begin{equation}
\kappa({\bf x})=\frac{\Sigma({\bf x})}{\Sigma_{\rm crit}},
\end{equation}

\noindent with $\Sigma_{\rm crit}$ defined as

\begin{equation}
\Sigma_{\rm crit}=\frac{c^2}{4\pi G}\frac{D_{\rm s}}{D_{\rm l} D_{\rm ls}}.
\end{equation}

\noindent Here $D_{\rm s}$, $D_{\rm l}$, and $D_{\rm ls}$ correspond
to the angular diameter distances between the observer and the source,
observer and the lens, and the lens and the source. Hence, the lensing
signal depends on both the redshifts of the lenses and the sources.
Note that (in particular) the sources are too faint to determine
spectroscopic redshifts, and photometric redshifts are used instead.
Early galaxy-galaxy lensing studies lacked redshift information for
both lenses and sources and average redshift distributions were used
to infer masses \citep[e.g.][]{Hoekstra04,Parker07}).  Redshift
information is particularly useful for the lenses, as it allows one to
study the lensing signal as a function of baryonic content and
environment. Often photometric redshifts are also used for the lenses
\citep{Hoekstra05,Kleinheinrich06}, with the notable exception
of the SDSS \citep{Mandelbaum06,vanUitert11}.

Redshifts for individual sources are not critical, provided their
redshift distribution is known. However, if such information is
lacking, faint satellite galaxies associated with the lens will dilute
the lensing signal, if left unaccounted for. Furthermore, if these
galaxies align their major axis towards the host galaxies, they will
bias the signal \citep{2006ApJ...644L..25A}.

The distortion matrix ${\cal A}$ can be written in terms of the second
derivatives of the deflection potential $\Psi$

\begin{equation}
{\cal A}=\delta_{ij}-\frac{\partial^2 \Psi}{{
\partial \theta_i\partial\theta_j}}=
\left(
    \begin{array}{cc}
        1-\kappa-\gamma_1 & -\gamma_2 \\
        -\gamma_2        & 1-\kappa+\gamma_1 \\
    \end{array}
\right),
\label{distort}
\end{equation}

\noindent where we used that $\kappa=\frac{1}{2}\nabla^2\Psi$ and introduced
the complex shear $\bm{\gamma}\equiv\gamma_1+i\gamma_2$, which are related
to the deflection potential through

\begin{equation}
\gamma_{1}=\frac{1}{2}(\Psi_{,11}-\Psi_{,22})\hspace{1em}{\rm and}\hspace{1em}
\gamma_2=\Psi_{,12},
\label{sheardef}
\end{equation}

If $\kappa$ and $\gamma\ll1$ (i.e., the weak lensing regime), the
effect of the remapping by ${\cal A}$ is to transform a circular source
into an ellipse, with axis ratio $\sim (1-|\gamma|)/(1+|\gamma|)$ and
position angle $\alpha=\arctan(\gamma_2/\gamma_1)/2$. In addition,
the source is magnified by a factor 

\begin{equation}
\mu=\frac{1}{\det \cal{A}}=\frac{1}{(1-\kappa)^2-|\gamma|^2},
\end{equation}

\noindent boosting the flux by the same amount. To first order, the
magnification depends on the convergence only. Both the shearing and
magnification of sources are observable effects, although both are
quite different in terms of techniques and systematics.

\subsection{Shear}

To study the dark matter distribution in the universe, the measurement
of the shearing of background galaxies is most commonly used, because
of the better signal-to-noise that can be achieved per lens-source
pair when compared to the effect of magnification. It involves the
measurement of the shapes of the faint background galaxies. Under the
assumption that galaxies are randomly oriented in the absence of
lensing, the strength of the tidal gravitational field can be inferred
from the measured ellipticities of an ensemble of sources.

If we consider an isolated lens, the effect of weak lensing is a
systematic (purely) tangential alignment of the images of the
background galaxies with respect to the lens galaxy. The average
tangential distortion, defined as
\begin{equation}
\gamma_T=-(\gamma_1\cos 2\phi + \gamma_2 \sin 2\phi),
\end{equation}
\noindent can then be used to quantify the lensing signal. Here $\phi$
is the azimuthal angle with respect to the lensing galaxy. For any
mass distribution the azimuthally averaged tangential shear can 
be interpreted as a mass contrast:

\begin{equation}
\langle\gamma_T\rangle(r)=\bar\kappa(<r)-\bar\kappa(r).
\end{equation}

A simple model to compare to the data is the singular isothermal
sphere (SIS) with $\rho(r)=\sigma^2/(2\pi G r^2)$, where $\sigma$ is
the line-of-sight velocity dispersion. For this mass distribution
we obtain

\begin{equation}
\gamma_T(r)=\kappa(r)=\frac{r_E}{2r},
\end{equation}

\noindent where $r_E$ is the Einstein radius, which can be expressed
in terms of $\sigma$ and $\beta=\langle D_{ls}/D_s\rangle$:

\begin{equation}
r_E=\beta\left(\frac{\sigma}{186{\rm km/s}}\right)^2~{\rm arcsec}.
\end{equation}

\noindent If we fit the model to the data from $R_0$ to $R$, the
corresponding error is

\begin{equation}
\sigma_{r_E}=\sqrt{\frac{1}{\pi\bar{n}\ln(R/R_0)}}\sigma_{\rm gal},
\end{equation}

\noindent where $\bar{n}$ is the number density of sources and
$\sigma_{\rm gal}$ their intrinsic ellipticity ($\sigma_{\rm gal}\sim
0.3$). For deep ground based observations $\bar n\sim 10-20$
arcmin$^{-2}$. As discussed below, we should only consider the signal
at $R<120''$ and $R>5''$ (because the lens light should not interfere
with our shape measurement). For a galaxy with a velocity dispersion
of 150 km/s, we obtain a typical signal-to-noise ratio of
$r_E/\sigma_{r_E}\sim 0.39/1.4\sim 0.26$ (adopting a typical value of
$\beta=0.6$). Even with the much higher source density in HST
observations the best we can achieve is $S/N\sim 1$.  Hence the signal
of an individual lens galaxy is by far too small to be detected.  Instead
we have to average the signals for an ensemble of lenses to improve
the signal-to-noise ratio.

Furthermore, the induced lensing signal is tiny, much smaller than the
typical observational distortions that affect the observed shapes of
the galaxies. The most relevant ones are the circularisation by the
PSF (seeing) and PSF anisotropy. The former lowers the signal (if
uncorrected for) and the latter can mimic a lensing signal. Much
effort has been spent on understanding and correcting these sources of
systematics. A major driver has been the study of lensing by
large-scale structure, a.k.a. cosmic shear, which is an important way
to study dark energy (see \cite{HJ08} for a review) and extremely
sensitive to residual systematics. In galaxy-galaxy lensing, however,
one averages the signal perpendicular to lines connecting many
lens-source pairs, which are randomly oriented with respect to the
direction of PSF anisotropy. As a result any residual systematics are
suppressed. The measurement of halo shapes is somewhat more sensitive,
but as shown in \cite{Hoekstra04} current results are not affected and
it is possible to reduce the impact further, albeit at the expense of
increasing the noise \citep{Mandelbaum06shape,vanUitert12}.

\subsection{Magnification}

The measurement of the magnification provides a complementary way to
study the mass distribution. The actual magnification cannot be
measured because the intrinsic fluxes of the sources are unknown.
Instead, the signal can be inferred from the change in the source
number counts. Such a change arises from the balance between two
competing effects. On the one hand the actual volume that is surveyed
is reduced, because the solid angle behind the cluster is enlarged.
However, the fluxes of the sources in this smaller volume are boosted,
thus increasing the limiting magnitude. As a consequence, the net
change in source surface density depends not only on the mass of the
lens, but also on the steepness of the intrinsic luminosity function
of the sources. If it is steep, the increase in limiting magnitude
wins over the reduction in solid angle, and an excess of sources is
observed. If the number counts are shallow a reduction in the source
number density is observed.

The uncertainty in the measurement is determined by variations in the
number density (i.e., a combination of Poisson noise and the
clustering of the sources). A correct interpretation of the results
only requires accurate photometry and knowledge of the (unlensed)
luminosity function. Therefore the requirements on the PSF are much
less stringent compared to the shear-based approach.

The magnification has been measured for quasars in the SDSS
(\cite{Scranton05,Menard08}) and Lyman-break galaxies in the CFHT
Legacy Survey by \cite{Hildebrandt09}. The latter study is of
particular interest, because these drop-out galaxies are readily
identified in deep wide-field imaging surveys. Furthermore, as the
sources are all at high redshift, this approach provides a unique way
to study the masses of high redshift ($z\sim 1$) galaxies using
ground-based data.

\begin{figure}[t]
\centering
\includegraphics[width=0.6\textwidth]{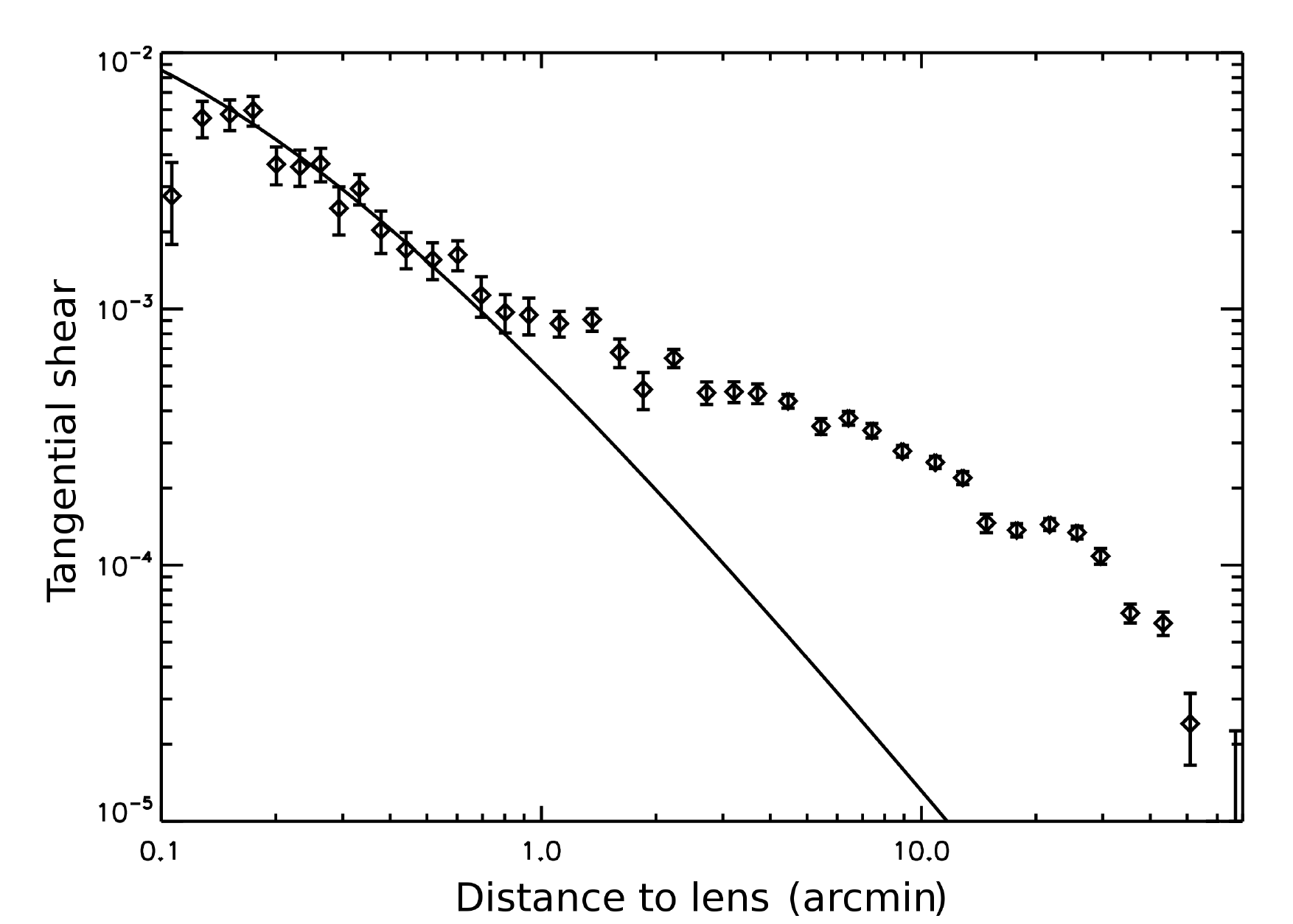}
\caption{Plot of the ensemble averaged tangential shear
    as a function of distance from the lens using 334 deg$^2$ of $r'$
    data from the RCS2 \citep{vanUitert11}. The lenses are selected
    to have apparent magnitudes $19.5<m_r<21.5$ and the sources
    $22<m_r<24$. For reference the best fit NFW profile is also drawn,
    which shows that on scales beyond 1\arcmin ($\sim 300$ kpc) the clustering
    of the lenses becomes important.}
\label{fig:gtprof}
\end{figure}

\subsection{Galaxy-Mass Cross-Correlation Function}

If galaxies are well separated, or randomly distributed, the observed
lensing signal can be directly related to the ensemble averaged dark
matter distribution. In the real universe, however, galaxies are
clustered. This complicates such a simple interpretation of the data.
On sufficiently small scales the lensing signal is dominated by
individual halos, but on larger scales we measure the combined signals
from many halos.  An example is shown in \Fig{gtprof}, which shows
results from the analysis of 334 deg$^2$ of data from the RCS2
(\cite{vanUitert11}; also see e.g., \cite{Fischer00, Hoekstra04,
  Sheldon04}). A significant signal is measured out to 30' from the
lenses, which corresponds to $\sim 9$~Mpc. \Fig{gtprof} also shows the
best fit NFW profile (to data between 0.2 and 0.6 arcmin), which drops
below the observations for scales larger than 1' ($\sim 300$
kpc). Hence on these larger scales the clustering of the lenses is
important to interpret the data.  \cite{vanUitert11} also compared
the data to a SIS model and found that it also fits the data well
out to $\sim 300$kpc, indicating that it is typically difficult to
distinguish between profiles. We note that \cite{Gavazzi07} used a
combination of strong and weak lensing measurements around massive
ellipticals and found that a SIS model provides a good fit to the
data. As discussed below, this does not imply that the density profile
is isothermal. Rather, it is believed to be the result of the clustering
of galaxies, which themselves have NFW density profiles.

It is therefore more appropriate to think of the galaxy-galaxy lensing
signal as a measurement of the cross-correlation between the galaxy
and mass distribution: the galaxy-mass cross-correlation function.
This observable provides additional constraints for models of galaxy
formation and can be used to study the bias parameter as a function of
scale \citep[e.g.][]{Waerbeke98}. In particular, it allows us to study
whether the (small scale) bias is non-linear and/or stochastic
\citep[e.g.][]{Pen98, Dekel99}).

The galaxy-mass cross-correlation function is closely related to the
galaxy two-point correlation function and the cosmic shear signal, as
they all provide ways to study the growth of structures via
gravitational instability. On large scales the biasing is (close to)
linear and the galaxy and dark matter distributions are well
correlated.  In this situation the galaxy power spectrum is $b^2$
times the matter power spectrum, $P(k)$ which can be measured through
cosmic shear studies \citep{HJ08, Fu08}). The value of the bias
parameter $b$ is not known a priori. Similarly, the galaxy-mass
cross-power spectrum will be $b\times r \times P(k)$, where $r$ is the
cross-correlation coefficient. The observed tangential shear
measurement can be expressed as a integral over the power spectrum
multiplied by a filter \citep[see][for more details]{Guzik01,Hoekstra02}.

The combination of the galaxy two-point correlation function and the
galaxy-galaxy lensing signal provides a direct measurement of the
ratio $b/r$ as a function of scale. This ratio was first measured by
\cite{Hoekstra01}, who later extended the analysis to the full RCS
data in \cite{Hoekstra02}.  They found that $b/r$ is constant out to
$\sim 7$~Mpc with an average value of $1.09\pm0.035$ for $\sim L_*$
galaxies.  \cite{Sheldon04} found similar results using SDSS
data. \cite{Hoekstra02} also included cosmic shear measurements from
the VIRMOS-Descart survey \citep{Waerbeke02} to study the scale
dependence of $b$ and $r$ separately. Although, \cite{Hoekstra02}
found tentative evidence for a variation of $b$ and $r$ with scale,
this results needs to be confirmed as residual systematics in the
cosmic shear signal may have affected the results \citep{Waerbeke05}.

Finally, \cite{Reyes10} recently showed how General Relativity can be
tested by combining the observed galaxy-galaxy lensing signal with
measurements of redshift-space distortions and the clustering of
galaxies. This measurement combines three probes of large-scale
structure to compare the two scalar potentials in the gravitational
metric ($\psi$ and $\phi$). In $\Lambda$CDM and GR, both scalar
potentials are equal. The lensing signal is sensitive to the sum
of these, whereas the clustering measurements are only sensitive to
the Newtonian potential $\phi$.  Although each of the observational probes
depends on the value of the bias and the normalization of the power
spectrum, the combination of these probes does not. \cite{Reyes10}
used data from the SDSS and found good agreement with GR on scales
ranging from $\sim 2-40$ Mpc.  The results cannot yet rule out $f(R)$
gravity models, but do disfavor some TeVeS models.

\subsection{Properties of Dark Matter Halos}

Although the study of the galaxy-mass cross-correlation function can
provide useful constraints for models of galaxy formation, one would
also like to learn more about the properties of the galaxy dark matter
halos themselves. This requires us to ``separate'' the contributions
from individual halos from the clustering of the lenses. There are a
number of ways this can be done. For instance, we can use the actual
positions of the lenses and make the simplifying assumption that the
observed signal arises only from the dark matter halos associated with
those lenses. Hence this approach does not describe well the situation
in clusters or the large-scale structure.  Furthermore, it is
computationally expensive, in particular if the model is extended to
include more parameters. An advantage, however, is that it uses the
two-dimensional shear field around the lenses, which is compared to
the observations in a maximum likelihood fashion. The maximum
likelihood method, however, has not been studied in detail using
numerical simulations and it is currently unclear to what extent the
simplifying assumptions bias the results. Such tests are needed before
this approach can be applied to modern, large data sets.

\begin{figure}[t]
\centering
\includegraphics[width=0.6\textwidth]{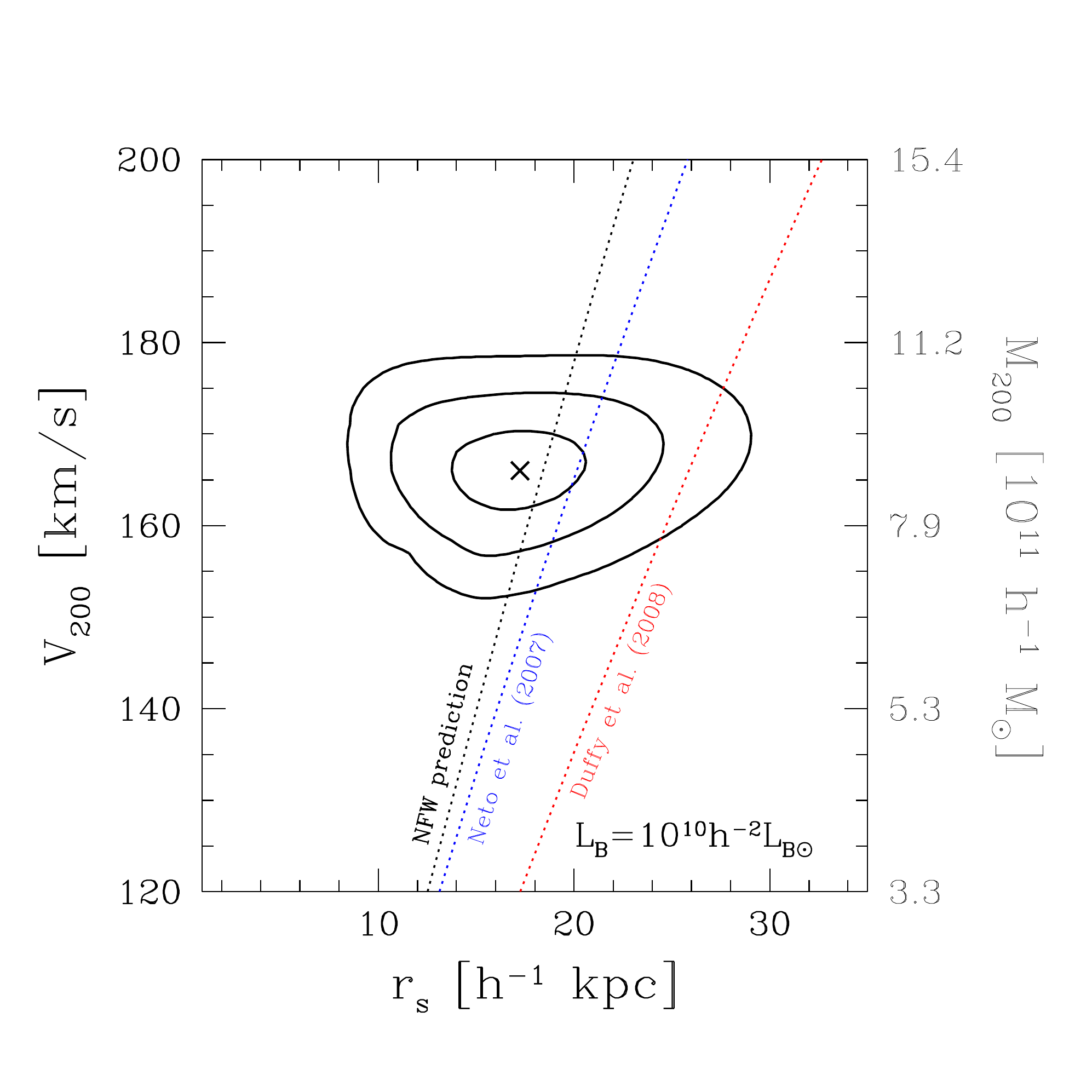}
\caption{Joint constraints on the scale radius $r_s$ and mass
  $M_{200}$ (and corresponding rotation velocity $V_{200}$ for a
  galaxy with an NFW profile and a fiducial luminosity
  $L_B=10^{10}h^{-2}L_{B\odot}$. The contours indicate the 68\%, 95\%
  and 99.7\% confidence on two parameters jointly. The dotted line
  indicates the prediction from the numerical simulations by NFW (from
  Hoekstra et al. 2004).  Our updated version of this figure shows
  expectations from works by Neto et al. (2007) and Duffy et al. (2008).}
\label{fig:halosize}
\end{figure}

The maximum likelihood method was used by \cite{Hoekstra04} to examine
the extent of dark matter halos around galaxies \citep[also see
e.g.][]{Brainerd96,Hudson98}.  The lack of color information limited
the analysis, but \cite{Hoekstra04} was able to constrain for the
first time the sizes of the dark matter halos. \Fig{halosize} shows
the result when an NFW model is assumed \citep{NFW97}. The mass and
scale radius $r_s$ are free parameters in the model, which are well
constrained. Numerical simulations of cold dark matter
\citep[e.g.][]{NFW97}) predict a correlation between these parameters
and the dashed line shows this prediction, which is in excellent
agreement with these measurements. However, the mass-concentration
relation depends on cosmology (in particular the normalization of the
matter power spectrum $\sigma_8$), which has changed over the years.
Furthermore, larger numerical simulations have been used to examine
the relations between halo properties and their evolution. The
original results presented in \cite{Hoekstra04} used $\sigma_8=0.85$,
but updated results with both lower and higher figures both yield larger
values for $r_s$.  We therefore also show in \Fig{halosize} the
expectations for \cite{Neto07} who used $\sigma_8=0.9$ and
\cite{Duffy08} who used $\sigma_8=0.8$; given the limited
investigations on the merits of the maximum likelihood method,
it is unclear whether or not there is tension between the data
and the predictions. It does suggest that this is an avenue worth
pursuing.

In high density regions, such as clusters of galaxies, the galaxy
dark matter halos are expected to be tidally stripped due to the
interaction with the tidal field of the smooth cluster mass
distribution. Galaxy-galaxy lensing studies provide the only way
to examine the sizes of the dark matter halos as a function of
cluster-centric radius \citep{Natarajan97,Natarajan98}.
A complication is that the signal arises from a combination of the
stripped halos and the global cluster mass distribution. The various
components can be modelled by a maximum likelihood
method\footnote{Which can also include strong lensing features to
constrain the model further.}). To minimize contamination by field
galaxies, with their much more extended halos, early work has been
confined to early type galaxies \citep{Natarajan98,Limousin07},
but recently \cite{Natarajan09} were able to study late type galaxies
as well. These studies have shown that dark matter halos of cluster
galaxies are tidally truncated \citep{Limousin07,Natarajan09}, which
is also observed in strong lensing studies \cite{Halkola07}.

The clustering of dark matter halos as a function of mass is
well-understood. This knowledge can be used to predict the galaxy-mass
cross-correlation function, by relating the dark matter distribution
statistically to the observable galaxies through a halo occupation
distribution (HOD; \cite{Seljak02}; for a review see \cite{Cooray02}).
This halo model approach is a powerful (and natural) way to interpret
the data as it provides a natural way to account for the fact that the
clustering depends on mass and that more massive halos host more than
one galaxy (i.e., groups and clusters of galaxies). A minor drawback
is that it only uses the tangential component of the shear. For
isolated lenses the signal is indeed purely tangential, but this is no
longer the case for an ensemble of lenses. The halo model was used by
\cite{Guzik02} and \cite{Mandelbaum06} to interpret the results from
the SDSS and constrain the fraction of satellite galaxies. A similar
study was carried out by \cite{vanUitert11} who complemented the SDSS
spectroscopic data with deep imaging data from the second Red-sequence
Cluster Survey (RCS2). This improved constraints for the massive
galaxies, which on average are at higher redshifts. 

The former studies focus only on the galaxy-mass cross-correlation
function, but a consistent model for galaxy formation also makes
predictions for the clustering of galaxies and the luminosity and/or
stellar mass functions. Combining the information of these
complementary probes can improve the constraints on the halo model
parameters \citep{Leauthaud11}. Such a joint analysis was performed
recently by \cite{Leauthaud12} using data from the
COSMOS survey. The high quality lensing data, in combination with
unprecedented wavelength coverage, allowed \cite{Leauthaud12} to study
the evolution of the stellar-to-halo mass relation from $z=0.2$ out to
$z=1$. They found that the halo mass scales with stellar mass $\propto
M_*^{0.46}$ for galaxies with $M_*<5\times 10^{10}$M$_{\odot}$ over
the redshift range studied, whereas the slope of the relation steepens
for higher masses. 

One can also attempt to 'avoid' the complication caused by the
clustering of the lenses by considering only relatively `isolated'
lenses.  In this case the observed lensing signal is dominated by only
a single galaxy. Such a selection, which requires (photometric)
redshift information, was used in \cite{Hoekstra05} to study the
relation between the virial mass and the luminosity. They limited the
sample to galaxies that are more than 30'' away from a brighter
galaxy. For these galaxies \cite{Hoekstra05} found that the virial
mass scales with luminosity as $\propto L^{1.5}$, in agreement with
the results from \cite{Guzik02,Mandelbaum06} who used the halo model
to interpret the SDSS data. For a galaxy with fiducial luminosity of
$L_B=10^{10} h^{-2}L_{B\odot}$ \cite{Hoekstra05} obtained a virial
mass $M_{\rm vir}=9.9^{+1.5}_{-1.3}\times 10^{11}M_\odot$, also in
good agreement with \cite{Mandelbaum06}.

If we assume that baryons do not escape the dark matter overdensity
they are associated with, the ratio of $M_{\rm b}$, the mass in
baryons, to the virial total mass of the halo is $M_{\rm b}/M_{\rm
  vir}=\Omega_b/\Omega_m$. Furthermore, the amount of cold gas is
negligible for massive galaxies. Therefore, by comparing the stellar
mass of the lenses to the virial mass determined by weak lensing, the
efficiency with which baryons are converted into stars can be
constrained. \cite{Hoekstra05} found that late types convert a $\sim
2$ times larger fraction of baryons into stars compared to early-type
galaxies.  The measurement of \cite{Hoekstra05} is mostly constrained
by relatively luminous galaxies. \cite{Mandelbaum06} find that the
conversion efficiencies are independent of morphological type for
stellar masses less than $\sim 7\times 10^{10}M_\odot$, but also find
that later type galaxies appear more efficient for higher stellar
masses. \cite{Heymans06} used GEMS data to study the mean virial to
stellar mass ratio for a complete sample of massive galaxies out to
$z\sim 0.8$. The results, which agreed well with \cite{Hoekstra05}
and \cite{Mandelbaum06} showed little evidence for evolution.
\cite{Leauthaud12} also study the stellar mass fraction as a function
of mass finding a minimum at a halo mass of $\sim 1.2\times10^{12}\Msol$.

These studies demonstrate the potential of weak lensing results for
the study of galaxy evolution. We note, however, the accuracy of the
halo model is limited, and that measurements soon will be limited by
this.

\subsection{Halo Shapes}

The average shape of dark matter halos can provide another way to
learn more about the nature of dark matter (and the interaction with
baryons). Numerical simulations of CDM have shown that the resulting
dark matter halos are triaxial with a typical ellipticity of $\sim
0.3$ \citep[e.g.][]{Dubinski91,Jing02,Hayashi07}. In the case of
self-interacting dark matter, the predicted halos are more spherical,
although this difference is more pronounced on small scales
\citep{Dave01}. We also note that hydrodynamic simulations suggest
that baryonic effects cause dark matter halos to evolve more oblate
configurations at all radii, even if the effect of baryons is most
prominent in the inner parts \citep[e.g.][]{Kazantzidis04,Kazantzidis10}.  

The small scales, which are baryon-dominated, are best probed by
strong lensing studies or dynamical studies. The latter approach has
been extended to larger scales through the study of streams of stars
in the Milky Way \citep{Helmi04, Koposov10}. On large scales, which
are best constrained by numerical simulations, only weak lensing
studies can provide observational constraints on the shapes of dark
matter halos. The measurement, however, is difficult: we now need to
measure an azimuthal variation in the, already small, galaxy-galaxy
lensing signal. The azimuthal variation is measured with respect to
the major axis of the light distribution, i.e. we assume that the
halos are aligned with the lens. If halos are flattened, but not
aligned with the light distribution, the resulting lensing signal will
be isotropic. Hence, any misalignment will reduce the amplitude of the
azimuthal variation and the weak lensing constraints are in effect
lower limits on the shapes. Such misalignments might result from
baryonic effects. For instance \cite{Bailin05} found in their
hydrodynamic simulations of disk galaxies that the outer part of the
halo is not well aligned with the inner regions, which do show a good
alignment between disk and inner halo \citep[also see
e.g.][]{Abadi10}.

Weak lensing studies of dark matter halo shapes are more sensitive to
systematic effects/errors compared to the measurement of the
galaxy-mass cross-correlation function. For instance, residual PSF
anisotropy leads to correlations between the lenses and sources,
biasing the dark matter halo shape determination
\citep[e.g.]{Hoekstra04,Mandelbaum06shape}). However, even in the
absence of residual systematics, lensing by lower redshift structures
can align the lens and the source, reducing the signature of an
anisotropic halo. This cosmic shear contribution is described in
detail in \cite{Mandelbaum06shape} (also see \cite{Brainerd10} and
\cite{Howell10} who also discuss this problem). This source of bias
becomes more prominent with increasing lens redshift and will need to
be taken into account when interpreting the current generation of
surveys. Fortunately, \cite{Mandelbaum06shape} provide a method to
suppress this signal, although this can only be applied reliably on
relatively small scales and with limited accuracy
\citep{vanUitert12}.

A successful measurement of the halo shapes requires a much larger
number of lens-source pairs, as the signal-to-noise ratio is about a
tenth of that of the tangential shear signal, and limited to small
scales \citep{vanUitert12}. In recent years a number of pioneering
studies have been carried out, but the results are still inconclusive.
The first claim of elliptical dark matter halos using weak lensing was
presented in \cite{Hoekstra04} using 42 deg$^2$ of data from the
RCS. The lenses were selected based on their apparent magnitude and
early type galaxies contribute most of the signal. \cite{Hoekstra04}
found that the halos are aligned with the light distribution and
estimated an ellipticity of $\langle e_{\rm
  halo}\rangle=0.33^{+0.07}_{-0.09}$. These results are in fair
agreement with the results from \cite{Parker07}, who used 22 deg$^2$
of deep $i'$ data from the CFHT Legacy Survey and also inferred an
ellipticity $\sim 0.3$.

The analysis of SDSS data by \cite{Mandelbaum06shape} did not detect
an azimuthal variation in the lensing signal when considering the full
sample of lens galaxies. However, when restricting the sample to
bright early type galaxies, the results of \cite{Mandelbaum06shape} do
suggest that the halos are aligned with the light. Recently,
\cite{vanUitert12} completed the analysis of 800 deg$^2$ of RCS2 data,
but did not detect a significant anisotropy signal. Part of the loss
of precision is caused by a careful accounting for possible systematic
effects. 

An interesting, still open question, is whether the alignments are the
same for different types of galaxies. The possible differences between
the various analyses needs to be investigated further, but it is clear
that significant progress will only be possible with the next
generation of deep, multi-color surveys which provide redshift
information for the lenses. The latter is important because of the
scale dependence of the anisotropy signal
\citep{Mandelbaum06shape,vanUitert12}.

\begin{figure}[t]
\centering
\includegraphics[width=0.5\textwidth]{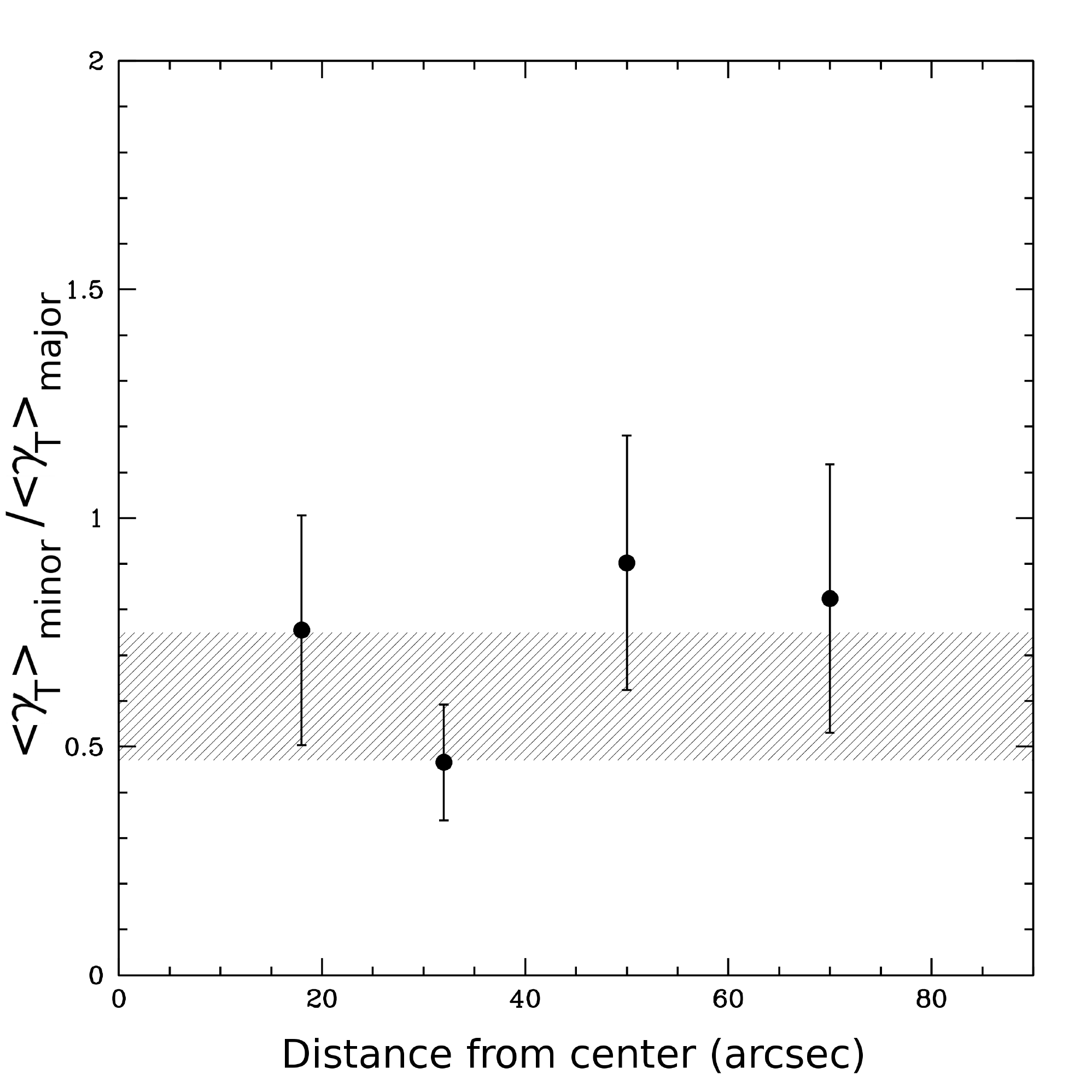}
\caption{Ratio of the mean shear experienced by sources
    closest to the minor axes of a foreground lens to that of sourcest
    closest to the major axes for lenses with axis ratios $b/a$
    between 0.5 and 0.8. This criterion preferentially selects early
    type galaxies.  The results are taken from \cite{Parker07} who
    analysed 22 deg$^2$ of $i'$ data from the CFHTLS. The weighted
    average shear ratio is $0.61\pm0.14$.}
\label{fig:haloshape}
\end{figure}

\subsection{Future Prospects}

Early weak galaxy-galaxy studies have already provided unique
constrains on the properties of dark matter halos, such as constraints
on their extent and shapes, as well as masses.  The SDSS results have
demonstrated the usefulness of multi-color data and redshift
information for the lenses. Of particular interest would be studies
that complement the SDSS with deeper imaging. The precision of
photometric redshifts limits their usefulness at low redshifts, but
for the study of lenses with $z>0.3$ the excellent statistics that can
be achieved from the next generation surveys are expected to outweigh
the limitations. For instance the KiloDegree Survey (KiDS) which
started observations in the fall of 2011 will cover 1500 deg$^2$ in
nine filters.  The extensive wavelength coverage will yield accurate
photometric redshifts for the lenses. As a result statistical errors
will be reduced by a factor $\sim 4$ over current results and provide
a first opportunity to study in detail the properties of dark matter
halos as a function of density and baryon contents. The Dark Energy
Survey, which recently started taking data, will image 5000 deg$^2$.

Another major step forward will come from the Large Synoptic Survey
Telescope (LSST) which plans to start surveying $\sim 20,000$ deg$^2$
around the turn of this decade and space-based dark energy projects,
such as the recently approved Euclid mission \cite{Laureijs11}, which
is scheduled for launch in 2020. The excellent statistics provided by
the latter projects will allow for the study the galaxy lensing signal
over a wide range in physical properties and redshift. Thanks to these
developments galaxy-galaxy lensing will continue to develop into an
important tool to study galaxy masses as a function of their
(observable) baryonic properties.


\newpage

\section{The Dark and Luminous Mass Distribution of Early-type Galaxies\\
using Strong Gravitational Lensing}\label{sec:strong}

\subsection{Introduction}\label{sec:level1}

Despite being a fundamental parameter required to test galaxy
formation models \citep{White78,Blumenthal84}, the measurement of
galaxy masses with few percent accuracy on any scale is notoriously
difficult.  This is particularly the case for early-type galaxies (ETGs)
which are not rotationally supported and generally lack gas-dynamical
tracers.  As we have seen in previous sections, a wide range of
methods, or their combinations, have been employed to measure galaxy
total masses on very different scales: e.g. stellar and gas dynamics,
hydrostatic equilibrium of X-ray emitting gas and weak gravitational
lensing.  Whereas each of these methodologies have their own
advantages and limitations, they also all have varying levels of
precision and varying scales within which a mass can be measured.

In general, however, none of these methods reach the percent-level
precision which is often required to accurately measure, for instance,
the contribution of dark matter to the inner regions of galaxies
where both baryons and dark matter interact and possibly play equal
partners in galaxy formation models
\citep[e.g.][]{Saglia92,Bertin92,Loewenstein99,Keeton01,Padmanabhan04}.
In addition to degeneracies, some methods also become problematic
beyond the local universe considering the limited signal-to-noise
ratio of observations with present-day telescopes.

In this section, we focus on strong gravitational lensing (plus
stellar dynamics) as a probe of the mass of galaxies out to tens of
kpc scales, covering their inner regions to several effective radii
(and beyond in combination with weak lensing). In addition, we shortly
discuss the use of gravitational lensing to quantify the level of mass
substructure in the dark matter haloes (e.g. CDM substructure or dwarf
satellites).  The basics of strong lensing theory and the ability to
measure galaxy masses to percent level accuracy on different scales
are shortly introduced.  This is exemplified with several recent
highlights, mostly based on the largest strong lensing survey to
date: the Sloan Lens ACS (SLACS) Survey \cite{Bolton06}.  This section
is neither complete nor unbiased and we refer to \cite{SchneiderP06}
or \cite{TreuARAA10} for more thorough theoretical and observational
overviews.

\subsection{Basic Lensing Theory}\label{sec:level2}

Strong gravitational lensing can, to very good approximation,
be regarded as geometric optics in curved space-time
\citep{SEF92,Kochanek06}, with the usual conservation of surface
brightness.  We can also assume in nearly all astrophysical
circumstances that the gravitational field that causes lensing
(weak or strong) satisfies $|\phi|/c^{2} \ll 1$ and that perturbations
from the FRW metric of the Universe are small.  As in geometric optics,
curved space-time can heuristically be associated with a refractive
index $n=1 + 2 |\phi| /c^2$ for each point in space.  A change in
refractive index leads to a deflection of the light-ray by
\begin{equation}
 \vec{\alpha} = \int \vec{\nabla}_{\perp} n \, dl = 
 \frac{2}{c^{2}} \int \vec{\nabla}_{\perp} \phi \,dl.
\end{equation}
Hence an observer sees the light ray from a different direction than
where it originated.  In summary, two parallel rays of light (or
wavefronts; \citealp{Kayser83}) originating from slightly different
positions will in general not remain parallel and can either diverge
or converge. Similarly, rays of light being emitted in different
directions from a single source can sometimes end up crossing each
other again. If an observer (e.g.\ on Earth) happens to be at that
crossing point, the emitting source will be seen multiple times. The
deflection can be used to learn more about the mass distribution of
the deflector (e.g. galaxy, cluster, stars, etc). It is worth noting
that the concepts of time-delay and deflection can be unified in the
generalization of the concept of ``Fermat's principle''
\citep{Schneider85,Blandford86}, where lensed images
form on extrema of a so-called time-delay surface.

\subsubsection{The Thin-Lens Approximation}\label{sec:thinlens}

Before coming up with a general equation for (strong) lensing,
we illustrate some of these aspects in case of a point-mass 
(e.g.\ a star or stellar remnant) with gravitational potential
\be
\phi(\xi,z) = \frac{G M}{\sqrt{\xi^{2} + z^{2}}},
\ee
where $\xi$ and $z$ are the distances perpendicular and parallel to the
line-of-sight from the point mass to the observer, respectively, and $G$
and $M$ are the gravitational constant and mass of the lens. One finds
(note that $\xi^{2} = \vec{\xi} \cdot \vec{\xi}$ and then the gradient
is carried out w.r.t. $\vec{\xi}$):
\be
\vec{\nabla}_{\perp} \phi = \frac{G M \vec{\xi}}{(\xi^{2} + z^{2})^{3/2}}.
\ee
Integrating this along the l.o.s., assuming $z$ goes from minus to plus
infinity, the deflection angle for a point mass is
\be
\hat{\alpha} = \frac{4 G M}{c^{2} \xi} = \frac{2 R_{\rm s}}{\xi},
\ee
where $R_{\rm s} = 2 G M/c^{2}$ is the Schwarzschild radius. In general
the impact parameter $b \gg R_{\rm s}$, hence the deflection angles
are far smaller than unity (weak deflection), justifying the approximations
that were made so far.

We now also note that in general the distance $\Delta z$ over which
light is substantially deflected is much smaller than the distance of
the lensed source of light to the deflector and the deflector to the
source. In that case, we can approximate any extended lens by the
``thin lens approximation'', where the density distribution ($\rho$)
is collapsed along the line of sight in to a surface density
$\Sigma(\vec{\xi}) = \int \rho(\vec{\xi}, z) dz$. The latter is often
the only mass-related quantity that can be determined. Deflection is
assumed to occur (effectively) instantaneously in the lens plane of
the deflector. The thin lens approximation is practically always
justified for describing the main deflector. However, whenever very
high accuracy is required it should be kept in mind that the universe
is not exactly homogeneous and isotropic on large scales and therefore
describing the intervening space between the source and the deflector
and between the deflector and the observer with a standard
Robertson-Walker metric is only an approximation. In reality, photons
will propagate through over and under-densities, resulting effectively
in additional distortion (shear) and focus/defocus in addition to the
one provided by the main deflector (see \se{weak}). This effect is
usually accounted for as external shear and convergence and result
in typical corrections of order a few percent to the strong lensing
inference \cite[e.g.][]{KKS97,Treu09,Suyu++10,Suyu++13}.

Extending now from the point-source deflector to a general (surface) mass
distribution, using $M \rightarrow \Sigma(\vec{\xi}) d^{2}\vec{\xi}$, one
readily finds that
\be
\vec{\hat{\alpha}}(\vec{\xi}) = \frac{4 G}{c^{2}} \iint \frac{\Sigma(\vec{\xi}) (\vec{\xi} - \vec{\xi}')}{|\vec{\xi} - \vec{\xi}'|^{2}} d^{2}\vec{\xi}.
\ee
For circularly symmetric lenses with $\xi = |\vec{\xi}|$ we have
\be
\vec{\hat{\alpha}}(\xi) = \frac{4 G M(\le \xi) \vec{\xi}}{c^{2} \xi^{2}}
\ee
with
\be
M(\le \xi) = 2 \pi \int_{0}^{\xi} \Sigma(\xi') \xi' d\xi'.
\ee
We are now ready to introduce the lens equation which forms the basis
of lensing theory.

\subsubsection{The Lens Equation}\label{sec:lens}

Now that the deflection for any general and circularly symmetric mass
distribution can be calculated, we can relate any point in plane (at
a distance $D_{s}$) where the emitting source is, to a point at
$\vec{\theta}=\vec{\xi}/D_{d}$ in
the plane of the deflector (at a distance $D_{d}$) as seen by the observer.
We also assume that the distance from the deflector to the source is $D_{ds}$,
which in GR is not necessarily equal to $D_{s}-D_{d}$.
In that case (\Fig{mosaic}), we readily find in scalar notation
$\theta D_{s} = \beta D_{s} + \hat{\alpha} D_{ds}$, assuming the
small-angle approximation, or equivalently
\be
\vec{\beta} = \vec{\theta} - \frac{D_{ds}}{D_{s}} \vec{\hat{\alpha}},
\ee
where $\vec{\beta}$ is the vector angle to the source as it would be seen
(w.r.t.\ some arbitrary coordinate origin, usually chosen to the deflector
centroid) if not lensed and $ \vec{\theta}$ is the vector angle
of the lensed image(s) as observed. Defining the reduced deflection angle
as $\vec{\alpha} \equiv \frac{D_{ds}}{D_{s}} \vec{\hat{\alpha}}$, we arrive
at the standard non-linear lens equation
\be
\vec{\beta}(\vec{\theta}) = \vec{\theta} - \vec{\alpha}(\vec{\theta}).
\ee
We note here that the non-linearity of $\vec{\alpha}(\vec{\theta})$ can
lead to multiple solutions of $\vec{\theta}$ of the lens equation for
a given source position $\vec{\beta}$, hence multiple imaging (strong
lensing) occurs (note that this equation holds for each position of an
extended source and that image surface brightness for each solution
$\vec{\theta}$ is identical to that of the source at $\vec{\beta}$.
The extreme case of multiple imaging is the creation of the
``Einstein Ring'' for circularly symmetric lenses, for which
\be
\beta = \theta - \frac{4 G M(\theta) D_{ds} }{c^{2} D_{d} D_{s}} \frac{1}{\theta}.
\ee
Defining the Einstein radius
\be
\theta_{E} \equiv \left[\frac{4 G M(\theta) D_{ds} }{c^{2} D_{d} D_{s}}\right]^{1/2},
\ee
the lens equation for $\beta=0$ has the solution $\theta=\pm \theta_{E}$.
Because of symmetry, a source aligned with the source-deflector line will
be imaged into a perfect Einstein ring.

We can take one more step to simplify the equations. We do this assuming
the deflector has constant density (a ``mass-sheet''). In that case
\be
\alpha(\theta) = \left[ \frac{4 \pi G \Sigma D_{d} D_{ds}}{c^{2} D_{s}} \right] \theta
		= \left[ \frac{\Sigma}{\Sigma_{\rm crit}} \right] \theta,
\ee
with $\Sigma_{\rm crit} \equiv c^{2} D_{s}/(4 \pi G D_{d} D_{ds})$.
We further define the so called ``convergence''
$\kappa \equiv  \Sigma/\Sigma_{\rm crit}$. Hence, the deflection angle
is linear and $\beta = (1-\kappa)\theta$. For $\kappa=1$, parallel
rays converge to a single focus, making the transition from $\kappa <1$
to $\kappa > 1$ special. The mass-sheet with $\kappa=1$ is a perfect
focusing lens. Whereas this is not the case for general lenses, it turns
out that lenses with $\kappa>1$ at any point, can (do not have to though)
create multiple images because of over-focusing. With this definition the
deflection angle becomes
\be
\vec{{\alpha}}(\vec{\theta}) = \frac{1}{\pi} \iint \frac{\kappa(\vec{\theta}) (\vec{\theta} - \vec{\theta}')}{|\vec{\theta} - \vec{\theta}'|^{2}} d^{2}\vec{\theta}.
\ee
It can be shown that the average convergence inside the Einstein radius
of {\sl any} circularly symmetric deflector is exactly equal to one.
Thus for lens systems with a (near) Einstein ring, the mass inside
the Einstein radius is $M_{E}\equiv \pi (D_{d}\theta_{E}) \Sigma_{\rm crit}$
independent of the density profile of the deflector. In fact, deviations
from symmetry are only secondary effects. Hence the masses of strong lenses
can be determined to rather exquisite accuracy if a reasonable
Einstein radius can be defined. 

\subsubsection{Axisymmetric Lenses}

Whereas lens modeling can be rather complex, in general axisymmetric
(in 2D) lenses give good insight into the processes that are important
in lensing. This is because many lenses are ETGs, which generally have
round mass distributions (and potential) with small ellipticities.
It is therefore useful to derive some properties for these
types of lenses, giving first-order results for other lenses as well.

We thus assume $\kappa(\theta) = \kappa(|\vec{\theta}|)$ and
$\alpha(\theta) = |\vec{\alpha}(|\vec{\theta}|)|$. It is then
easy to show that
\be
\alpha(\theta) = \frac{M(<\theta)}{\theta} = \langle \kappa \rangle(\theta) \theta,
\ee
with
\be
M(<\theta) \equiv 2 \int_{0}^{\theta} d\theta' \theta' \kappa(\theta')
\ee
and
\be
\langle \kappa \rangle(\theta) = \frac{M(<\theta)}{\theta^{2}}.
\ee
Hence the deflection angle then reduces to
\be
\beta = (1-\langle \kappa \rangle(\theta)) \theta.
\ee
This immediately shows that for an Einstein ring with $\beta=0$,
that $\langle \kappa \rangle(\theta) = 1$. Hence for any axisymmetric
lens the average convergence inside the Einstein radius is unity. The
enclosed mass can thus be inferred independent of the density profile.
In physical units
\be
M_{\rm E} = \pi \theta_{\rm E}^{2} D_{d}^{2} \Sigma_{\rm crit},
\ee
or conversely
\be
\theta_{\rm E} = \left(\frac{4 G M_{\rm E} D_{ds}}{c^{2} D_{d} D_{s}}\right)^{1/2}.
\ee
Thence if the Einstein radius can be determined, the mass can be determined.

\subsubsection{Lensing \& Stellar Dynamics}

A powerful complementary constraint that is worth mentioning at this point
is the combination of the precise total mass measurement using strong
gravitational lensing with stellar kinematic measurements. Whereas this
combinations can becomes rather complex for two or three integral
(non-spherical) models, here we illustrate the basic idea assuming
spherical symmetry and power-law density and luminosity density models
\cite{TK02,2004bdmh.confE..66K, 2006PhRvD..74f1501B}. Despite these
simplifications, these toy-models give rather robust results for the
density slopes of ETGs.

Let us suppose that the stellar component has a luminosity
density $\nu_l(r) = \nu_{l,o} r^{-\delta}$ and is a trace component
embedded in a total (i.e. luminous plus dark-matter) mass distribution
with a density $\nu_\rho(r) = \nu_{\rho,o} r^{-\gamma'}$. In addition,
let us assume that the anisotropy of the stellar component $\beta =
1-(\overline{\sigma^2_\theta}/\overline{\sigma^2_r})$ is constant with
radius. For a lens galaxy with a projected mass $M_{\rm E}$ inside the
Einstein radius $R_{\rm E}$, the luminosity weighted average
line-of-sight velocity dispersion inside an aperture $R_{\rm A}$ is
given, after solving the spherical Jeans equations, by
\be
\label{eq:sigav}
  \langle \sigma_{||}^2\rangle (\le R_{\rm A}) = \frac{1}{\pi}
  \left[ \frac{G M_{\rm E}}{R_{\rm E}} \right] f(\gamma',
  \delta,\beta) \times \left(\frac{R_{\rm A}}{R_{\rm
  E}}\right)^{2-\gamma'}
\ee
with
\begin{eqnarray}
  f(\gamma', \delta,\beta) = 2 \sqrt{\pi}\,\left(\frac{\delta -3}{(\xi
  - 3)(\xi - 2\beta)} \right)&\times \\
  \left\{\frac{\Gamma[(\xi-1)/2]}{\Gamma[\xi/2]} - \beta
  \frac{\Gamma[(\xi+1)/2]}{\Gamma[(\xi+2)/2]} \right\} \times
  \nonumber\\ \left\{\frac{\Gamma[\delta/2]\Gamma[\gamma'/2]}
  {\Gamma[(\delta-1)/2]\Gamma[(\gamma'-1)/2)]}\right\}
\end{eqnarray}
with $\xi = \gamma'+\delta -2$.  Similarly,
\be
  \sigma_{||}^2(R) = \frac{1}{\pi} \left[ \frac{G M_{\rm
  E}}{R_{\rm E}} \right] \left(\frac{\xi - 3}{\delta -3}
  \right) f(\gamma', \delta,\beta) \times \left(\frac{R}{R_{\rm
  E}}\right)^{2-\gamma'}.
\ee
In the simple case of a SIS with $\gamma'=\delta=\xi=2$ and
$\beta=0$, we recover the well-known result
\be
\sigma_{||}^2(R) = \frac{1}{\pi} \left[ \frac{G M_{\rm
  E}}{R_{\rm E}} \right]~~~({\rm SIS}).
\ee
From \Eq{sigav}, one sees that the radial dependence of the stellar
velocity dispersion depends on $\gamma'$ only. All other parameters
(i.e.\ $\delta$, $\beta$, etc.) only enter into the normalization.
Since the luminosity density (i.e.\ $\delta$) and $M_{\rm E}$ are
measured with little uncertainty, the measurement of $\langle
\sigma_{||}^2\rangle (\le R_{\rm A} \neq R_{\rm E})$ immediately gives
the density slope $\gamma'(\beta)$ (where $\beta$ in general plays
only a minor role). This is the basis of combining stellar dynamics
with gravitational lensing to obtain not only the mass but also the
density-slope of ETGs.

\newpage
\centerline{\bf Figure ``hugeslacsmosaic.eps'' to be shown here,
 using the full page.}
\bigskip
\begin{figure*}
\centering
\caption{A sub-sample of SLACS lenses (credit: Adam Bolton;
see www.slacs.org) in false-color. Each panels shows the data
on the left and a model of the system on the right.}
\label{fig:mosaic}
\end{figure*}

\newpage

We can estimate the change $\delta \gamma'$ from the observables.
One finds to first order (assuming fixed values of $\beta$ and $\delta$):
\begin{eqnarray}\label{eq:err1}
  \frac{\delta \sigma_{||}}{\sigma_{||}}(\le R_{\rm A}) & = &\frac{1}{2}
  \frac{\delta M_{\rm E}}{M_{\rm E}} + \frac{1}{2}\left(\frac{\partial
  \log f}{\partial \log \gamma'} - \gamma' \, \log\left[\frac{R_{\rm
  A}}{R_{\rm E}}\right]\right) \nonumber \\ &\times&  \frac{\delta \gamma'}{\gamma'}
  \equiv \frac{1}{2} \left( \frac{\delta M_{\rm E}}{M_{\rm E}} + \alpha_g
  \frac{\delta \gamma'}{\gamma'}\right) .
\end{eqnarray}
The second term in this equation was already derived by \cite{TK02}.
If we further assume the errors on $M_{\rm E}$
and $\sigma_{||}$ to be independent,
\be
  \left\langle \delta_{\gamma'}^2\right\rangle \approx \alpha_g^{-2}
  \left\{\left\langle \delta_{M_{\rm E}}^2\right\rangle + 4
  \left\langle \delta_{\sigma_{||}}^2\right\rangle \right\},
\ee
where $\delta_{\dots}$ indicate fractional errors. Since in general
$\delta_{M_{\rm E}}\ll \delta_{\sigma_{||}}$, one finds the simple
rule of thumb that the error $\delta_{\gamma'} \sim
\delta_{\sigma_{||}}$ for close-to-isothermal mass models, since
$\alpha_g \sim 2$. This estimate is in very good agreement with the
results from properly solving the Jeans equations for two-component
mass models and justifies neglecting the mass errors \citep{Treu04}.

\subsection{Observational Results}

In this section we highlight some recent results on the study
of early-type galaxies using strong gravitational lensing.

We focus on two aspects of strong gravitational lensing that have
recently progressed rapidly and that we think have great promise in
future galaxy structure and evolutions studies: (i) The combination
of strong lensing, stellar kinematics and stellar populations to
constrain the inner stellar and dark-matter mass profiles of ETGs
as function of their mass and redshift and (ii) the use of
simply-parameterized and 
grid-based modeling of strong lenses to constrain the level mass
substructure in the inner regions of ETGs
We illustrate how lensing can address these two science drivers based
mostly on recent results from the Sloan Lens ACS Survey (SLACS).

\subsubsection{Sloan Lens ACS Survey (SLACS)}\label{sec:slacs}

The SLACS gravitational lenses \citep{Bolton06,Treu06,Koopmans06,Gavazzi07,
Bolton08,Gavazzi08,2008ApJ...684..248B,Auger09,Treu09,Auger10X,New++11}
were selected from the spectroscopic database of the SDSS based on the
presence of absorption-dominated galaxy continuum at one redshift and
nebular emission lines (Balmer series, [O\textsc{ii}]\,3727,
or [O\textsc{iii}]\,5007) at a higher redshift.  The spectroscopic
lens survey technique was first envisioned by
\citet{1998MNRAS.299.1215W,2000ASPC..195...94H}
following the serendipitous discovery of the gravitational lens
0047$-$2808 through the presence of high-redshift Lyman-$\alpha$
emission in the spectrum of the targeted lower redshift
elliptical galaxy.
Further details of the SLACS approach are provided
in \cite{2005ApJ...624L..21B, 2004AJ....127.1860B}.
The SLACS Survey includes candidates from the SDSS MAIN galaxy sample
\cite{2002AJ....124.1810S} in addition to candidates from the SDSS luminous
red galaxy (LRG) sample \cite{2001AJ....122.2267E}.  Most candidates were
selected on the basis of multiple emission lines, though several lens
candidates were observed on the basis of secure \oii\,3727 line
detections alone.  By virtue of this spectroscopic selection method, all
SLACS lenses and lens candidates have secure foreground (``lens'') and
background (``source'') redshifts from the outset.  Accurate redshifts
such as these are essential for most quantitative scientific
applications of strong lensing, as they are required to convert
angles into physical lengths.
\begin{figure}
\centering
\includegraphics[width=0.5\textwidth]{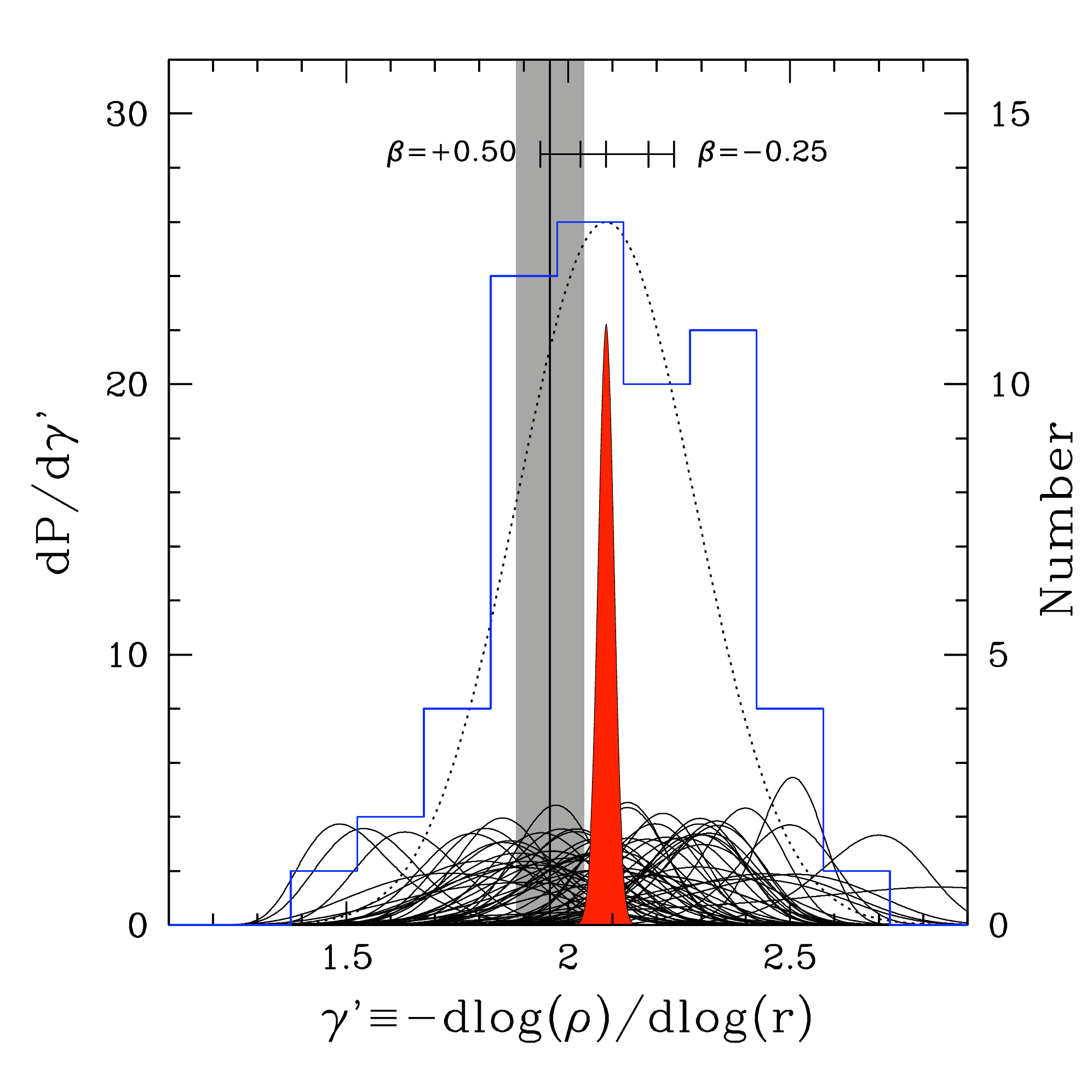}
\caption{The logarithmic density slopes of 58 SLACS early-type galaxies
(thin solid curves).  The filled red curve is the joint posterior
probability distribution of the {\sl average} density slope of the
sample. The histogram indicates the distribution of median values of
the density slopes. The dotted Gaussian curve indicates the intrinsic
scatter in $\gamma'_{\rm LD}$ (see text).  We assume a Hernquist
luminosity-density profile and no anisotropy (i.e.\ $\beta_r=0$).
The small dashes indicate the shift in average density slope for
$\beta_r=+0.50, +0.25, -0.50 -0.25$ (left to right),
respectively. Note the reversal of the $\beta_r=-0.50$ and $-0.25$
dashes.  The vertical solid line and gray region indicates the
best-fit value and 68\% CL interval, respectively, of the average
density derived from scaling relations.}
\label{fig:slopes}
\end{figure}

\begin{figure*}
\centering
\includegraphics[scale=0.41]{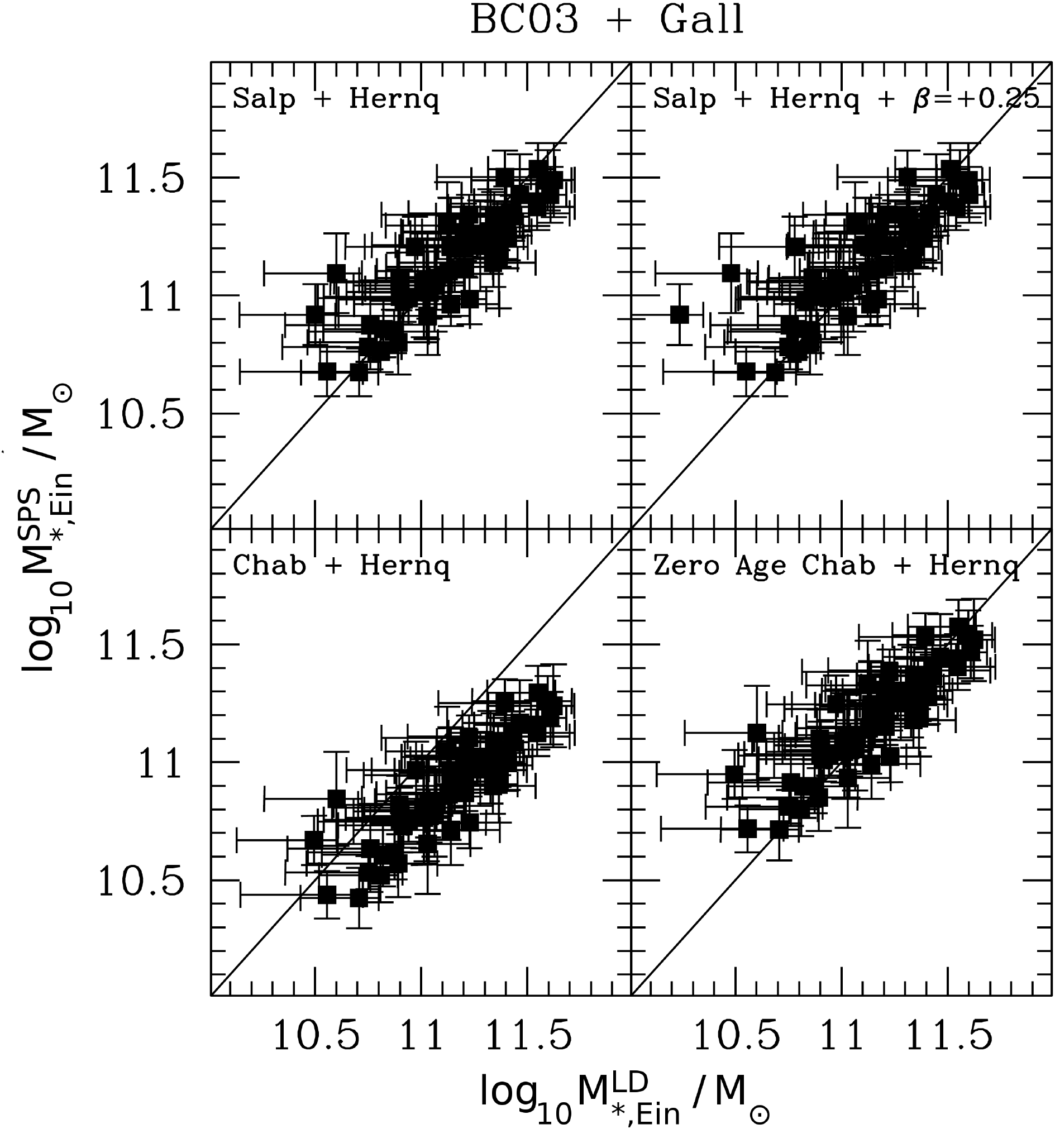}
\includegraphics[scale=0.38]{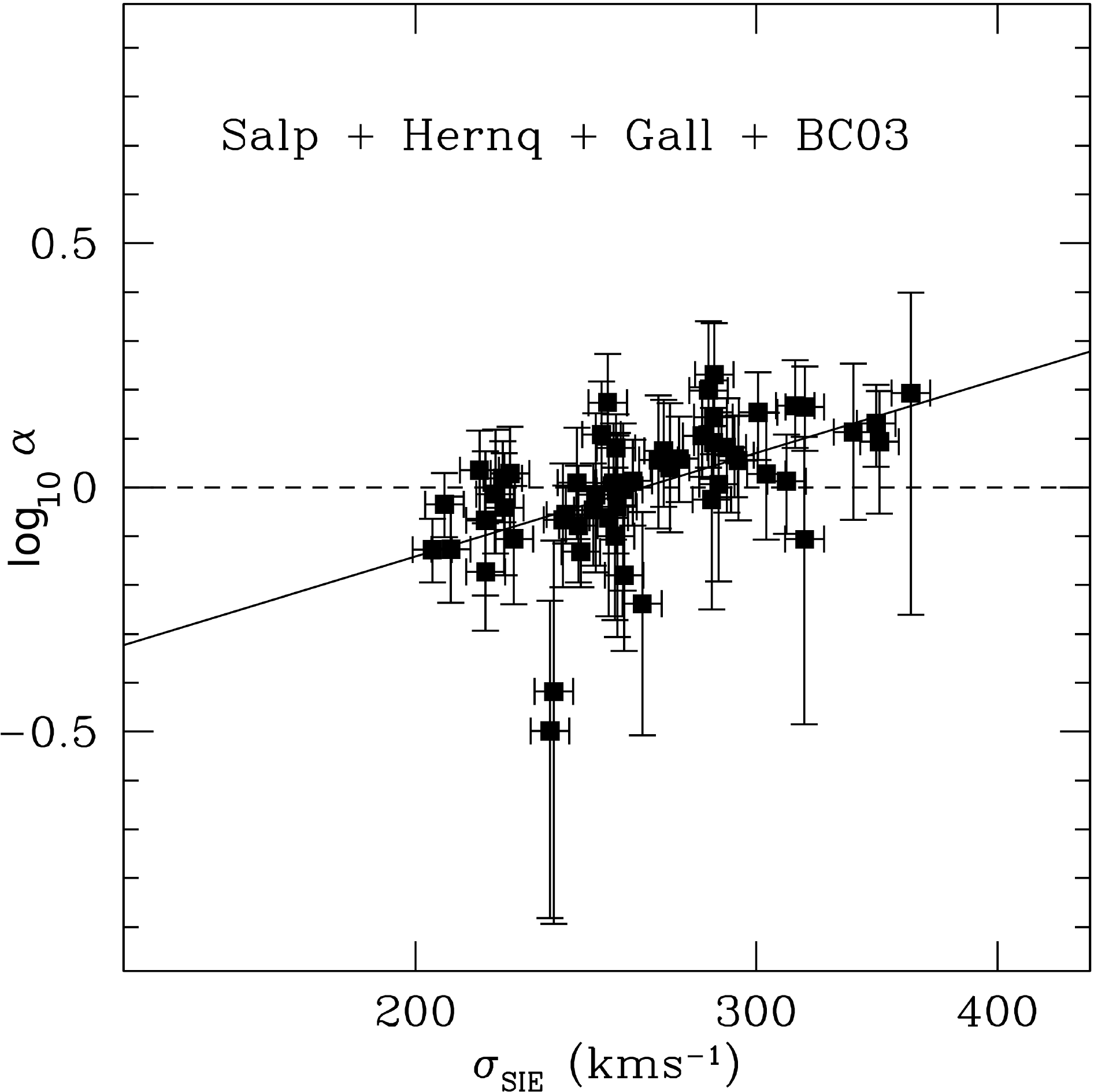}
\caption{{\it Top figure}: Comparison between stellar mass in the
cylinder of radius equal to the Einstein Radius as inferred from
lensing and dynamical models (x-axis) and that inferred from fitting
stellar populations synthesis models to the observed spectral energy
distribution (y-axis).  The solid line indicates the identity. Stellar
populations synthesis models by \cite{brucha03} are assumed together
with an informative metallicity prior \cite{Gallazzi05}.
{\it Bottom figure}: Template mismatch parameter
$\alpha\equiv${\mseld}/{\msesed} for Salpeter IMF as a function of
lensing velocity dispersion (left), stellar velocity dispersion
(center) and V-band luminosity corrected to $z=0.2$.  A tentative
positive trend with velocity dispersion is observed (solid line).
The dashed line represents the trend expected for a universal Salpeter IMF.}
\label{fig:IMF}
\end{figure*}

\subsubsection{The Density Profiles of Early-Type Galaxies (ETGs)}

When combining the total mass inside the Einstein radius given by
detailed lensing modeling, usually accurate to a few percent
\citep{Kochanek91}, with measurements of stellar kinematics (either
inside an aperture, along a slit or through 2D IFU measurement), a
powerful constraint can be set on the average (kinematically weighted)
density profile of ETGs inside the Einstein radius (or effective
radius which ever is larger).  This methodology, shortly outlined
in the previous section, has been successfully applied initially by
the Lenses Structure \& Dynamics (LSD) survey
\cite[e.g.][]{KT02,TK02,2002MNRAS.337L...6T,
2003ApJ...583..606K,Treu04} and more recently by the SLACS survey,
the BELLS Survey \citep{Bol++12,Brownstein12}, the SL2S Survey
\citep{Ruff11,Gavazzi12}, and the SWELLS Survey \citep{Treu11}
for the case of spiral deflectors. Whereas the quality of the
kinematic profiles in general can not compete with that obtained
for local ETGs (\se{gaspoor}), the combination of these data with
strong lensing at higher redshifts has several major advantages.

First, even ``low quality'' lensing information combined with a single
measurement of the stellar velocity dispersion can often be obtained
out to $z\sim 1$ without major telescope investment. This allows their
inner mass profiles to be determined even at half the age of the
Universe. Hence, evolution (in the ensemble average properties)
of ETGs can be studied \citep[e.g.][]{Ruff11,Bol++12}. Second, the
additional use of strong lensing masses (plus constraints on its density
slope near the Einstein radius) helps break the mass-sheet and
mass-anisotropy degeneracies.

The results of the SLACS survey based on the combination of lensing
and kinematic constraints for $\sim60$ ETGs lenses are described in
a number of papers \citep{Koopmans06, 2008MNRAS.384..987C,
2009ApJ...703L..51K, 2009MNRAS.399...21B, 2010MNRAS.406.2339B,Auger10}.
Regarding the total mass density profile,
\citet{2009ApJ...703L..51K} and \citet{Auger10} found that inside one
effective radius massive elliptical galaxies with $M_{\rm eff} \ge 3\cdot
10^{10}$\,M$_{\odot}$ are well-approximated by a power-law ellipsoid
with an average logarithmic density slope of $\langle \gamma'_{\rm LD}
\rangle \equiv -d\log(\rho_{\rm tot})/d\log(r)=2.078\pm0.027$ (random
error on mean) with an intrinsic scatter of $0.16\pm0.02$ (for isotropic
orbits; results change as shown in \Fig{slopes} for reasonable amounts
of anisotropy).  Whereas this result is based on a separate analysis
of the lensing and stellar kinematics and assumes spherical symmetry
(or simple scaling relations), it has been confirmed by more
sophisticated joint and self consistent lensing and dynamical
analysis methods based on axisymmetric mass distributions and
two integral Schwarzschild modeling of the full lensing data
and two dimensional velocity fields
\citep{2008MNRAS.384..987C,2009MNRAS.399...21B}. Based on a subset
of 16 lens ETGs with deep integral field spectroscopy \citet{Barnabe11}
find $\langle \gamma'\rangle>=2.074^{+0.043}_{-0.041}$ with an intrinsic
scatter of $0.143^{+0.054}_{-0.014}$. Overall the internal structure
of the SLACS ETGs at $z=0.1-0.4$ is found to be perfectly consistent
with that found for their nearby counterparts, as described in
\se{gaspoor} of this review.

The addition of weak gravitational lensing data to the strong lensing
and dynamics analysis allows one to extend the measurement of the
total mass density profile well beyond the effective radii in an
ensemble sense. With exquisite HST data, the weak lensing signal is
measurable for sample of just a few tens of ETGs in the redshift range
$z=0.1-0.8$ \citep{Gavazzi07,Lagattuta10,Auger10}. The two main
results of the combined weak, strong and dynamics analyses are that
SLACS lenses have average virial mass $\sim 2\cdot 10^{13} M_\odot$
and that their {\it total} mass density profile is well described by a
single isothermal sphere $\gamma'\approx2$ out to $\sim100$ effective
radii. This result is remarkable because neither the stellar component
nor the dark matter halo are well described by single power laws, and
yet their sum is. This total mass profile is well reproduced by the
combination of a stellar component and a standard NFW halo \citep{Gavazzi07}
for sensible values of stellar mass to light ratio. This is very
different with what is found at higher and lower masses (clusters and
dwarfs) where typically a single isothermal sphere is not a good
description of the total mass density profile. The simplicity of the
total mass density profiles of ETGs has been sometimes called the
``bulge-halo'' conspiracy \citep{Dutton13}, and it
provides important constraints on theoretical models of ETGs
formation, especially on parameters that drive the star formation
efficiency like supernovae and nuclear feedback
\citep{Remus13,Dubois13}.

The conclusion that can be drawn from these analyses is that ETGs on
average have density profiles that are close to isothermal. However,
one needs to keep in mind that there is an intrinsic scatter of $\sim
10\%$ in the logarithmic density slope between galaxies (i.e.\ they do
not all have similar density profiles), which could be due to their
formation history.  This intrinsic scatter is comparable to studies of
nearby galaxies \citep[e.g.][]{gerhard01} based on stellar
kinematics alone. The only dependence on third parameters identified
so far is that between the slope and the stellar mass density inside
the effective radius, where higher stellar mass-density ETGs have
steeper density slopes \citep{Dutton13}. This tantalizing result is
confirmed by the self-consistent axisymmetric modeling technique
\citep{2007ApJ...666..726B,Barnabe12}, and proves that ETGs are
at least a two parameter family even when it comes to their internal
mass structure.

\subsubsection{The Stellar IMF and Dark Matter Fraction in ETGs}
\label{sec:DMfraction}

Strong lensing can constrain the mass inside the Einstein radius very
accurately.  In combination with the luminosity inside the Einstein
radius this yields a firm upper limit on the stellar mass-to-light
ratio inside that radius \cite[e.g.][]{Brewer12}. As
discussed in Chapter~2, given an optical infrared spectral energy
distribution, modern stellar population synthesis models are believed
to provide estimates of the stellar mass to light ratio that are
accurate to within roughly for old stellar populations like the one
found in massive ETGs. In this case, the main source of uncertainty
is the shape of the stellar initial mass function, which is needed
to convert the observed luminosity - dominated by a small range of
stellar masses - to the total mass in stars and stellar remnants.
Thus, by combining gravitational lensing, stellar kinematics and
stellar population synthesis modeling, powerful new constraints
can be set on the stellar IMF and the fraction of dark matter
in the inner regions of ETGs. By means of additional information,
like spatially resolved kinematics and/or simple assumptions on
the functional form of the dark matter density profile, one can
break in part the degeneracy between the stellar IMF and the dark
matter fraction and derive very realistic limits on the either
one \citep[e.g.][for limits on the dark matter fraction inside
the Einstein Radius out to $z\sim1$]{Treu04}.

\citet{Treu10} studied the stellar initial mass function
of ETGs by comparing the stellar mass fraction inside one effective
radius determined solely from lensing and stellar dynamics with that
inferred from stellar population synthesis models. Whereas these
limits are rather weak on a system-to-system basis, the combination of
56 SLACS ETGs allows a rather detailed comparison.  The main result is
that bottom-heavy IMFs such as that measured by \citet{Salpeter55} are
strongly preferred over light-weight IMFs such as that proposed by
\citet{chabrier03}, assuming standard NFW dark matter density
profiles. This result is further strengthened by \cite{Auger10} who
modeled these systems in detail, including adiabatic contraction and
weak-lensing constraints, and found that only heavy Salpeter-type IMF
are consistent with the observed properties of ETGs. In combination
with standard results based on spiral galaxy rotation curves (see
\se{gasrich}) and dynamical measurements of early-type galaxies (see
\se{gaspoor}), these results indicated the stellar initial mass
function cannot be universal.

The lensing and kinematic studies by themselves \citep{Treu10} also
suggest that the IMF normalization varies with galaxy mass \Fig{IMF}
within the sample of SLACS lenses, if NFW halos are allowed. However,
the mass dependency within the SLACS sample becomes insignificant if
the halos are allowed to contract in response to baryonic physics
\citep{Auger10}. 

Several subsequent studies, discussed elsewhere in his review, also
point toward non-universal IMFs using independent techniques.  For
example, based on detailed modeling of weak stellar absorption
features \cite{vanDokkum10}, confirmed the lensing result that
the IMF of massive early-type galaxies are inconsistent with
Chabrier. In addition, they provide the crucial suggestion that the
extra mass is to be attributed to low mass stars, with overall IMF
shape similar to Salpeter's. Detailed stellar dynamical modeling of
spatially resolved velocity fields of ETGs also adds important
information. First, it provides an independent confirmation of the
initial lensing results that Chabrier-like IMFs are disfavored for
massive ETGs. Second, the local galaxy samples cover a large enough
range in stellar mass to detect a trend in stellar mass-to-light
normalization within ETGs themselves, assuming the inner mass density
profiles of their dark matter halos can be modelled as power laws
which are allowed to vary within a fixed range across the sample
\citep{Cappellari+12,Cappellari13}. 

Overall there is good agreement between the dynamical, lensing
and stellar population probes \citep{DuttonMMS13}. Given its broad
implications, it is reassuring that many independent lines of evidence
\citep[e.g.][]{Zaritsky12} contradict the simple hypothesis of a
universal IMF, which has been a central tenet of extragalactic
astronomy for the few past decades. Much work is currently under
way to determine the exact form of the IMF, clarify systematic
uncertainties, and investigate possible variations with morphology
or other parameters \citep[e.g.][and elsewhere in this
review]{Dutton12,Conroy12,Conroy13,Sonnenfeld12,
Spiniello11,Spiniello12,Spiniello13,Smith13,Ferreras13,Goudfrooij13}.

\subsubsection{Mass Substructure in ETGs}

Whereas the results described above are concerned with the smooth mass
distributions of ETGs, gravitational lensing can also measure the
level of mass-density fluctuations, and in particular on the amount
of substructure in their inner regions. Thus, gravitational lensing
provides an opportunity to measure directly the mass function of
subhalos, irrespective of their stellar content. This is a stringent
test of the nature of dark matter, since cold dark matter predicts
that the subhalo mass function should go as $dN/nM\propto M^{-1.9}$
down to very small masses \citep{Springel+08}.

The first lensing studies of this topic, based on the so-called
flux-ratio anomalies \cite{1998MNRAS.295..587M,Metcalf01} of
radio-loud lenses quasars from the CLASS survey
\citep{2002ApJ...572...25D} indicated a level of substructure broadly
consistent with the expectations of CDM cosmology.  In the following
decade, much work has been devoted to understanding the systematic
uncertainties associated with this method
\citep[e.g.][]{Koopmans03,Kochanek04,Dobler06,Shin08,Metcalf12},
but overall progress has been limited mostly by the small number of
known quadruply imaged radio loud quasars. Detailed comparisons with
cosmological numerical simulations are challenging, owing to the need
for high resolution and approximations related to the implementation
of baryonic physics
\citep[e.g.][]{Kochanek04,Mao04,Maccio06,Xu09,Xu12}.
It also remains an open question whether some
of these anomalies are caused by dark or luminous substructure
\cite[e.g.][]{More08,More09,Jackson10,Nierenberg12}.

To overcome some of the limitations of flux-ratio anomaly systems
(e.g.\ the position and mass of the substructure are highly degenerate
so only statistical constraints can really be placed currently), and
exploit the large samples of galaxy-galaxy strong lens systems,
\citet{2005MNRAS.363.1136K} developed a new method based on the 
entire surface brightness distribution of the (extended) lens
source. The information contained in thousands of pixels allows one to
reconstruct the surface mass density of the deflector, pin-point the
position of possible substructures, and determine their masses (see
also \citep{2009MNRAS.392..945V} for a complete Bayesian extension of
this method). Alternative methods have been developed by other groups
to reconstruct on a grid the surface mass density of gravitational
lenses, or their gravitational potential, albeit mostly with the goal
of studying cosmology from gravitational time delays. The method
developed by \citet{Suyu09} is very similar in spirit to that of
\citet{2009MNRAS.392..945V}, while the one developed and applied
by \citet{Saha06} differs substantially. In the latter method the
mass distribution is reconstructed on a grid using only multiple
image positions as a constraints. Thus the amount of freedom in
the models is substantially larger and the choice of geometric
priors becomes more important. Putting this class of model into
a statistical framework is challenging although efforts are underway
\citep{Coles08}.

\begin{figure}
\centering
\includegraphics[scale=0.45]{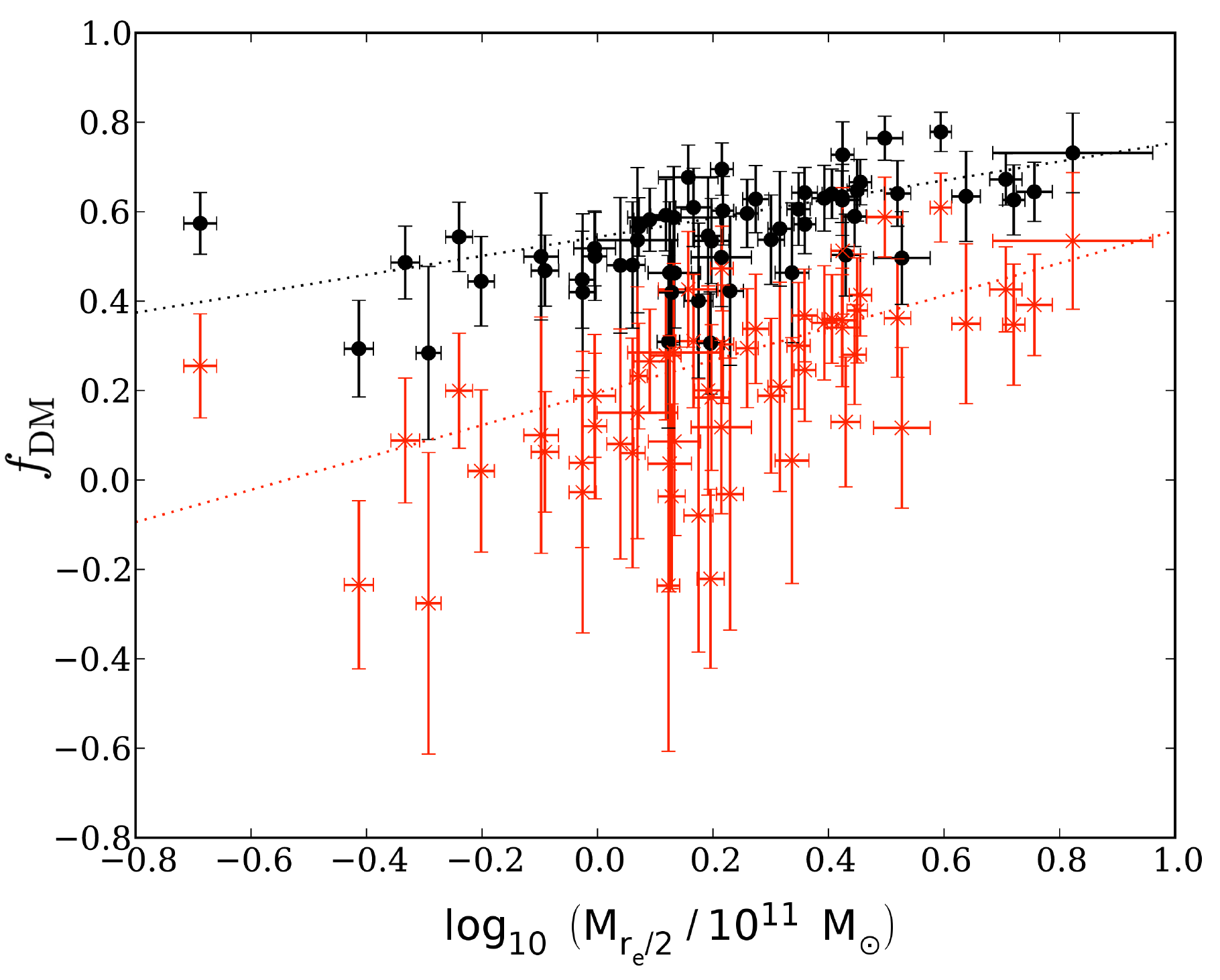}
\caption{Relations between the projected dark matter fraction within
 half of the effective radius and $\mre$. Red points are for a
 Salpeter IMF and black points are for a Chabrier IMF.}
\label{fig:DM_fraction}
\end{figure}

\subsubsection{Luminous Dwarf Galaxies}

\begin{figure}[t]
\centering
\includegraphics[scale=1.4]{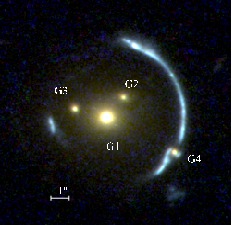}
\caption{Overview of the lens system the ``Clone''.
This false-colour image was created from HST/WFPC2 images
through filters F450W, F606W and F814W.}
\label{fig:cloneHST}
\end{figure}

\begin{figure*}
\centering
\includegraphics[scale=0.8]{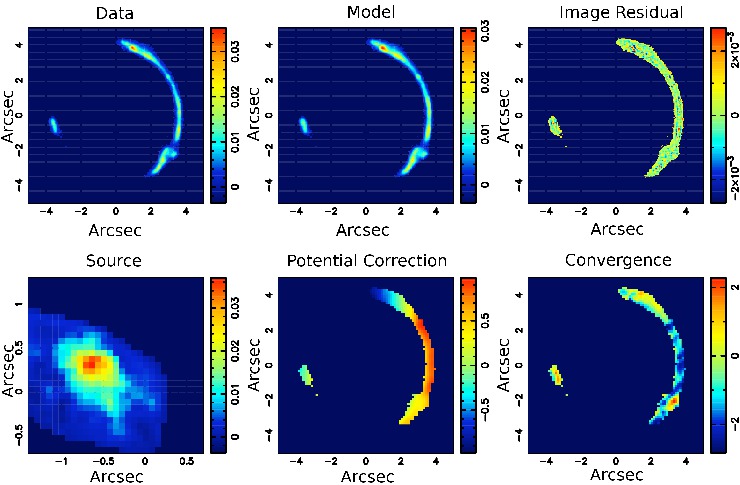}
\caption{Illustration of pixelized reconstruction of the source surface
brightness and lens potential corrections of the lens system shown in
\Fig{cloneHST} using the gravitational imaging technique. The top-left
panel shows the data, consisting of the surface brightness of a highly
distorted lensed source. The top-middle panel shows the model surface
brightness, while the top-right one shows the image residuals (data-model).
The bottom-left panel shows the source surface brightness distribution
reconstructed in the source plane (i.e. after ``delensing''; note the
zoomed-in angular scale). The bottom-middle panel shows the corrections
to the gravitational potential, with respect to a smooth simply parameterized
mass distribution. The bottom-right panel shows the inferred convergence,
i.e., the projected surface mass distribution. Note the peak at the lower
right corner of the image corresponding to the satellite responsible
for the curvature in the arc.}
\label{fig:clonemodel}
\end{figure*}

A nice demonstration of these methodologies is provided by known
luminous substructures in gravitational lens systems. For example,
the system shown in \Fig{cloneHST} was discovered by \cite{Lin09} and
shows a bright arc with a dwarf galaxy (G4) splitting the giant arc on
sun-arcsec scale.  Whereas this anomaly of the arc is caused by the
dwarf galaxy, similar anomalies could in principle also be caused by
dark substructure and be used to reconstruct their mass and position.
Not all such cases however are as obvious as this case.  The best
reconstruction of the lensed arcs is shown in \Fig{clonemodel} (a
galaxy surface brightness model has been subtracted) and a grid-based
reconstruction of the potential and surface density (lower-right
panel) has been constructed \citep[e.g.][]{Vegetti10}.  A
high over-density is clearly visible at the position of the anomaly.
Replacing this object by a tidally-truncated pseudo-Jaffe mass model,
a mass of mass of M$_{\rm sub} = (2.75 \pm 0.04) \times 10^{10}$
M$_{\odot}$ inside its tidal radius of $r_{t} = 0.68$ arcsec is
found. This result is robust against changes in the lens model. The
satellite luminosity is $L_{\rm B}= (1.6 \pm 0.8) \times 10^{9}$
L$_{\odot}$, leading to a total mass-to-light ratio within the tidal
radius of $(M/L)_{\rm B} = (17.2 \pm 8.5)$ M/L$_{\odot}$/L$_{\odot}$.
While this mass-to-light ratio is high compared to early-type dwarfs,
it is also an upper limit since the extended emission is hard to
measure due to the arc. Another demonstration of the power of this
method is given by the analysis of the system SL2SJ08544-0121 by
\citet{Suyu10}.

\begin{figure*}
\centering
\includegraphics[scale=0.8]{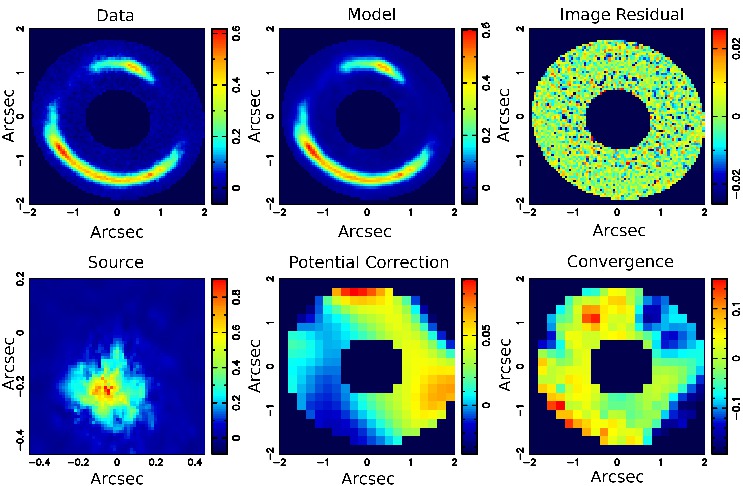}
\caption{Gravitational imaging analysis of the Jackpot gravitational
lens system. The panels are as in \Fig{clonemodel}. Note the
convergence peak in the top right portion of the bottom-right
panel, corresponding to the detected substructure.}
\label{fig:DRrecon}
\end{figure*}

\subsubsection{Dark Substructures}

The method is currently being applied to the SLACS lenses with the
goal of quantifying the abundance of substructures independent of
their luminosity. Two detections have been reported so far.

The first substructure detected via gravitational imaging is in the
Jackpot system \citep{Gavazzi08}, which shows two concentric
rings of sources at two redshifts. The inner ring of this system, even
though rather smooth, has a very high signal-to-noise ratio and is
therefore quite suitable for the grid-based analysis method.

A simply-parameterized elliptical power-law density model plus external
shear provides a good fit to the data, but for a rather structured source
model \citep{2010MNRAS.408.1969V}. \Fig{DRrecon} shows a reconstruction
of the system down to the noise level where the source is more smooth,
but a perturbation of the lensing potential is required at 4.3 kpc
projected distance from the lens center (the red feature in the
upper-left of the lower-right panel shows the corresponding over
density).

Whereas there is a tradeoff between the complexity of the source and
that of the lens potential, this can objectively be assessed through
the Bayesian evidence (i.e.\ the probability of the data when
marginalizing over the full posterior probability function) and the
smooth-source plus more complex lens model is preferred at a (rough)
equivalent of 16--$\sigma$ significance \citep{2010MNRAS.408.1969V}.

This detection is confirmed by modeling the substructure with a
tidally truncated pseudo-Jaffe density profile
\citep{2010MNRAS.408.1969V}. The substructure mass is
$M_{\rm{sub}}=(3.51\pm 0.15)\times 10^9\msun$. A lower limit of ${\rm
(M/L)}_{{\rm V},\odot}\ge 120~ \msun/{\rm L}_{{\rm V}\odot}$
(3--$\sigma$) is set inside a sphere of 0.3 kpc centred on the
substructure ($r_{\rm tidal}$=1.1\,kpc). This implies a projected dark
matter mass fraction in substructure at the radius of the inner
Einstein ring of $f=2.15^{+2.05}_{-1.25}$ percent (68\% C.L) in
the mass range $4\times10^6\msun$ to $4\times10^9\msun$, assuming
$\alpha=1.9\pm0.1$ (with $dN/dm\propto m^{-\alpha}$). Assuming
a flat prior on $\alpha$, between 1.0 and 3.0, increases this to
$f=2.56^{+3.26}_{-1.50}$ percent (68\% C.L). The likelihood ratio
is $\sim$0.5 between these fractions and that from simulations ($f
_{\rm{N-body}} \approx 0.003$). More recently, a second detection
has been reported in the CLASS gravitational lens system B1938+666 
based on HST and Keck adaptive optics images \citep{Vegetti12}.
Remarkably the satellite mass is only $2 \times 10^8 M_\odot$ and
yet it is detected at redshift 0.881.

The inference on the substructure mass function based on just two
systems are clearly very uncertain \citep{Vegetti12}, but so
far the results are broadly consistent with those expected from
numerical simulations. Effort is under way to refine those
measurements by applying the gravitational imaging technique to larger
samples of lenses (Vegetti et al. 2013, in preparation).  Proving
that the substructure is inside the ETG and not along the line of
sight is actually challenging \citep{Chen03,chenj11}.
At the moment the level of line of sight contamination is ill-constrained.

\newpage
\subsection{Future Prospects}

The method of strong gravitational lensing has progressed significantly
in the past decade, proving to be fundamental tool for precision
astrophysics and cosmology. Progress in the field has come from
new observations of unprecedently large samples of lensed systems,
a growing synergy with other techniques (stellar kinematic and stellar
population studies), and the development of new methodologies such
as self-consistent lensing and dynamics and grid-based strong lensing.
Whereas this research area is just too extensive for an exhaustive
review in this short a space, we have tried to illustrate the progress
and potential of strong lensing by highlighting some recent results from
the largest galaxy-scale strong lens survey to date, the SLACS survey.
These examples demonstrate that valuable constraints can be set on
the inner density profiles of ETGs as well as on their dark-matter mass
fraction as function of galaxy mass and cosmic time, their stellar
IMF and the level of mass substructure.

One obvious concern is that strong gravitational lenses are rare in
the sky (approximately fewer than 1/100-1/1000 massive ETGs can be
detected as strong lenses, depending on resolution and depth).
However, even in an era when exquisite data can be gathered for much
larger samples of non-lens galaxies, strong lensing still brings
unique and extremely precise measurements of mass (typically to a few
percent) which are independent of the standard assumptions and
uncertainties of other more traditional methods, as discussed at some
length in this review.  By combining strong gravitational lensing
information with that inferred from other methods one may break many
of the traditional degeneracies (e.g. mass-anisotropy, IMF versus
stellar mass) and achieve new insights into the formation and
evolution of early-type galaxies.
Furthermore strong lensing thrives at cosmological distances where
other methods suffer from the inevitable loss in sensitivity and
angular resolution.  For example, as discussed in this review,
only by using strong lensing information can one determine, as
a function of cosmic time accurate mass profiles, the normalization
of the stellar IMF and the abundance of dark substructures.
Thus, strong lensing is an essential tool for any evolutionary
study of the mass structure of ETGs.

Furthermore, the upcoming decade will see a revolution in the study
of strong gravitational lens systems. At the moment, most strong
lensing applications are limited by the number of known strong lens
systems suitable for that particular application.  The current
samples, limited to only a few hundred galaxy-scale lenses, are
insufficient to explore detailed trends in mass, redshifts, and
other potentially illuminating parameters. However, the current
and next generation of wide field sky surveys
(e.g.\ from Herschel-ALMA, DES, LSST, PanSTARRS, LOFAR, Euclid,
KiDS, SKA, etc.) will enable the discovery of 10$^3$-10$^5$ galaxy-scale
lens systems thus removing the limitations stemming from sample size
once and for all. With only limited resources for detailed follow-up
of individual sources at present, strong lenses will clearly become
a high-priority target given the high density of information
that they provide.

\newpage 

\section{Acknowledgments}
We are most grateful to Lia Athanassoula, Giuseppina Battaglia,
Matt Bershady, James Bullock, Laurent Chemin, Enrico Maria Corsini,
Nathan Deg, Ken Freeman, Dimitri Gadotti, Ortwin Gerhard, Stacy
McGaugh, Kristine Spekkens, and Piet van der Kruit for useful
discussions.  Laurent Chemin, Octavio Valenzuela and Giuseppina
Battaglia are also thanked for providing Figures
\ref{fig:chemintilt}, \ref{fig:Valenzuela}, and \ref{fig:battaglia}
respectively.  Cory Wagner very kindly improved the presentation
of various figures, and two diligent referees provided valuable
comments that improved various parts of this review.

SC and LMW acknowledge the support of the Natural Sciences and
Engineering Research Council of Canada through respective Discovery
grants.  MC acknowledges support from a Royal Society University
Research Fellowship.  AAD acknowledges support from the Canadian
Institute for Theoretical Astrophysics (CITA) National Fellows program.
HH and LVEK acknowledge support by NWO VIDI grants, while TT acknowledges
support from the Packard Foundation through a Packard Research Fellowship.

\newpage


\end{document}